\documentclass[12pt]{iopart}

\usepackage{graphicx}       
\usepackage{dcolumn}        
\usepackage{bm}             
\usepackage{amstext}
\usepackage{amssymb}
\usepackage{amsbsy}
\usepackage{amscd}
\usepackage{amsfonts}

\newcommand{\nn}{\nonumber}

\newcommand{\sig}{\sigma}

\newcommand{\eps}{\epsilon}

\newcommand{\Aout}{B^\text{(refl)}_{lm\omega}}
\newcommand{\Ain}{B^\text{(inc)}_{lm\omega}}
\newcommand{\Bout}{A^\text{(refl)}_{lm\omega}}
\newcommand{\Bin}{A^\text{(inc)}_{lm\omega}}
\newcommand{\Rout}{R^\text{(refl)}_{lm\omega}}
\newcommand{\Rin}{R^\text{(inc)}_{lm\omega}}

\newcommand{\Rinc}{B^{\text{(inc)}}_{lm\omega}}
\newcommand{\Rref}{B^{\text{(refl)}}_{lm\omega}}
\newcommand{\Rtrans}{B^{\text{(trans)}}_{lm\omega}} 

\newcommand{\Pin}{P^{\text{(inc)}}_{lm\omega}}
\newcommand{\Pout}{P^{\text{(refl)}}_{lm\omega}}

\newcommand{\Sspher}{{}_{-2}S_l^2}
\newcommand{\Yspher}{{}_{s}Y}
\newcommand{\rstar}{r_\ast}

\newcommand{\lc}{l_{\text{c}}}
\newcommand{\lm}{l_{\text{max}}}
\newcommand{\as}{a_\ast}
\newcommand{\astar}{{a_\ast}}
\newcommand{\aw}{z}

\newcommand{\diffop}{ \hat{\mathcal{L}}_x }
\newcommand{\diffopm}{ \hat{\mathcal{L}}_{-x} }
\newcommand{\alpco}{\alpha_{lm}^{(2)}}
\newcommand{\betco}{\beta_{lm}^{(2)}}

\begin{document}


 \title{Scattering and absorption of gravitational plane waves by rotating black holes}

\author{Sam R. Dolan}
 \address{
 School of Mathematical Sciences,
 University College Dublin, Belfield, Dublin 4, Ireland
}
\ead{sam.dolan@ucd.ie}

\date{\today}
 
\begin{abstract}
This is a study of the scattering and absorption of planar gravitational
waves by a Kerr black hole in vacuum. We apply
the partial wave method to compute cross sections for the special case of radiation incident along the
rotation axis. A catalogue of numerically-accurate cross sections is presented, for a range of
incident wavelengths $M\omega \le 4$ and rotation rates $a \le 0.999M$. Three effects are studied in detail: polarization, helicity-reversal and glory scattering. 
First, a new approximation to the polarization in the long-wavelength limit is derived. We show that black hole rotation distinguishes between co- and counter-rotating wave helicities, leading to a term in the cross section proportional to $a\omega$. Second, we confirm that helicity is not conserved by the scattering process, and show that superradiance amplifies the effect. For certain wavelengths, the 
back-scattered flux is enhanced by as much as $\sim 35$ times for a rapidly-rotating hole (e.g. for $a = 0.999M$ at $M\omega = 0.945$). Third, we observe regular glory and spiral
scattering peaks in the numerically-determined cross sections. We show that the angular width and intensity of the peaks may be estimated via a semi-classical approximation. 
We conclude with a discussion of the observable
implications of our results.



\end{abstract}

\maketitle

\section{\label{sec:introduction}Introduction}
Gravitational waves are propagating ripples in spacetime that are predicted by General Relativity (GR). There is strong indirect evidence for their existence, for example, from thirty-five years of pulsar timing measurements \cite{Lorimer-1998}. Due to the tiny amplitude of waves reaching Earth (with a dimensionless strain $h \sim 10^{-21}$), gravitational waves are yet to be detected directly. Many experimentalists are optimistic that first-light observations will occur within the decade. Ground-based detectors such as LIGO \cite{Frey-2007} are reaching sufficient sensitivities such that even non-detection will constrain astrophysical theories (for example, on the origin of gamma ray bursts \cite{LIGO-2007}). Plans for a space-based detector (LISA) are well advanced \cite{Danzmann-2003}.

Gravitational waves are of interest to astronomers because they are generated by some of the most energetic astrophysical processes, such as binary mergers, supernovae and galaxy collisions. Electromagnetic radiation carries relatively little information about the most energetic regions of such processes, because photons are strongly scattered, absorbed and thermalized by intervening matter. On the other hand, gravitational waves are only weakly coupled to matter, and carry information about the dynamics at the heart of such processes. 

Black holes are likely to be powerful sources of gravitational waves. For example, at the end of the inspiral phase, the merger of a pair of black holes creates a distinctive ``chirp'' signal. Rapid progress has been made in numerically modelling the merger process \cite{Pretorius-2005}, providing template signals for experimentalists seeking to separate a weak gravitational-wave signal from a noisy background. 

In addition to emitting radiation, black holes will scatter and absorb radiation that impinges upon them. To some extent, this is a secondary effect. However, scattering processes have long been of foundational interest to physicists, and this process is no exception. In this paper, we seek to answer the following question: how are gravitational plane waves scattered and absorbed by rotating (Kerr) black holes?  In our scenario, we assume a monochromatic wave produced by some far-distant source impinges on an isolated black hole. We assume the incident wave is essentially planar, long-lasting, and weak enough to be treated through the linearized approximation. Here, we consider only the special case of on-axis incidence. Our principle aim is to determine the absorption cross section $\sigma_a$ and the differential scattering cross section $d \sigma / d \Omega$ as a function of scattering angle $\theta$. Inevitably, the long-range $1/r$ ``Newtonian'' component of the potential leads to a divergent scattering cross section in the direction of incidence. In this respect, a black hole is no different to any generic massive source. More interestingly though, the signal scattered through large angles carries information about the near-horizon structure of the hole. In this paper we compute this signal and examine in detail the effect of black hole rotation.

Many studies of black hole scattering have been made over the last forty years. The most relevant for our purposes are studies of time-independent scattering, where it is assumed that the incident wave is monochromatic and long-lasting. Pioneering contributions include the works of Hildreth \cite{Hildreth-1964}, Matzner and co-workers \cite{Matzner-1968, Chrzanowski-1976, Matzner-1977, Matzner-1978, Handler-1980}, Sanchez \cite{Sanchez-1976, Sanchez-1977, Sanchez-1978a, Sanchez-1978b} and Mashhoon \cite{Mashhoon-1973,Mashhoon-1974, Mashhoon-1975}. Early studies focussed on the simplest setup: scalar (Klein-Gordon) waves incident on a non-rotating hole. Later works extend the analysis to higher spin: fermionic \cite{Dolan-2006}, electromagnetic \cite{Fabbri-1975} and gravitational \cite{Futterman-1981} fields. A number of authors have considered the polarizing effect of a rotating hole on waves with spin \cite{Mashhoon-1973, Mashhoon-1974, Mashhoon-1975, DeLogi-1977, Guadagnini-2002, Barbieri-2004, Barbieri-2005, Guadagnini-2008}. The physical origin of regular oscillations in cross-section-versus-scattering-angle plots was clarified by DeWitt-Morette \cite{DeWitt-Morette-1984}, Zhang \cite{Zhang-1984} and others \cite{Matzner-1985, Anninos-1992}. Such oscillations are a diffraction effect due to interference of null rays passing close to the unstable photon orbit. In other words, the large-angle scattering cross section contains information about the near-horizon structure of the hole. 

The efforts of Matzner and co-workers \cite{Matzner-1968, Matzner-1977, Matzner-1978, Handler-1980, Futterman-1981} culminated in the publication of a monograph \cite{Futterman-1988}, which set out in detail the partial-wave approach to computing black hole scattering cross sections. In this work, a range of plots of numerically-determined cross sections were presented. We believe that the plots for the rotating black hole are not accurate, due to numerical difficulties arising from the long-range nature of the gravitational field. Our belief is supported by more recent work by Glampedakis and Andersson \cite{Andersson-1995} and \cite{Glampedakis-2001}, who examined the scattering of scalar waves by a Kerr hole. A primary aim of this paper is to combine the numerical accuracy of Glampedakis and Andersson with the partial wave methods of Matzner \emph{et al.} for the gravitational wave case.

A further objective is to examine in detail the coupling between the rotation of the hole and the helicity of the incident wave. The black hole is expected to have a polarizing effect \cite{Mashhoon-1973}. A key result of this paper is that the polarizing effect is present even in the long-wavelength limit. A further observation is that parity-dependence of the scattering potential gives rise to a non-zero amplitude for helicity reversal. The helicity-reversed flux is maximal in the antipodal direction. We find that, close to the extremal rotation limit, the flux scattered in the backward direction can be significantly enhanced by the effects of superradiance.

This paper is structured as follows. In Section \ref{sec-background} we outline basic results for black hole wave scattering and quote the key analytic results of this work. In Section \ref{sec-partial-wave-method} we introduce the partial-wave method of Matzner \emph{et al.} (\ref{subsec-partialwave}), Teukolsky's equations (\ref{subsec-teuk}), and the spin-weighted spheroidal harmonics (\ref{subsec-spheroidal}). In Section \ref{sec-long-wavelength} we apply the partial wave method to derive a new approximation for the scattering cross section and polarization in the long wavelength regime. In the process, we derive asymptotic results for the phase shifts (\ref{subsec-mst}), the spheroidal harmonics (\ref{subsec-spheroidal-expansion}) and the scattering amplitudes (\ref{subsec-amplitudes}). Section \ref{sec-nummeth} describes methods for the numerical calculation of cross sections. The numerical task splits into three parts. First (\ref{subsec-num-spheroidal}) we compute the spin-weighted spheroidal harmonics using a spectral decomposition method. Second (\ref{subsec-phaseshifts}) we calculate phase shifts using a Sasaki-Nakamura transformation. Third (\ref{subsec-reduction}) we introduce a series reduction method for improving the convergence properties of partial wave series. In Section \ref{sec-results} we present numerical results for the absorption and scattering cross sections and the polarization. We conclude in Section \ref{sec-discussion} with a discussion of the physical implications of our results. Note that units $G = c = 1$ are employed throughout this paper.

\section{Scattering by Black Holes\label{sec-background}}
The scattering scenario is characterised by two dimensionless parameters,
\begin{equation}
M |\omega| = \pi r_S / \lambda 
,  \quad \quad 
\text{ and } 
\quad \quad 
\as = a / M = J / M^2
\end{equation}
The coupling $M\omega$ expresses the ratio of Schwarzschild horizon $r_S$ to incident wavelength $\lambda$, and $0 \le \as < 1$ is a measure of the rotation rate of hole. Here, $M$ and $J$ are the mass and angular momentum of the hole, and $\omega$ is the angular frequency of the incident wave. Note that the frequency $\omega$ may take either sign, depending on the polarization of the incident wave. Positive $\omega > 0$ corresponds to a circularly-polarized incident wave co-rotating with the black hole, whereas negative $\omega < 0$ corresponds to a counter-rotating circular polarization. Approximations for the scattering cross section are available in two limiting regimes, $M |\omega| \ll 1$ and $M |\omega| \gtrsim 1$. 


 \subsection{The Long-Wavelength Regime, $M |\omega| \ll 1$\label{subsec-approx}}
If the wavelength is much larger than the horizon size, $M |\omega| \ll 1$, then the black hole interaction may be treated through perturbative methods. Various authors have shown \cite{DeLogi-1977, Westervelt-1971, Peters-1976, Doran-2002, Dolan-2008} that, in the non-rotating case ($a=0$), the limiting cross section depends on the spin $s$ of the (massless) wave as follows,
\begin{eqnarray}
\lim_{M \omega \rightarrow 0} \, \left( \frac{1}{M^2} \, \frac{d \sigma}{d \Omega} \right)  &\approx  
\left\{ 
\begin{array}{ll} 
  \frac{1}{\sin^4(\theta / 2)}  & \quad s = 0   \\ 
  \frac{\cos^2(\theta /2)}{\sin^4(\theta / 2)}  & \quad s = 1/2 \\
  \frac{\cos^4(\theta /2)}{\sin^4(\theta / 2)}  & \quad s = 1   \\
  \frac{\cos^8(\theta /2)}{\sin^4(\theta / 2)} + \sin^4(\theta/2) & \quad s = 2   ,
\end{array} \right.
\label{csec-low-approx}
\end{eqnarray}
where $\theta$ is the scattering angle. Regardless of spin, all cross sections diverge in the same manner in the forward direction, $d \sigma / d \Omega \sim 16 M^2 / \theta^4$. The divergence is due to the long-range nature of the gravitational potential. Through semiclassical arguments, it may be argued that the divergence is a consequence of the weak-field Einstein deflection (i.~e.~ in the far-field, an incoming geodesic with an impact parameter $b$ is bent through an angle $\theta \approx 4M / b$).

It is interesting that all cross sections in (\ref{csec-low-approx}) obey the general rule $d\sig/d\Omega \sim M^2 \cos^{|4s|}(\theta/2) / \sin^4(\theta /2)$, except for the gravitational wave cross section \cite{Peters-1976, Dolan-2008}, which includes an extra term. The extra term is due to the parity-dependence of the interaction of the linearised wave with the black hole. Odd-parity (axial or Regge-Wheeler) and even-parity (polar or Zerilli) partial waves with the same $l$ pick up different phase shifts on scattering. This leads to the non-conservation of helicity in the scattering of a (linearized) gravitational wave.

If the black hole is rotating, then an unpolarized incident wave will acquire a partial polarization. It is shown in Section \ref{sec-long-wavelength} that the cross section for a long-wavelength gravitational wave impinging along the axis of incidence of a rotating hole is
\begin{equation}
\fl\quad\quad M^{-2} \frac{d \sigma}{d \Omega} \approx \frac{\cos^8 (\theta / 2)}{\sin^4 (\theta/2)}  \left[ 1 - 4 a \omega \sin^2(\theta/2) \right]   +  \frac{\sin^8 (\theta / 2)}{\sin^4 (\theta / 2)}  \left[ 1 + 4 a \omega \sin^2(\theta/2) \right].  \label{csec-final}
\end{equation}
The cross section (\ref{csec-final}) depends on the sign of $\omega$, so scattering from a rotating black hole induces a partial polarization,
\begin{equation}
\fl\quad\quad \mathcal{P} \equiv \frac{ {\frac{ d\sig}{ d\Omega }}{(\omega>0)} - {\frac{d\sig}{d\Omega}}{(\omega<0)} }
{{\frac{ d\sig}{ d\Omega }}{(\omega>0)} + {\frac{d\sig}{d\Omega}}{(\omega<0)}  } = - 4 a |\omega| \sin^2(\theta/2) \left( \frac{\cos^8(\theta/2) - \sin^8(\theta/2)}{\cos^8(\theta/2) + \sin^8(\theta/2)} \right).
\label{pol-final}
\end{equation}
In the small-angle limit, the polarization $\mathcal{P} \approx -  |\omega| (J/M) \theta^2$ is half that predicted by Guadagnini and Barbieri \cite{Guadagnini-2002, Barbieri-2004, Barbieri-2005, Guadagnini-2008} for scattering from classical rotating matter.

\subsection{Glory and Spiral Scattering, $M |\omega| \gtrsim 1$ \label{subsec-glory}}

When $M |\omega| \gtrsim 1$, interference effects arise in the scattering cross sections \cite{Matzner-1985}. Rays which pass in opposite senses around the black hole interfere to create constructive or destructive fringes. In the backward direction, a `glory' will arise. This is a phenomenon familiar from optical physics \cite{Laven-2005}. Strictly speaking, a glory is a bright spot in the backward-scattering direction ($\theta = \pi$). However, we will use the term more loosely to describe the interference pattern near $\theta \sim \pi$. Using semi-classical arguments, DeWitt-Morette, Zhang and others \cite{DeWitt-Morette-1984, Zhang-1984, Matzner-1985, Anninos-1992} derived an approximation for the scattering cross section close to the backward direction,
\begin{eqnarray}
\frac{d \sigma}{d \Omega} &\approx  2 \pi M \omega b_g^2 \left|\frac{d b}{d \theta} \right|_{\theta=\pi} {J_{2|s|}}^2(b_g \omega \sin \theta) ,
\label{glory-approx}
\end{eqnarray}
where $b$ is the impact parameter, and $b_g$ is the impact parameter for which the deflection angle is $\theta = \pi$. 
Here $J_{2|s|}$ is a Bessel function and $s$ is the spin of the field. It is worth noting that (\ref{glory-approx}) is zero in the backward direction if the perturbing field has spin (hence, a bright spot on-axis is absent). The gravitational-wave glory appears as a ring around $\theta = \pi$, with an angular radius proportional to $1 / M |\omega|$.
%
%
In addition to a glory ring, we expect to see ``spiral scattering'' oscillations at intermediate angles \cite{Anninos-1992}. These oscillations arise from the interference between overlapping classical paths that are scattered through the angles $\theta$, $2\pi - \theta$, $2 \pi + \theta$, etc. 

To extend the glory-scattering approximation (\ref{glory-approx}) to the rotating hole it is necessary to compute the glory impact parameter $b_g$ and the derivative $db/d\theta$ for scattering geodesics. In the Schwarzschild case, these quantities may be computed by numerically solving the orbit equations, or via the deflection-angle approximation of Darwin \cite{Darwin-1959}, 
\begin{equation}
\theta(b) \approx - \ln \left( \frac{ b - b_c }{ 3.48 M } \right) .  \label{eq-darwin}
\end{equation}
This approximation is valid for impact parameters $b$ sufficiently close the critical value $b = b_c$. We find the exact values are $b_g = 5.3570M$ and $b_g^2 |db/d\theta| = 4.896M^3 $ whereas the Darwin approximation gives $b_g = 5.3465M$ and $b_g^2 |db/d\theta| = 4.30 M^3$.

Let us now consider incoming geodesics that are initially parallel to the rotation axis of the hole. The deflection angle $\theta$ is determined by the integral equation \cite{Chandrasekhar-1983}
\begin{equation}
\int_0^{(\theta+\pi)/2} \frac{d\theta}{\sqrt{\Theta}} = - \int_{\infty}^{r_3} \frac{dr}{\sqrt{R}} \label{R-Theta}
\end{equation}
where
\begin{eqnarray}
R &= r^4 + (2a^2 - b^2) r^2 + 2b^2 r - a^2(b^2 - a^2) , \\ 
\Theta &= b^2 - a^2 \sin^2 \theta .
\end{eqnarray}
For $b > b_c$, the quartic $R$ has four distinct real roots $r_0 < r_1 < r_2 < r_3$, of which $r_3$ (the radius of closest approach) is the largest. For the critical case, $b = b_c$, the roots $r_2$ and $r_3$ coincide at the photon orbit radius ($r_c = r_2 = r_3$ and $dR/dr|_{r=r_c} = 0$). Explicitly, 
\begin{eqnarray}
b_c^2 &= \frac{1}{r_c^2 - a^2} \left( (27M^2-a^2)r_c^2 - 12 a^2 M r_c - (9a^2M^2 + a^4) \right), \label{bc-eq} \\
r_c &= v/3 + (3M^2-a^2)/v + M, \\
v^3 &= 27M (M^2-a^2) + 3a \sqrt{3a^4 + 54a^2M^2 - 81M^4}  .
\end{eqnarray}
In \ref{appendix-darwin} we derive a logarithmic approximation similar to (\ref{eq-darwin}) for the polar orbits that pass close to the critical radius. We find that
\begin{equation}
\theta \approx -C(a) \ln \left( \frac{b - b_c}{D(a) M} \right)   \label{log-approx}
\end{equation}
where $C(a)$ and $D(a)$ are dimensionless constants which depend on the rotation rate of the hole. Closed-form expressions (\ref{Ca-eq}) and (\ref{Da-eq}) for $C(a)$ and $D(a)$ are somewhat complicated; for convenience, numerical values are listed in Table \ref{tbl-rcbc} for a range of $\as$.

\begin{table}
\begin{tabular}{l|lllllll}
$\as = $ \quad \quad & $0.0$ & $0.25$ & $0.5$ & $0.7$ & $0.9$ & $0.99$ & $1.0$ \\
\hline
$r_c/M$ & $3.00000$ \quad & $2.97190$ \quad & $2.88322$ \quad & $2.75791$ \quad & $2.56000$ \quad & $2.43100$ \quad & $2.41421$ \\
$b_c/M$ & $5.19615$ & $5.17791$ & $5.12053$ & $5.04015$ & $4.91606$ & $4.83828$ & $4.82843$ \\
$C(a)$ & $1.000$ & $1.010$ & $1.042$ & $1.094$ & $1.200$ & $1.293$ & $1.307$ \\
$D(a)$ & $3.482$ & $3.461$ & $3.384$ & $3.247$ & $2.942$ & $2.665$ & $2.623$ 
\end{tabular}
\caption{Numerical values of the unstable orbit radius ($r_c$), the critical impact parameter ($b_c$) and the coefficients in the logarithmic approximation to the scattering angle ($C(a)$ and $D(a)$).}
\label{tbl-rcbc}
\end{table}

We may use the logarithmic approximation (\ref{log-approx}) to estimate $b_g$ and $db_g / d\theta$ in Eq. (\ref{glory-approx}). For $\as = 0.99$, the logarithmic approximation gives $b_g = 5.0609M$, $db_g/d\theta = -0.1722M$, and $b_g^2 |db_g/d\theta| = 4.4109M^3$ which may be compared with exact values $b_g = 5.0925M$, $db_g / d\theta = -0.2209M$ and $b_g^2 |db_g/d\theta| = 5.7286M^3$, obtained by solving (\ref{R-Theta}) numerically. Hence (\ref{log-approx}) is a reasonable but not precise approximation for our purposes. In Fig. \ref{fig-glory-approx} the semi-classical approximation is compared with exact cross sections calculated via the partial wave approach.

\section{The Partial Wave Method\label{sec-partial-wave-method}}

To compute absorption and scattering cross sections, we invoke the partial wave methods of Matzner \emph{et al.} \cite{Chrzanowski-1976, Matzner-1978, Handler-1980, Futterman-1988}. Since the full analysis is somewhat involved and is presented in full elsewhere \cite{Futterman-1988} we just recap the key results here. We restrict attention to the special case of a gravitational wave impinging along the rotation axis of a black hole. 

 \subsection{Partial Wave Series\label{subsec-partialwave}}
The scattering cross section is found from the sum of the square of two amplitudes,
\begin{equation}
\frac{d \sigma}{d \Omega} = \left| f(\theta) \right|^2 + \left| g(\theta) \right|^2   \label{csec-defn}
\end{equation}
The amplitudes may be expressed as partial wave series,
\begin{eqnarray}
\fl\quad\quad f(\theta) &= \frac{\pi}{i \omega} \sum_{P=\pm1} \sum_{l=2}^\infty \, \Sspher(0; a\omega) \, \Sspher(\theta; a\omega) \left[ \exp \left(2 i \delta_{l2\omega}^P \right) - 1 \right] ,  \label{f-def} \\
\fl\quad\quad g(\theta) &= \frac{\pi}{i \omega} \sum_{P=\pm1} \sum_{l=2}^\infty \, P (-1)^{l} \, \Sspher(0; a \omega) \, \Sspher(\pi - \theta; a\omega) \left[ \exp \left(2 i \delta_{l2\omega}^P \right) - 1 \right] . \label{g-def}
\end{eqnarray}
In these expressions, $\exp(2i\delta_{lm\omega}^\pm )$ are phase factors which must be determined from a radial equation, $\Sspher(\theta; a \omega)$ are spin-weighted spheroidal harmonics \cite{Berti-2005}, and $\theta$ is the scattering angle. Note the presence of a sum over parities, $P = \pm 1$.

In the far-field limit, the metric perturbation $h^{\text{(up)}}_{\mu \nu}$ expressed in the outgoing transverse-traceless gauge takes the simple asymptotic form \cite{Chrzanowski-1976}
\begin{equation}
h^{\text{(up)}}_{\mu \nu} \bar{m}^{\mu} \bar{m}^{\nu} \sim \frac{f(\theta) e^{i \omega (\rstar - t)} e^{2i \phi}}{r}  +  \frac{g^\ast(\theta) e^{-i \omega (\rstar - t)} e^{-2i \phi}}{r}  ,
\end{equation}
where $\rstar$ is the radial tortoise coordinate defined in (\ref{tortoise}), and $m^\nu$ is a (complex) null vector lying on the surface of a unit sphere. $f(\theta)$ represents a helicity-conserving amplitude, whereas $g(\theta)$ represents a helicity-reversing amplitude. The helicity-reversing amplitude is non-zero in general, because the phase shifts depend on parity $P$.

The absorption cross section $\sigma_a$ may also be computed from the phase shifts. It is given by a sum over angular modes,
\begin{equation}
\sigma_a = \frac{4 \pi^2}{\omega^2} \sum_{l=2}^\infty \left| \Sspher(0; a\omega) \right|^2 \mathbb{T}_{l2}, \label{abs-csec}
\end{equation}
where $0 \le \mathbb{T}_{lm} \le 1$ is the transmission factor for a given mode,
\begin{equation}
\mathbb{T}_{lm} = 1 - \left|e^{2i \delta_{lm\omega}^{\pm}}\right|^2 .  \label{T-trans}
\end{equation}
Hence absorption is related to the imaginary component of the phase shift $\delta_{lm\omega}^P$.

  \subsection{Teukolsky Equations\label{subsec-teuk}}
In the early 1970s, Teukolsky and Press \cite{Teukolsky-1972, Press-1973, Teukolsky-1973, Teukolsky-1974} applied the Newman-Penrose formalism \cite{Newman-1962} to derive perturbation equations for the Kerr black hole. Somewhat unexpectedly, they found that a complete separation of variables was possible for the Weyl scalars $\Psi_i$. For instance,
\begin{equation}
\rho^{-4} \Psi_4 = e^{-i \omega t} \, e^{im\phi} \, {}_{-2}S_l^m (\theta; a\omega) \, {}_{-2} R_{lm\omega}(r).
\end{equation}
With this ansatz, the Newman-Penrose equations decouple into radial and angular ODEs, 
\begin{equation}
\Delta^{-s} \frac{d}{dr} \left( \Delta^{s+1} \frac{dR}{dr} \right) + \left(\frac{K^2 - 2is(r - M) K}{\Delta} + 4is \omega r - \lambda_{lm} \right)R = 0 \label{radial-eq}
\end{equation}
and
\begin{eqnarray}
\frac{1}{\sin \theta} \frac{d}{d\theta} \left(\sin \theta \frac{d S}{d \theta} \right) + & \left( a^2 \omega^2 \cos^2 \theta - \frac{m^2}{\sin^2 \theta} - \frac{2 m s \cos \theta}{\sin^2 \theta} \right. \nn \\ 
& \left. \quad \quad -2a\omega s \cos \theta - s^2 \cot^2 \theta + s + A_{lm} \right) S = 0 \label{angular-eq}
\end{eqnarray}
where
\begin{eqnarray}
\Delta &= r^2 - 2Mr + a^2, \\
K &= (r^2 + a^2) \omega - a m,  \label{K-defn} \\
 \lambda_{lm} &= A_{lm} + a^2 \omega^2 - 2am\omega, 
\end{eqnarray}
and $s = -2$ in this case. 

The angular equation (\ref{angular-eq}), together with boundary conditions of regularity at $\theta= 0$ and $\theta = \pi$ constitutes a Sturm-Liouville eigenvalue problem. Hence the eigenfunctions --- the spin-weighted spheroidal functions that appear in (\ref{f-def}) and (\ref{g-def}) --- form a complete basis. A method for their calculation is outlined in Section \ref{subsec-spheroidal}.

The Kerr black hole has two horizons, at $r_\pm = M \pm \sqrt{M^2 - a^2}$, where $\Delta=0$. The $s = -2$ radial equation is solved subject to the condition that the radiation flux is purely ingoing at the outer horizon $r_+$, 
\begin{equation}
\lim_{r \rightarrow r_+} \, {}_{-2} R_{lm\omega} \sim \Delta^{2} \, \exp(-i (\omega - m \Omega_h) \rstar)  ,  \quad \text{ where } \quad \Omega_h = a / (2 M r_+) ,
\end{equation}
and $\rstar$ is the tortoise coordinate defined by
\begin{equation}
\frac{d \rstar}{d r} = \frac{(r^2 + a^2)}{\Delta} \label{tortoise}.
\end{equation}
Towards spatial infinity, there are two linearly-independent solutions, $\Rin$ and $\Rout$. To lowest order, 
\begin{equation}
{}_{-2} R_{lm\omega} \sim \left\{ \begin{array}{ll} \Rin &= r^{-1} \, \exp(- i\omega \rstar) \left[1 + \mathcal{O}(r^{-1}) \right],  \\
 \Rout &= r^{3} \, \exp( i \omega \rstar ) \left[1 + \mathcal{O}(r^{-1}) \right]. \end{array} \right.    \label{Rinout}
\end{equation}
Hence the ingoing solution obeys the boundary conditions 
\begin{equation}
{}_{s}R_{lm\omega}(r) \sim \left\{ \begin{array}{l l} \Rtrans \Delta^{-s} e^{- i (\omega - m \Omega_h) \rstar} & \rstar \rightarrow -\infty  \\ \Rinc r^{-1} e^{-i\omega \rstar}  + \Rref r^{3} e^{+i\omega \rstar}  & \rstar \rightarrow +\infty  \end{array} \right.  .
\label{teuk-bc}
\end{equation}
where $\Rtrans$, $\Rinc$ and $\Rref$ are complex constants. The phase shifts appearing in (\ref{f-def}) and (\ref{g-def}) are defined by \cite{Matzner-1978, Futterman-1988}
\begin{equation}
\exp\left( 2 i \delta_{lm\omega}^P \right) = (-1)^{l+1}  \left( \frac{\text{Re}(C) + 12iM\omega P}{16 \omega^4} \right) \frac{\Rref}{\Rinc}  \label{eq-phaseshift1}
\end{equation}
where
\begin{eqnarray}
\left[\text{Re}(C)\right]^2 = & ((\lambda+2)^2 + 4am\omega - 4a^2 \omega^2) \left[\lambda^2 + 36am\omega - 36a^2 \omega^2 \right] \nn \\ & \quad + (2\lambda + 3)(96a^2\omega^2 - 48a\omega m) - 144 \omega^2 a^2.
\end{eqnarray}
Note the explicit parity-dependence in (\ref{eq-phaseshift1}).

A method for finding analytic solutions to Teukolsky's radial equation (\ref{radial-eq}) in the long-wavelength regime ($M\omega \ll 1$) is described in section \ref{sec-long-wavelength}. Outside this regime, we will use numerical methods.
In principle, to compute phase shifts all we need to do is integrate the radial equation out from near the horizon to large $r$, where we may read off the coefficients $\Ain$ and $\Aout$. In practice, it is not so straightfoward, since the `peeling' behaviour \cite{Newman-1962} --- the factor of $r^4$ difference between ingoing and outgoing asymptotics --- causes great numerical difficulties. A resolution to this problem is given in Section \ref{sec-nummeth}. 

\subsection{Spheroidal Harmonics\label{subsec-spheroidal}}
A range of methods \cite{Berti-2005} exist for solving the angular equation (\ref{angular-eq}). For instance, to compute the eigenvalues $\lambda_{lm}$ we might employ either small-$a\omega$ \cite{Seidel-1989} or large-$a\omega$ \cite{Casals-2004} approximations, a continued-fraction method \cite{Leaver-1985}, or a finite difference method \cite{Sasaki-1982}. Here we make use of a spectral decomposition approach described in Appendix A of \cite{Hughes-2000}. 

Following the spectral decomposition approach, each spheroidal harmonic is expressed as a sum of spherical harmonics,
\begin{equation}
{}_{s} S_l^m(\theta; a\omega) = \sum_{j = \text{max}(|m|,|s|)}^\infty b_{j}^{(l)} \, \Yspher_{j}^m(\theta) ,  \label{S-expansion}
\end{equation}
where $b_{j}^{(l)}$ are expansion coefficients to be determined. Note we have suppressed the azimuthal dependence of the spherical harmonics $\Yspher_j^m(\theta)$. The spherical harmonics satisfy the equation
\begin{equation}
\frac{1}{\sin \theta} \frac{d}{d\theta} \left(\sin \theta \frac{d \Yspher_j^m}{d \theta} \right) + \left[ j(j+1) - \frac{m^2 + 2ms \cos\theta + s^2}{\sin^2 \theta} \right] \Yspher_j^m =0.
\end{equation}
Hence, by substituting ansatz (\ref{S-expansion}) into the angular equation (\ref{angular-eq}) one gets
\begin{equation}
\fl\quad\quad
\sum_{j = \text{max}(|m|,|s|)} b_{j}^{(l)} \left[ a^2 \omega^2 \cos^2 \theta - 2 a \omega s \cos \theta - j(j+1) + E_{lm} \right] {}_{s}Y_j^m(\theta) = 0. \label{eq-subin}
\end{equation}
where $E_{lm} = A_{lm} + s(s+1)$. 
Now, we take advantage of the orthonormality of the spherical harmonics, $\int d\Omega \, {}_s\bar{Y}_{k}^{m}(\theta) {}_sY_{j}^{m}(\theta) = \delta_{kj}$. Multiplying (\ref{eq-subin}) by ${}_s\bar{Y}_{k}^{m}(\theta)$ and integrating yields
\begin{eqnarray}
\fl \left[a^2 \omega^2 c^{(2)}_{k, k-2} \right]  b_{k-2}^{(l)} + \left[a^2 \omega^2 c^{(2)}_{k, k-1} - 2 a \omega s c^{(1)}_{k, k-1} \right] b_{k-1}^{(l)} + \left[ a^2 \omega^2 c_{kk}^{(2)} - 2 a \omega s c_{kk}^{(1)} - k(k+1) \right] b_{k}^{(l)} \nn \\
+ \, \left[a^2 \omega^2 c^{(2)}_{k, k+1} - 2 a \omega s c^{(1)}_{k, k+1} \right] b_{k+1}^{(l)}  + \left[a^2 \omega^2 c^{(2)}_{k, k+2} \right] b_{k+2}^{(l)} \quad = - E_{lm}  b_{k}^{(l)}  \label{eq-quindiag}
\end{eqnarray}
where
\begin{eqnarray}
\fl c_{k j}^{(2)} = \int d\Omega \, {}_s{\bar{Y}}_k^m (\theta) \cos^2 \theta  \, {}_sY_j^m(\theta) &= \frac{\delta_{kj}}{3}  + \frac{2}{3} \sqrt{\frac{2j+1}{2k+1}} \left< j,2,m,0 | k,m \right> \left< j, 2, -s, 0 | k, -s \right> ,   \\
\fl c_{k j}^{(1)} = \int d\Omega \, {}_s{\bar{Y}}_k^m (\theta) \cos \theta \, {}_sY_j^m(\theta)  &= \sqrt{\frac{2j+1}{2k+1}} \left<j, 1, m, 0| k, m\right>\left< j, 1, -s, 0 | k, -s \right> .
\end{eqnarray}
The numbers $\left< j_1, j_2, m_1 , m_2 | j, m \right>$ are Clebsch-Gordan coefficients \cite{Abramowitz-1965}. 

Equation (\ref{eq-quindiag}) can be rewritten as a matrix equation, with the numbers $b_{j}^{(l)}$ representing the coefficients of the matrix's eigenvectors, and $-(A_{lm} + s(s+1))$ representing the matrix's eigenvalues. The matrix is band-diagonal, so the task of finding the eigenvalues and eigenvectors is straightforward and numerically efficient. 

To calculate the spheroidal harmonics from (\ref{S-expansion}), we need a method for computing the spherical harmonics of spin-weight $s = -2$. A suitable recursive method is described in \ref{appendix-spherical-harmonics}.


\section{Polarization of Long-Wavelength Waves\label{sec-long-wavelength}}

The influence of rotation on the scattering of long-wavelength waves ($M|\omega| \ll 1$) has been considered by a number of authors \cite{Matzner-1977, Mashhoon-1973, Mashhoon-1974, Mashhoon-1975, DeLogi-1977, Guadagnini-2002, Barbieri-2004, Barbieri-2005, Guadagnini-2008}. It seems that no clear consensus has yet emerged. Considering electromagnetic radiation, Mashhoon \cite{Mashhoon-1973} noted that ``one must expect partial polarization in the scattered light when an unpolarized wave is incident on a Kerr black hole", and went on to observe that ``the polarizing property of a Kerr black hole is probably maintained for very low frequencies". However, the corollary that incoming gravitational waves will also be polarized at low frequencies was thrown into doubt by a study \cite{DeLogi-1977} in which the scattering amplitude was computed using Feynman-diagram techniques. The authors concluded that ``the angular momentum of the scatterer has no polarizing effect on incident, unpolarized gravitational waves" (even though they found that unpolarized electromagnetic waves \emph{were} polarized). Some doubts about the gauge-invariance of the results in \cite{DeLogi-1977} have been raised. To clarify the issue, the Feynman-diagram approach was recently revisited and improved in \cite{Guadagnini-2008}. In a series of papers, Guadagnini and Barbieri \cite{Guadagnini-2002, Barbieri-2004, Barbieri-2005, Guadagnini-2008} have argued that the net polarisation $\mathcal{P}$ induced by rotating classical matter is
\begin{equation}
\mathcal{P} \equiv \frac{ {\frac{ d\sig}{ d\Omega }}{(\omega>0)} - {\frac{d\sig}{d\Omega}}{(\omega<0)} }
{{\frac{ d\sig}{ d\Omega }}{(\omega>0)} + {\frac{d\sig}{d\Omega}}{(\omega<0)}   }  \approx  - |s| \omega (J / M) \theta^2   \label{pol-guadagnini}
\end{equation}
where $\theta$ is the scattering angle, which is assumed to be small.

In this section, we apply the partial wave method described in the preceding section to derive a long-wavelength approximation to the polarization of gravitational waves impinging along the axis of a Kerr black hole. The key results of this analysis were presented in Eq. (\ref{csec-final}) and (\ref{pol-final}). Curiously, we find the polarization at small angles to be exactly half that given by (\ref{pol-guadagnini}).  

Our objective is to compute the scattering amplitudes $f$ (\ref{f-def}) and $g$ (\ref{g-def}) to lowest order in $M \omega$ and $a\omega$. In \cite{Dolan-2008} it was shown that, to lowest order in $M\omega$, the Schwarzschild ($a=0$) amplitudes may be written as
\begin{equation}
f_{\text{Schw}} (x) = M e^{i \Phi} \frac{\Gamma(1-i\epsilon)}{\Gamma(1 + i\epsilon)} \frac{ [\frac{1}{2} (1+x)]^2 }{[ \frac{1}{2} (1 - x)]^{1-i\epsilon}}   \label{f-Schw}
\end{equation}
and
\begin{equation}
g_{\text{Schw}} (x) = M e^{i \Phi} \frac{\Gamma(1-i\epsilon)}{\Gamma(1 + i\epsilon)} [\frac{1}{2} (1-x)]
\label{g-Schw}
\end{equation}
where $\Phi = 2\eps \ln | 2\eps |$, $x = \cos\theta$, $\eps = 2M\omega$ and $\Gamma(y)$ is the Gamma function. In the following analysis, we show that black hole rotation ($\as > 0$) introduces a new term in each amplitude which is proportional to $a \omega$. 

\subsection{Phase shifts via the Mano-Suzuki-Takasugi method\label{subsec-mst}}
Low-frequency analytic results for the phase shifts (\ref{eq-phaseshift1}) may be obtained from the formalism developed by Mano, Suzuki and Takasugi \cite{MST, MT} (MST). In the MST approach, reviewed in \cite{Sasaki-2003}, solutions to the Teukolsky radial equation satisfying boundary conditions (\ref{teuk-bc}) are expressed as infinite series of special functions. Two series are used. The `horizon' series of ${}_2F_1$ hypergeometric functions is convergent at all radii up to (but not including) spatial infinity. The `far-field' series of Coulomb wavefunctions is convergent up to (but not including) the outer horizon. The complex constants  $\Rinc$, $\Rref$ and $\Rtrans$ defined in (\ref{teuk-bc}) are determined by matching the two series. It is found that
\begin{eqnarray}
\Rinc &= A_s e^{i \epsilon \kappa} \omega^{-1} \left[ K_\nu(s) - i e^{-i \pi \nu} \frac{\sin \left(\pi(\nu-s+i\eps) \right)}{\sin \left(\pi(\nu+s-i\eps) \right)} K_{-\nu-1}(s) \right] A_+^\nu , \\
\Rref &= A_s e^{i \epsilon \kappa} \omega^{-1-2s} \left[  K_\nu(s)  + i e^{i \pi \nu} K_{-\nu-1}(s) \right] A_-^\nu .
\end{eqnarray}
Here $A_s$ is just a normalization constant, and we follow the conventions of \cite{MST} by defining
\begin{equation}
\eps = 2M\omega \quad \quad \text{ and  } \quad \quad \kappa = \sqrt{1- \astar^2}.
\end{equation}
Note that $\epsilon$ must be positive in the MST expressions; the $\omega < 0$ results may be found via the symmetry of the radial function ${}_sR_{lm\omega} = {}_sR_{l-m-\omega}^\ast$ from which it follows that $B^{\text{(inc/refl)}}_{lm -\omega} = B^{\text{(inc/refl)} \ast}_{l -m \omega}$.

The parameter $\nu$ is known as the ``renormalized angular momentum'' and has the low-frequency expansion
\begin{equation}
\fl  \nu = l  +  \frac{\eps^2 }{2l + 1} \left[ -2 - \frac{s^2}{l (l+1)} + \frac{\left[ (l+1)^2 - s^2 \right]^2}{(2l+1)(2l+3)(2l+3)} - \frac{(l^2 - s^2)^2}{(2l-1)(2l)(2l+1)} \right] + \mathcal{O}(\eps^3).
\end{equation}
Note the absence of a linear term in $\epsilon$. 

The coefficients $K_\nu$ and $K_{-\nu-1}$ may be computed via a complicated series expansion, detailed in \cite{MST}. This is not necessary for our purposes, because $  K_{-\nu - 1} / K_{\nu}  \sim \mathcal{O}(\epsilon^{2l-1}) $. Hence $  K_{-\nu - 1}$ may be neglected, and the $K_\nu$ terms will cancel upon taking the ratio of $\Rinc$ and $\Rref$.

The coefficients $A_+^\nu$ and $A_-^\nu$ are given by 
\begin{eqnarray}
A_+^{\nu} &= 2^{-1+s} \eps^{-i\eps} e^{i (\pi/2) (\nu + 1 - s)} e^{-\pi \eps / 2} \frac{\Gamma(\nu + 1 - s + i\eps)}{\Gamma(\nu + 1 + s -i\eps)} \sum_{n=-\infty}^{\infty} a_n^\nu (s) ,  \label{Ap-eq} \\
A_-^{\nu} &= 2^{-1-s} \eps^{+i \eps} e^{-i (\pi/2) (\nu + 1 + s)} e^{-\pi \eps / 2} \sum_{n=-\infty}^{\infty} (-1)^n \frac{ (\nu + 1 + s - i\eps)_n }{ (\nu + 1 -s + i \eps)_n } a_n^\nu (s), 
\label{Am-eq}
\end{eqnarray}
where $a_n^\nu \sim \mathcal{O}( \eps^{|n|} )$. Here, we wish to conduct an expansion accurate to second order in $M\omega$; hence we require explicit formulae for $a_{-2}^\nu,  \ldots, a_{2}^\nu$. These were computed in \cite{MST} and are listed in \ref{appendix-mst}. 

After inserting equations (\ref{a-nu-m2}--\ref{a-nu-2}) into (\ref{eq-phaseshift1}), and taking some care with series expansions in the small parameter $\eps$, we find
\begin{eqnarray}
\fl\quad\quad  \exp({2 i \delta_{lm\omega}^-}) = e^{2i\eps \ln 2\eps} e^{-i \eps \kappa} e^{- i \pi (\nu - l)} \frac{\Gamma(l + 1 - i\eps)}{\Gamma(l + 1 + i\eps)} e^{4 i \eps / l(l+1)} \left[ 1 +  \alpco \eps^2  + \mathcal{O}(\eps^3) \right]  \label{phase-shift-negative}
\end{eqnarray}
for positive $\omega$. 
Note that there is no first-order term in the square brackets; the linear terms cancel exactly. The second-order coefficient is
\begin{equation}
\alpco = - \frac{i m \astar}{l(l+1)} - \frac{ 12 i m \astar }{(l-1) l^2 (l+1)^2 (l+2)} ,
\end{equation}
where $m$ is the azimuthal number. Note that the $e^{-i \pi (\nu - l)}$ and $e^{4 i \eps / l(l+1)}$ factors also give an $l$-dependent contribution at second order in $\epsilon^2$. 

Via the symmetry $B^{\text{(ref/inc)}}_{lm -\omega} = B^{\text{(ref/inc)} \ast}_{l -m \omega}$ it follows that $\exp(2i \delta^{-}_{l m -\omega} ) = \exp(2i \delta^{-}_{l -m \omega})^\ast$. Hence the phase shift may be written more generally as 
\begin{eqnarray}
\fl\quad\quad \exp({2 i \delta_{lm\omega} ^-}) = e^{2i\eps \ln |2\eps|} e^{-i \eps \kappa} e^{- i \, \pi (\nu - l) \omega / |\omega| } \frac{\Gamma(l + 1 - i\eps)}{\Gamma(l + 1 + i\eps)} e^{4 i \eps / l(l+1)} \left[ 1 +  \alpco \eps^2  + \mathcal{O}(\eps^3) \right]  .  \label{phase-shift-negative-both}
\end{eqnarray}
This expression is valid for either sign of $\eps = 2M\omega$. 

In a previous study \cite{Dolan-2008}, the low-$M\omega$ approximations of Poisson and Sasaki \cite{Poisson-1995} were used to show that the Schwarzschild ($\astar = 0$) phase shifts are
\begin{eqnarray}
\exp( 2 i \delta_{lm\omega}^- )= e^{2i\eps \ln 2\eps} e^{-i \eps} \frac{\Gamma(l + 1 - i\eps)}{\Gamma(l + 1 + i\eps)} e^{4 i \eps / l(l+1)} \left[ 1 +  \mathcal{O}(\eps^2) \right] .  \label{schw-phase}
\end{eqnarray}
This is consistent with (\ref{phase-shift-negative}) since $\kappa = 1$ in the non-rotating case. Moreover, it is remarkable that, when $\astar = 0$, equation (\ref{schw-phase}) actually holds to one order higher in $\eps$ than first supposed. 

\subsection{Series Expansion of Spheroidal Harmonics\label{subsec-spheroidal-expansion}}

In order to compute low-frequency scattering amplitudes (\ref{f-def}, \ref{g-def}) we must first expand the spheroidal harmonics to second order in the spheroidicity parameter $z = a\omega$. Noting that $b_{l \pm n}^{(l)} \sim \mathcal{O}( \aw^n )$, let us make the expansion
\begin{eqnarray}
b_{l}^{(l)}    &= 1 + \aw^2 d_{0}^{(2)} ,  &   \\ 
b_{l-1}^{(l)} &=  \aw d_{-1}^{(0)} + \aw^2 d_{-1}^{(1)} , \quad \quad & 
b_{l+1}^{(l)} =  \aw d_{+1}^{(0)} + \aw^2 d_{+1}^{(1)} , \\ 
b_{l-2}^{(l)} &=  \aw^2 d_{-2}^{(0)} ,  & 
b_{l+2}^{(l)} =  \aw^2 d_{+2}^{(0)} .  \label{bdef}
\end{eqnarray}
The normalisation condition implies that $(d_0)^{2} = -\frac{1}{2}\left[ (d_{-1}^{(0)} )^2 + (d_{+1}^{(0)})^2 \right]$. The remaining six unknowns $\{ d_{-2}^{(0)},  d_{-1}^{(0)}, d_{-1}^{(1)}, d_{0}^{(2)}, d_{+1}^{(0)}, d_{+1}^{(1)}, d_{+2}^{(0)}  \}$ are determined from the equations 
\begin{eqnarray}
(l+1) d_{+1}^{(0)} &= - s c_{l+1, l}^{(1)} ,  \label{deqs1} \\
l d_{-1}^{(0)} &=  s c_{l-1, l}^{(1)},   \\
2(l+1) d_{+1}^{(1)} &= (-2s c_{l+1,l+1}^{(1)} + E^{(1)}_{lm}) d_{+1}^{(0)} + c_{l+1,l}^{(2)} , \\
-2l d_{-1}^{(1)} &= (-2s c_{l-1,l-1}^{(1)} + E^{(1)}_{lm} ) d_{-1}^{(0)} + c_{l-1,l}^{(2)} , \\ 
2(2l+3) d_{+2}^{(0)} &= c_{l+2,l}^{(2)} - 2s c_{l+2, l+1}^{(1)} d_{+1}^{(0)} , \\
-2(2l-1) d_{-2}^{(0)} &= c_{l-2,l}^{(2)}  - 2s c_{l-2, l-1}^{(1)} d_{-1}^{(0)} ,   \label{deqs2}
\end{eqnarray}
where the angular eigenvalue has been expanded in powers of $\aw = a \omega$ as
\begin{equation}
E_{lm}  =  l (l + 1) + \sum_{k = 1}^\infty E_{lm}^{(k)} \aw^k 
\end{equation}
and the first few coefficients are determined by the identities
\begin{eqnarray}
E_{lm}^{(1)} - 2 s c_{ll}^{(1)} &= 0 \, , \\
E_{lm}^{(2)} + c_{ll}^{(2)} - 2s c_{l,l-1}^{(1)} d_{-1}^{(0)} - 2s c_{l, l+1}^{(1)} d_{+1}^{(0)} &= 0 \, , \\
E_{lm}^{(3)} + c_{l,l-1}^{(2)} d_{-1}^{(0)} + c_{l,l+1}^{(2)} d_{+1}^{0} - 2s c_{l,l-1}^{(1)} d_{-1}^{(1)} - 2s c_{l,l+1}^{(1)} d_{+1}^{(1)} &= 0 .
\end{eqnarray}

For this calculation, we need only the $m = 2$, $s = -2$ harmonics. The required Clebsch-Gordan coefficients are listed in \ref{appendix-mst}, (\ref{cleb1}--\ref{cleb2}). Substituting the Clebsch-Gordon coefficients into (\ref{deqs1}--\ref{deqs2}) yields explicit expressions for the expansion coefficients $\{ d_{-2}^{(0)},  d_{-1}^{(0)}, d_{-1}^{(1)}, d_{0}^{(2)}, d_{+1}^{(0)}, d_{+1}^{(1)}, d_{+2}^{(0)}  \}$, which are listed in \ref{appendix-mst}, (\ref{d-explicit-1}--\ref{d-explicit-2}). 

To compute the spheroidal harmonics explicitly we require expressions for the spin-weighted spherical harmonics ${}_{-2}Y_l^2(x)$, where $x = \cos \theta$. These may be found by acting on spherical harmonics of spin-weight zero, ${}_{0}Y_l^0(x) \equiv \sqrt{\frac{2l+1}{4 \pi}} \, P_l(x)$, 
with ladder operators \cite{Goldberg-1967}. The spin-weight is lowered with the operator $\check{\delta}$, and the azimuthal number is raised with $L^+$. These operators are defined by
\begin{eqnarray}
\fl\quad \check{\delta} \, {}_sY_l^m(x) = \left( \sqrt{1-x^2} \, \partial_x - \frac{m + s x}{\sqrt{1-x^2}} \right) {}_sY_l^m(x) = -\sqrt{(l+s)(l-s+1)} \, {}_{s-1}Y_l^m(x), \\
\fl\quad  L^+ \, {}_sY_l^m(x) = -\left(\sqrt{1-x^2} \, \partial_x + \frac{s + mx}{\sqrt{1-x^2}} \right) {}_sY_l^m(x) = \sqrt{(l-m)(l+m+1)} \, {}_sY_l^{m+1}(x).
\end{eqnarray}
Here, $\partial_x$ is shorthand for the partial derivative with respect to $x = \cos\theta$. 
By acting with $\check{\delta} L^+ \check{\delta} L^+$ on ${}_{0}Y_l^0(x)$, it is straightforward to show that the spin-weighted harmonics in (\ref{f-def}) and (\ref{g-def}) can be written
\begin{eqnarray}
{}_{-2}Y_l^2 (x) &= \sqrt{\frac{2l+1}{4 \pi}} \frac{ \diffop \, P_l(x) }{ (l-1)l(l+1)(l+2) }  ,  \\ 
  \diffop P_l(x)  &= (1+x)^2 \, \partial_x (1-x) \partial_x \partial_x (1-x) \partial_x P_l(x)  \label{spin-2}.
\end{eqnarray}
Their values in the forward and backward directions are particularly simple,
\begin{equation}
{}_{-2}Y_l^2(x=1) = \sqrt{\frac{2l + 1}{4 \pi}} \, , \quad \quad {}_{-2}Y_l^2(x=-1) = 0 .
\end{equation}
This implies that the values of the spheroidal harmonics in the forward direction are
\begin{equation}
{}_{-2}S_l^m(x=1)  =  \sqrt{\frac{2 l + 1}{4 \pi}} \left( 1 + \mathcal{S}_1 z  + \mathcal{S}_2 z^2 + \mathcal{O}(z^3) \right)  \label{forward-direction}
\end{equation}
where
\begin{eqnarray}
\mathcal{S}_1 &= \frac{8}{(l+1)^2 l^2} , \\
\mathcal{S}_2 &= \frac{-3375}{(2l+3)^2(2l-1)^2} + \frac{16}{l^2(l+1)^2} + \frac{192 (l^2 + l + 1)^2}{(l+1)^4 l^4} - \frac{32}{(l+1)^4 l^4} .
\end{eqnarray}

\subsection{\label{subsec-amplitudes}Scattering Amplitudes}
Let us now calculate the helicity-conserving and helicity-reversing amplitudes ($f$ and $g$) in the long-wavelength limit.
\subsubsection{Helicity-conserving amplitude.}
First, consider the helicity-conserving amplitude $f$ defined in (\ref{f-def}). To begin, note that the sum of the positive and negative-parity phase terms can be written
\begin{equation}
\frac{1}{2}\left( e^{2i\delta^-_{lm\omega}} +  e^{2i\delta^+_{lm\omega}}  \right) = e^{i\chi} \, \frac{\Gamma(l - 1 - i\eps)}{\Gamma(l + 3 + i\eps)} \frac{\Gamma(l + 3)}{\Gamma(l - 1)} \left[ 1 +  \betco \eps^2  + \mathcal{O}(\eps^3) \right] , \label{plus-minus}
\end{equation}
where $
e^{i\chi} =  e^{2i\eps \ln |2\eps|} e^{-i \eps \kappa}
$
and the second-order coefficient is
\begin{equation}
\betco =  - i \pi \, \frac{\omega}{|\omega|} \left( \frac{\nu - l}{\epsilon^2} \right) - i \frac{m \astar}{l(l+1)} + \frac{2}{l(l+1)} - \frac{15}{(2l+3)(2l-1)} .   
\end{equation}
Inserting the expansions of the spheroidal harmonics (\ref{S-expansion}, \ref{forward-direction}), and using (\ref{plus-minus}), we may write the $f$ amplitude as
\begin{equation}
f(x) = \frac{e^{i\chi}}{2 i \omega} \diffop F(x) ,  
\end{equation}
where
\begin{equation}
\fl\quad\quad F(x) = \sum_{l=2}^{\infty} (2l+1) \frac{\Gamma(l - 1 - i\epsilon)}{\Gamma(l+3+i\epsilon)} \left( 1 + \betco \eps^2 + \mathcal{O}(\eps^3) \right) \left(1 + \mathcal{S}_1 \aw +  \mathcal{S}_1 \aw^2 + \mathcal{O}(\aw^3) \right)  V_l , \label{f-F}
\end{equation}
and
\begin{eqnarray}
\fl \quad\quad V_l =&  \sqrt{ \frac{2l-3}{2l+1} }  \frac{(l+1)(l+2)}{(l-3)(l-2)} b_{l-2}^{(l)} P_{l-2}(x) +  \sqrt{ \frac{2l-1}{2l+1} } \frac{(l+2)}{(l-2)} b_{l-1}^{(l)} P_{l-1}(x) + b_l^{(l)} P_l(x) \nn  \\
 &+  \sqrt{ \frac{2l+3}{2l+1} }  \frac{(l-1)}{(l+3)} b_{l+1}^{(l)} P_{l+1}(x) + \sqrt{ \frac{2l+5}{2l+1} } \frac{(l-1)l}{(l+3)(l+4)} b_{l+2}^{(l)} P_{l+2}(x)
\end{eqnarray}
and $b_k^{(l)}$ are the expansion coefficients defined in (\ref{S-expansion}).

To compute the higher-order terms in the sum $F(x)$, we substitute in the explicit forms for the coefficients $b_k^{(l)}$ calculated in subsection \ref{subsec-spheroidal-expansion}. Next, we rewrite the sum so all terms have a common factor of $P_l$. To demonstrate this process, let us begin by considering just the linear term in $\omega$, which we denote $F_{(\omega)}$. We find
\begin{eqnarray}
\fl F_{(\omega)}(x)  &\fl\quad\quad\quad=  a \omega \sum_{l = 2}^{\infty} (2l+1) \frac{\Gamma(l - 1 - i\epsilon)}{\Gamma(l+3+i\epsilon)} \left( \sqrt{\frac{2l-1}{2l+1}} \frac{(l+2)}{(l-2)} d_{-1}^{(0)}(l) P_{l-1}(x) +  \mathcal{S}_1 P_l(x) \right. \nn \\ & \quad\quad\quad\quad\quad\quad\quad\quad\quad\quad\quad\quad \left. + \sqrt{\frac{2l+3}{2l+1}} \frac{(l-1)}{(l+3)} d_{+1}^{(0)}(l) P_{l+1}(x) \right) \nn \\ 
\fl &\fl\quad\quad\quad= a \omega \sum_{l = 2}^{\infty} (2l+1) \frac{\Gamma(l - 1 - i\epsilon)}{\Gamma(l+3+i\epsilon)} \left( \sqrt{\frac{2l+3}{2l+1}} d_{-1}^{(0)}(l+1) + \mathcal{S}_1 + \sqrt{\frac{2l-1}{2l+1}} d_{+1}^{(0)}(l-1) \right) P_l(x)  + \mathcal{O}(\omega^2) \label{F-linear-resum} \nn \\ 
\fl &\fl\quad\quad\quad= 0 + \mathcal{O}(\omega^2) .
\end{eqnarray}
It is straightforward to verify that the terms in parantheses on the second line cancel exactly, and so a linear-in-$\omega$ term in $F(x)$ is not present.

Note that to obtain $P_l(x)$ as a common factor in (\ref{F-linear-resum}) we redefined the summation variable (i.e. $l \rightarrow l \pm 1$). Care must be taken in this process, since it changes the lower limit of the sum. In the above, we added an `extra' $P_2(x)$ term to the series without consequence, because the coefficient $d_{+1}^{(0)}(l-1)$ is zero for $l=2$. We also neglected a $P_1(x)$ term (present in the top line but not subsequently). This is justified for our purposes because, to obtain the final amplitude $f$, we act on $F$ with $\diffop$, and $\diffop P_1(x) = 0$. 

Let us now repeat this process and focus only on the part which is linear in $a$. This time, we will keep terms up to second order in $\omega$. Such terms arise from (i) the $\astar$-dependent part of the phase $\betco \eps^2$, and (ii) the linear term $\mathcal{S}_1$ coupled to an $\eps$ term arising from the effect of redefining the summation variable. We may split the result into two parts: a sum which turns out to be zero, and an $l=2$ term which arises from the change of summation variable. That is,
\begin{equation}
F(x) \approx F_{0}(x) + \aw \eps \left(  F_{1}^{(\Sigma)} +   F_{1}^{(l=2)}  \right) + \mathcal{O}(\aw^2, \eps^2) + \mathcal{O}(\omega^3) .
\end{equation}
The lowest-order term
\begin{equation} 
F_{0} = \sum_{l=2}^\infty (2l+1) \frac{\Gamma(l-1-i\eps)}{\Gamma(l+3+i\eps)} P_l(x) \label{F0} 
\end{equation}
was defined and summed in \cite{Dolan-2008}. The additional terms are 
\begin{eqnarray}
\fl F_{1}^{(\Sigma)} (x) = i \sum_{l=2}^\infty  (2l+1)  \frac{\Gamma(l - 1 - i\epsilon)}{\Gamma(l + 3 + i\epsilon)} \times \nn \\ 
\fl \quad  \quad   \quad   \quad     \left( \frac{-2(l+1)}{(l-1)(l+3)} \sqrt{\frac{(2l+3)}{(2l+1)}} d_{-1}^{(0)}(l+1) - \frac{4}{l(l+1)}  + \frac{2l}{(l-2)(l+2)} \sqrt{\frac{(2l-1)}{(2l+1)}} d_{+1}^{(0)}(l-1)  \right) P_l(x)  \nn \\
\fl \quad \quad \quad= 0 + \mathcal{O}(\eps) ,
\end{eqnarray}
and 
\begin{eqnarray}
F_{1}^{(l=2)} (x) &= -2i \frac{\Gamma(1-i\epsilon)}{\Gamma(5+i\epsilon)} P_2(x) .  \label{F1-l2}
\end{eqnarray}
The term $\aw \epsilon F_{1}^{(l=2)} (x)$  is responsible for the lowest-order polarizing effect. It leads to a term in the scattering amplitude which depends on the sign of $\omega$. That is, $f \approx f_{\text{Schw}} + f_{\text{pol}}$ where
\begin{eqnarray}
f_{\text{pol}} = -2M a\omega e^{i \chi} \left[ \frac{1}{2}(1+x) \right]^2 \frac{\Gamma(1-i\epsilon)}{\Gamma(1+i\epsilon)} \left( 1 + \mathcal{O}(\epsilon) \right) .
\end{eqnarray}
and $f_{\text{Schw}}$ was defined in (\ref{f-Schw}).
Since $f_{\text{pol}}$ is in phase with $f_{\text{Schw}}$, this gives a first-order contribution to the cross section,
\begin{eqnarray}
2 \left| f_{\text{Schw}}^\ast f_{\text{pol}}   \right|
= -4 a \omega M^2 \frac{ \cos^8(\theta/2) }{ \sin^2(\theta/2)} .  \label{f-result}
\end{eqnarray}

\subsubsection{Helicity-reversing amplitude.}
We now repeat the analysis for the helicity-reversing amplitude, $g(\theta)$. To simplify matters, we only expand the relevant terms to first-order to recover the polarizing correction.
Let us begin by writing the amplitude as
\begin{equation}
g(\theta) = \frac{e^{i\chi}}{2 i \omega} \diffopm G(x),  \quad \text{where} \quad G(x) = \sum_{l=2}^\infty (-1)^l W_l V_l (1 + \mathcal{S}_1 \aw + \mathcal{O}(\aw^2) ) , 
\end{equation}
and
\begin{eqnarray}
\fl W_l &= 6i \eps  \frac{\Gamma(l - 1 - i \epsilon)}{\Gamma(l + 3 + i\epsilon)} \, \frac{\Gamma(l - 1)}{\Gamma(l + 3)}  \left( 1 + \frac{2 \astar m}{l(l+1)} \epsilon + \mathcal{O}(\epsilon^2) \right)  , \\
\fl V_l &= P_l(-x) + \aw \sqrt{\frac{2l-1}{2l+1}}\frac{l+2}{l-2} d_{-1}^{(0)}(l) P_{l-1}(-x) + \aw  \sqrt{\frac{2l+3}{2l+1}} \frac{l-1}{l+3}  d_{+1}^{(0)}(l) P_{l+1}(-x) + \mathcal{O}(z^2) .
\end{eqnarray}
As before, we may move terms of the series up or down ($l \rightarrow l \pm 1$) to obtain a common factor of $P_l(-x)$. As before, we find that, to lowest-order, the terms in the sum cancel out, leaving only an $l=2$ term which arises from redefining the summation variable. That is, 
\begin{equation}
G(x) \approx G_{0}(x) + \aw \eps \left( G_{1}^{(\Sigma)} +  G_{1}^{(l=2)} \right) + \mathcal{O}(\aw^2, \eps^2) + \mathcal{O}(\omega^3),
\end{equation}
and
\begin{eqnarray}
\fl G_{1}^{(\Sigma)} (x) &= 6 i \sum_{l=2}^\infty (-1)^l \frac{\Gamma(l - 1 - i\eps)}{\Gamma(l + 3 + i\eps)} \, \frac{\Gamma(l-1)}{\Gamma(l+3)} \left( (2l+1) \frac{16}{l(l+1)} - \frac{2(l+2)^2}{l^2} + \frac{2(l-1)^2}{(l+1)^2} \right) P_l(-x)  \nn \\
\fl &= 0 + \mathcal{O}(\eps), \\  
\fl G_{1}^{(l=2)} (x) &= 2 i \frac{\Gamma(1 - i\epsilon)}{\Gamma(5 + i\epsilon)} \, P_{2}(-x) .
\end{eqnarray}
The $z \epsilon G_{1}^{(l=2)} (x) $ term is responsible for the lowest-order polarizing effect. It leads to a term in the scattering amplitude which depends on the sign of $\omega$. That is, $g \approx g_{\text{Schw.}} + g_{\text{pol}},$ where
\begin{eqnarray}
g_{\text{pol}} = 2M a\omega e^{i \chi} \left[ \frac{1}{2}(1-x) \right]^2 \frac{\Gamma(1-i\epsilon)}{\Gamma(1+i\epsilon)} \left( 1 + \mathcal{O}(\epsilon) \right) .
\end{eqnarray}
and $g_{\text{Schw.}}$ was defined in (\ref{g-Schw}). 
The first-order contribution to the helicity-reversal cross section is
\begin{eqnarray}
2 \left| g_{\text{Schw}}^\ast g_{\text{pol}}   \right|
= 4 a \omega M^2  \sin^6(\theta/2).   \label{g-result}
\end{eqnarray}
Combining results (\ref{f-result}) and (\ref{g-result}) leads to the cross section (\ref{csec-final}), and hence the polarization (\ref{pol-final}). The consequences of polarization are discussed further in Section \ref{sec-discussion}. In the next section we outline a numerical method for moving beyond the long-wavelength regime.




\section{Numerical Method\label{sec-nummeth}}
The task of computing scattering and absorption cross sections numerically may be
divided into three steps. First, we must solve the angular equation
(\ref{angular-eq}) to determine the spin-weighted spheroidal harmonics
$\Sspher(\theta; a\omega)$ and their eigenvalues $\lambda_{lm}$. Second, we
must solve the radial equation (\ref{radial-eq}) to determine the
phase shifts $e^{2i\delta_{l2\omega}^P}$. Third, we must find a way of
improving the convergence properties of the partial wave series
(\ref{f-def}--\ref{g-def}) for a numerical calculation. These steps
are described in sections \ref{subsec-num-spheroidal},
\ref{subsec-phaseshifts} and \ref{subsec-reduction} below.

\subsection{Spheroidal Harmonics\label{subsec-num-spheroidal}}
The the spheroidal harmonics ${}_{-2}S_l^2(\theta, a \omega)$ and angular separation constants ${}_{-2}\lambda_{lm}$ were found using the matrix method detailed in Section \ref{subsec-spheroidal}.  The accuracy of the eigenvectors of coefficients $b_k^{(l)}$ was checked by back-substitution into (\ref{eq-quindiag}). The spherical harmonics ${}_{-2}Y_l^2(\theta)$ calculated via the method in \ref{appendix-spherical-harmonics} were verified to be highly accurate up to $l \sim 100$. The spheroidal harmonics ${}_{-2}S_l^2(\theta, a \omega)$ were tested against the results of Leaver's expansion \cite{Leaver-1985}.  The eigenvalues were tested against Leaver's continued fraction method, and against low-$a\omega$ expansions \cite{Berti-2005}.

\subsection{Phase Shift Calculation\label{subsec-phaseshifts}}
As we saw in (\ref{Rinout}), the outgoing part of Teukolsky's radial function ${}_{-2}R_{lm\omega}$ dominates over the ingoing part by a factor of $r^4$ in the far field. This behaviour is due to the long-ranged nature of Teukolsky's equation. A solution to this problem was supplied by Sasaki and Nakamura \cite{Sasaki-1982, Sasaki-2003}, who found a transformation that maps the Teukolsky function $R(r)$ to a function $X(r)$ governed by an equation with a short-ranged potential.

To compute phase shifts (\ref{eq-phaseshift1}), we take the following numerical approach. First, we find a power series representation for the Teukolsky function near the outer horizon. Then, starting at $r = r_+ + \eta$, where $\eta \sim 10^{-3}M$, we integrate the Teukolsky radial equation (\ref{radial-eq}) out to $r = r_m$. Typically, $r_m \sim 5M$. At $r_m$ we transform to the Sasaki-Nakamura formalism and integrate (\ref{sas-eqn}) out into the far field, up to $r = r_\infty$, where $r_\infty \sim 50M$. There, we read off the ingoing and outgoing coefficients and thus compute the phase shifts. The phase shifts are insensitive to changes in $\eta$, $r_m$ and $r_{\infty}$. The details of this method are outlined below.

\subsubsection{Horizon expansion of the Teukolsky function.\label{subsubsec-horizon}}
Teukolsky's radial equation (\ref{radial-eq}) has a regular singular point at the outer horizon, $r = r_+$. Using the method of Frobenius, the solution near this point may be expressed a power series in the distance from the (outer) horizon. Introducing the radial variable $\eta = r - r_+$, the radial equation (\ref{radial-eq}) may be rewritten
\begin{equation}
\frac{d^2 R}{d \eta^2} - \frac{A(\eta)}{\eta} \frac{dR}{d\eta} - \frac{B(\eta)}{\eta^2} R = 0 ,
\end{equation}
where $R = {}_{-2}R_{lm\omega}$. The functions $A(\eta)$ and $B(\eta)$ are 
\begin{eqnarray}
A(\eta) &= \frac{2 (\eta + \nu) }{\eta + 2 \nu}, \\
B(\eta) &= - \frac{K^2 + 4 i (\eta + \nu) K - \left(8i \omega (M+\nu+\eta)  + \lambda\right) \eta (\eta + 2\nu)}{(\eta + 2\nu)^2} ,
\end{eqnarray}
where $\nu = (r_+ - r_-) / 2$. Note that $A(\eta)$ and $B(\eta)$ are regular at $\eta = 0$, so may be expanded in Maclaurin series,
\begin{equation}
A(\eta) = \sum_{k=0}^\infty A_k \, \eta^k ,  \quad \quad B(\eta) = \sum_{k=0}^\infty B_k \, \eta^k
\end{equation}
The coefficients $A_k$ are straightforward to compute: $A_0 = 1$ and $A_k = (-1)^{k+1} / (2\nu)^{k}$ for $k > 0$. The coefficients $B_k$ are less simple, but are easily found with the help of a symbolic algebra package such as Maple.   
The radial function may be expressed as a power series, 
\begin{equation}
R(\eta) = \eta^\gamma \, \sum_{k = 0}^\infty R_k \, \eta^k,
\end{equation}
where $\gamma$ and $R_k$ are constants to be determined.
The index $\gamma$ satisfies an indicial equation, $\gamma(\gamma-1) - \gamma A_0 - B_0 = 0$, which has two solutions,
\begin{equation}
\gamma = i \sigma , \quad \text{and} \quad \gamma = 2 - i \sigma, \quad \text{where} \quad \sigma = \frac{2M\omega r_+ - a m}{r_+ - r_-} .
\end{equation}
The latter choice corresponds to an ingoing wave. The coefficients $R_j$ may be found from the recurrence relation
\begin{equation}
\fl \quad \quad \left[ (\gamma + j) (\gamma + j - 1) - (\gamma + j) A_0 - B_0 \right] R_j = 
\sum_{k = 1}^{j} \left[ (\gamma + j - k) A_k + B_k \right] R_{j-k} .
\end{equation}
Hence the series expansion may be taken to the desired order.

\subsubsection{Sasaki-Nakamura transformation.\label{subsubsec-sasaki}}
The Teukolsky function $R(r)$ is transformed to a Sasaki-Nakamura function $X(r)$ with the rule
\begin{equation}
X(r) = \frac{\sqrt{r^2 + a^2}}{\Delta} \left( \alpha(r) R + \frac{\beta(r)}{\Delta} \frac{d R}{d r} \right) 
\end{equation}
where
\begin{eqnarray}
\alpha(r) &= -i K(r) \beta / \Delta^2 + 3i dK/dr + \lambda + 6\Delta / r^2, \\
\beta(r) &= 2 \Delta \left[- iK(r) + r - M - 2\Delta / r \right].
\end{eqnarray}
The function $X(r)$ satisfies the equation \cite{Sasaki-1982},
\begin{equation}
\frac{d^2 X}{d\rstar} - F(r) \frac{d X}{d \rstar} - U(r) X = 0 ,  \label{sas-eqn}
\end{equation}
where the functions $F(r)$ and $U(r)$ are defined in \ref{appendix-sas}. In the non-rotating limit $\as = 0$, the Sasaki-Nakamura reduces to the Regge-Wheeler radial equation \cite{Regge-1957, Sasaki-2003}. 

Like the Teukolsky equation, the Sasaki-Nakamura equation admits two independent solutions at infinity,
\begin{equation}
X(r) \sim \Bin \Pin(r) e^{-i \omega \rstar} + \Bout \Pout(r) e^{i \omega \rstar}, \quad \quad r \rightarrow \infty .   \label{sas-asymp}
\end{equation}
The principle advantage of the Sasaki-Nakamura formalism is that $\Pin(r)$ and $\Pout(r)$ tend to unity as $r \rightarrow \infty$. Hence it is straightforward to numerically determine the relative magnitude and phase of the ingoing and outgoing coefficents $\Bin$ and $\Bout$, which are related to $\Ain$ and $\Aout$ in (\ref{teuk-bc}) by
\begin{equation}
\Bin = -4 \omega^2 \Ain , \quad \quad  \Bout = -\frac{c_0}{4 \omega^2} \Aout ,
\end{equation}
where $c_0$ is defined in (\ref{c-co-defn}) in \ref{appendix-sas}.
The phase shifts follow immediately from (\ref{eq-phaseshift1}). 

The function $\Pin(r)$ has the series expansion 
\begin{equation}
\Pin(r) = 1 + \frac{\mathcal{A}_\text{in}}{\omega r} + \frac{\mathcal{B}_\text{in}}{(\omega r)^2} + \frac{\mathcal{C}_\text{in}}{(\omega r)^3} + \mathcal{O}\left(\frac{1}{(\omega r)^{4}}\right) 
\end{equation}
where \cite{Hughes-2000}
\begin{eqnarray}
\fl \mathcal{A}_\text{in} = -\frac{i}{2} \left(\lambda + 2 + 2 a m \omega \right) \\
\fl \mathcal{B}_\text{in} = -\frac{1}{8} \left[(\lambda + 2)^2 - (\lambda+2)(2 - 4am\omega) - 4\left( a m \omega + 3 i M \omega - a m \omega (a m \omega + 2i M \omega) \right) \right] \\
\fl \mathcal{C}_\text{in} = - \frac{i}{6} \left[ 4 a m \omega + \mathcal{B}_\text{in} (\lambda - 4 + 2 a m \omega + 8 i M \omega) + 12 M^2 \omega^2 \right. \nn  \\ 
 \quad \quad \quad  \left. - 2 \mathcal{A}_\text{in} \lambda M \omega - (a \omega)^2 (\lambda - 3 + m^2 + 2 a m \omega) \right].
\end{eqnarray}
Whilst it has been stated (e.g.~\cite{Hughes-2000}, eq. (4.21)) that $\Pout(r) = {\Pin}^\ast(r)$, this is not the case because $F(r)$ and $U(r)$ are complex. We find instead that
\begin{equation}
\Pout(r) = 1 + \frac{\mathcal{A}_\text{out}}{\omega r} + \frac{\mathcal{B}_\text{out}}{(\omega r)^2} + \frac{\mathcal{C}_\text{out}}{(\omega r)^3} + \mathcal{O}\left(\frac{1}{(\omega r)^{4}}\right)  
\end{equation}
where
\begin{eqnarray}
\fl \mathcal{A}_\text{out} = \mathcal{A}_\text{in}^\ast + \omega c_1 / c_0 , \\
\fl \mathcal{B}_\text{out} = \mathcal{B}_\text{in}^\ast + \left[ \omega^2 c_2 - \omega c_1 (\mathcal{A}_\text{in} + i/2) \right] / c_0 , \\
\fl \mathcal{C}_\text{out} = \mathcal{C}_\text{in}^\ast \nn \\ \fl\quad\quad\quad + \left[ \omega^3 c_3 - (\mathcal{A}_\text{in} + i ) \omega^2 c_2 + \left(\mathcal{B}_\text{in} + i \mathcal{A}_\text{in} / 2 - \frac{1}{2} + 2 i M\omega(a \omega m - 1) \right) \omega c_1 \right] / c_0 ,
\end{eqnarray}
and $c_0$, $c_1$ and $c_2$ are defined in (\ref{c-co-defn}) in \ref{appendix-sas}. This result is confirmed in a recent erratum \cite{Hughes-2000-erratum}.

   \subsection{Partial Wave Series Reduction\label{subsec-reduction}}
The partial wave series (\ref{f-def}) is divergent on-axis at $\theta = 0$. Physically, this is because the gravitational interaction is long-ranged; geodesics in the far-field are deflected through the Einstein angle $\theta \sim 4M/b$. Since each term in the series is finite, an infinite number of terms is required to reproduce the on-axis divergence. The magnitude of the series coefficients grows with $l$, and the series does not converge.

Our computing power is finite, hence we must terminate the series at some $l = \lm$. However, a naive truncation introduces unwanted `noise' with a magnitude roughly proportional to $2 \lm + 1$ and an angular width proportional to $1 / \lm$. To avoid this problem, we employed a \emph{series reduction} technique, inspired by a numerical method developed in the 1950s \cite{Yennie-1954} to compute Coulomb scattering series. 

Before applying the method, we first rewrite the partial wave series in terms of associated Legendre polynomials (and their derivatives). Series (\ref{f-def}) is a sum over spheroidal harmonics; for brevity, let us write
\begin{equation}
f(\theta) = \sum_{l = 2}^\infty f_l \,\, \Sspher(\theta ; a \omega) . \label{reduct1}
\end{equation}
Using the results of Section \ref{subsec-spheroidal}, series (\ref{reduct1}) can be rewritten as a sum over spherical harmonics, 
\begin{equation}
f(\theta) = \sum_{j = 2}^\infty F_j \, {}_{-2}Y_j^2(\theta) \quad \quad \text{where} \quad \quad  F_j =  \sum_{l = 2}^\infty f_l \, b_j^{(l)} ,
\end{equation}
and $b_{j}^{(l)}$ are coefficients found from the eigenproblem described in Sec. \ref{subsec-spheroidal}. The spherical harmonics of spin-weight $-2$ may be expressed in terms of associated Legendre polynomials $P_{lm}(\cos\theta)$ and their first and second derivatives. This decomposition is described in detail in \ref{appendix-spherical-harmonics}. The series (\ref{reduct1}) may be written
\begin{equation}
f(x) = \sum_{j = 2}^\infty c_j^{(0)} (1-x^2) \frac{d^2 P_{j2}}{d x^2} + d_j^{(0)} \frac{ d P_{j2} }{d x} + e_j^{(0)} \frac{P_{j2}(x)}{1-x^2}  
\end{equation}
where $x = \cos \theta$ and
\begin{eqnarray}
c_j^{(0)} =  A_j F_j , \quad \quad 
d_j^{(0)} = -4 A_j F_j, \quad \quad   
e_j^{(0)} = 4 (1 - x) A_j F_j ,  \label{cde0} \\
A_j = \sqrt{\frac{2j+1}{4\pi}} \frac{1}{(j-1)j(j+1)(j+2)} .
\end{eqnarray}

The idea behind the series reduction method is to improve the convergence properties of the series by removing the divergence at $x = 1$. We define a new series
\begin{equation}
(1-x)^{k} f(x) = \sum_{j = 2}^\infty c_j^{(k)} (1-x^2) \frac{d^2 P_{j2}}{d x^2} + d_j^{(k)} \frac{ d P_{j2} }{d x} + e_j^{(k)} \frac{P_{j2}(x)}{1 - x^2}  ,   \label{reduced-series}
\end{equation}
The functions $c_j^{(k)},d_j^{(k)}$ and $e_j^{(k)}$ are computed from (\ref{cde0}) with the following recurrence relations,
\begin{eqnarray}
c_j^{(k+1)} &= c_j^{(k)} - \frac{(j-2)}{2j-1} c_{j-1}^{(k)} - \frac{j+3}{2j+3} c_{j+1}^{(k)}   \label{recur-c} \\
d_j^{(k+1)} &= d_j^{(k)} - \frac{(j-2)}{2j-1} d_{j-1}^{(k)} - \frac{j+3}{2j+3} d_{j+1}^{(k)} + 2(1+x) c_j^{(k+1)} \\ 
e_j^{(k+1)} &= e_j^{(k)} - \frac{(j-2)}{2j-1} e_{j-1}^{(k)} - \frac{j+3}{2j+3} e_{j+1}^{(k)} + (1+x) d_j^{(k+1)} \label{recur-e} .
\end{eqnarray}
These relations were calculated with the aid of the recursive formulae (\ref{LegP-1}, \ref{LegP-d}, \ref{LegP-dd}). In \ref{appendix-spherical-harmonics} we outline a stable recursive method for computing $(1-x^2)\frac{d^2P_{lm}}{dx^2}$,  $\frac{dP_{lm}}{dx}$ and $P_{lm} / (1-x^2)$.

The reduced series are found to converge more quickly as $l \rightarrow \infty$. Typically, we apply two iterations of (\ref{recur-c}--\ref{recur-e}). In other words, we calculate the amplitude $f(\theta)$ from (\ref{reduced-series}) with $k = 2$. The series is terminated at $\lm \sim 60$.

\section{Numerical Results\label{sec-results}}

In this section we present a range of numerically-determined absorption and scattering cross sections for a gravitational wave impinging along the axis of a rotating black hole. 

   \subsection{Absorption}

\begin{figure}
\begin{center}
\includegraphics[width=12cm]{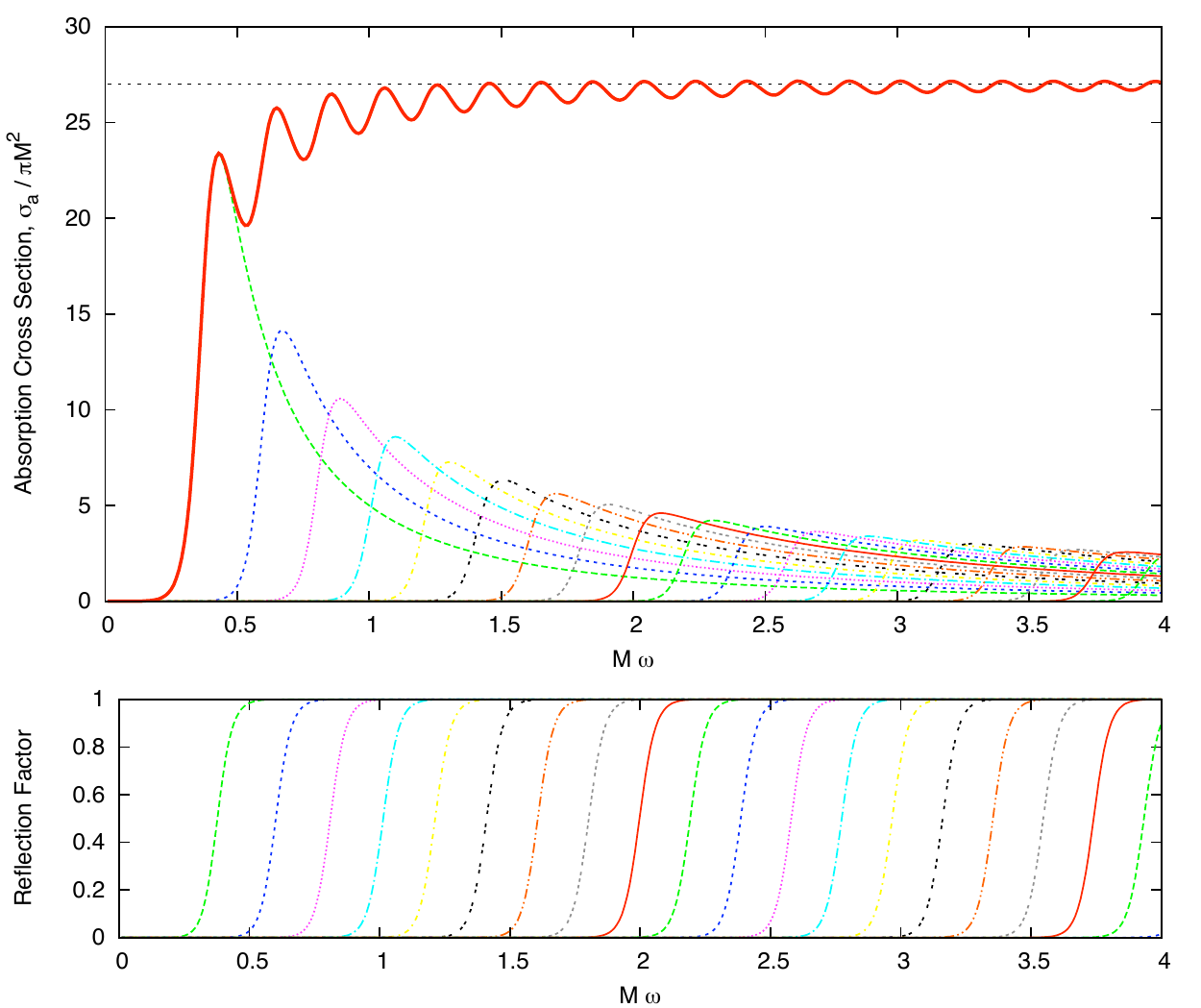}
\end{center}
\caption[]{\emph{Gravitational wave absorption cross section of the Schwarzschild black hole.} The top plot shows the gravitational absorption cross section $\sig_a$ as a function of gravitational coupling $M\omega$. The line at $\sig_a = 27 \pi M^2$ shows the classical limit. The broken curves show the contributions $\sig_l$ of each angular mode, from $l=2$ onwards. The lower plot shows the relection factor for each angular mode.}
\label{fig-abs-schw}
\end{figure}

Figure \ref{fig-abs-schw} shows the absorption cross section $\sig_a$ (eq. \ref{abs-csec}) for a non-rotating black hole. The cross section is plotted as a function of gravitational coupling $M\omega$. If the wavelength of the incoming gravitational wave is much larger than the horizon size (that is, $M\omega \ll 1$), then absorption is negligible. Conversely, if the wavelength is much smaller than the horizon (i.e. $M\omega \gg 1$) then the cross section tends towards the geometric-optics limit of $\sigma_a = 27 \pi M^2$. In the intermediate regime $M\omega \sim 1$, the contributions from successive angular modes ($l \ge 2$) create a regular oscillatory pattern. 

\begin{figure}
\begin{center}
\includegraphics[width=11cm]{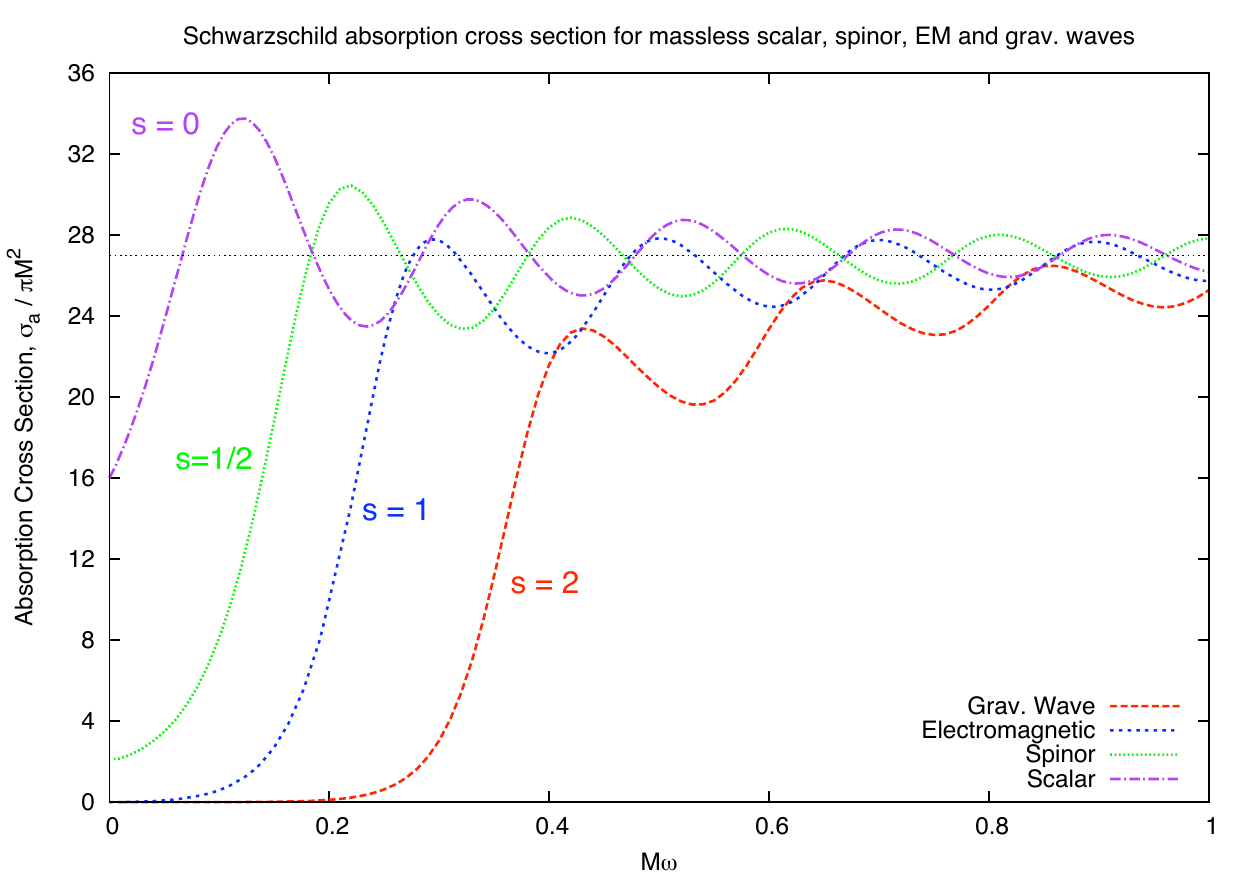}
\end{center}
\caption[]{\emph{Absorption cross section of scalar, spinor, EM and gravitational waves up to $M\omega = 0.5$.} This plot compares the Schwarzschild absorption cross section of massless scalar $s = 0$, spinor $s = 1/2$, electromagnetic $s = 1$ and gravitational $s = 2$ waves. The electromagnetic cross section is reproduced from a paper by Crispino, Oliveira, Higuchi and Matsas \cite{Crispino-2007}, with many thanks to the authors.}
\label{fig-abs-spins}
\end{figure}

Figure \ref{fig-abs-spins} compares Schwarzschild absorption cross sections for massless waves of spin $0$, $1/2$, $1$ and $2$. The spin-$0$ and spin-$1/2$ cross sections approach $\sig_a = 16\pi M^2$ and $\sig_a = 2 \pi M^2$ as $M\omega \rightarrow 0$ \cite{Unruh-1976-absorption}, whereas the $s = 1$ and $s=2$ fields tend to zero in this limit. The cross section at low $M\omega$ is dominated by the lowest allowed angular modes (i.e. the $l = |s|$ modes). 

\begin{figure}
\begin{center}
\includegraphics[width=12cm]{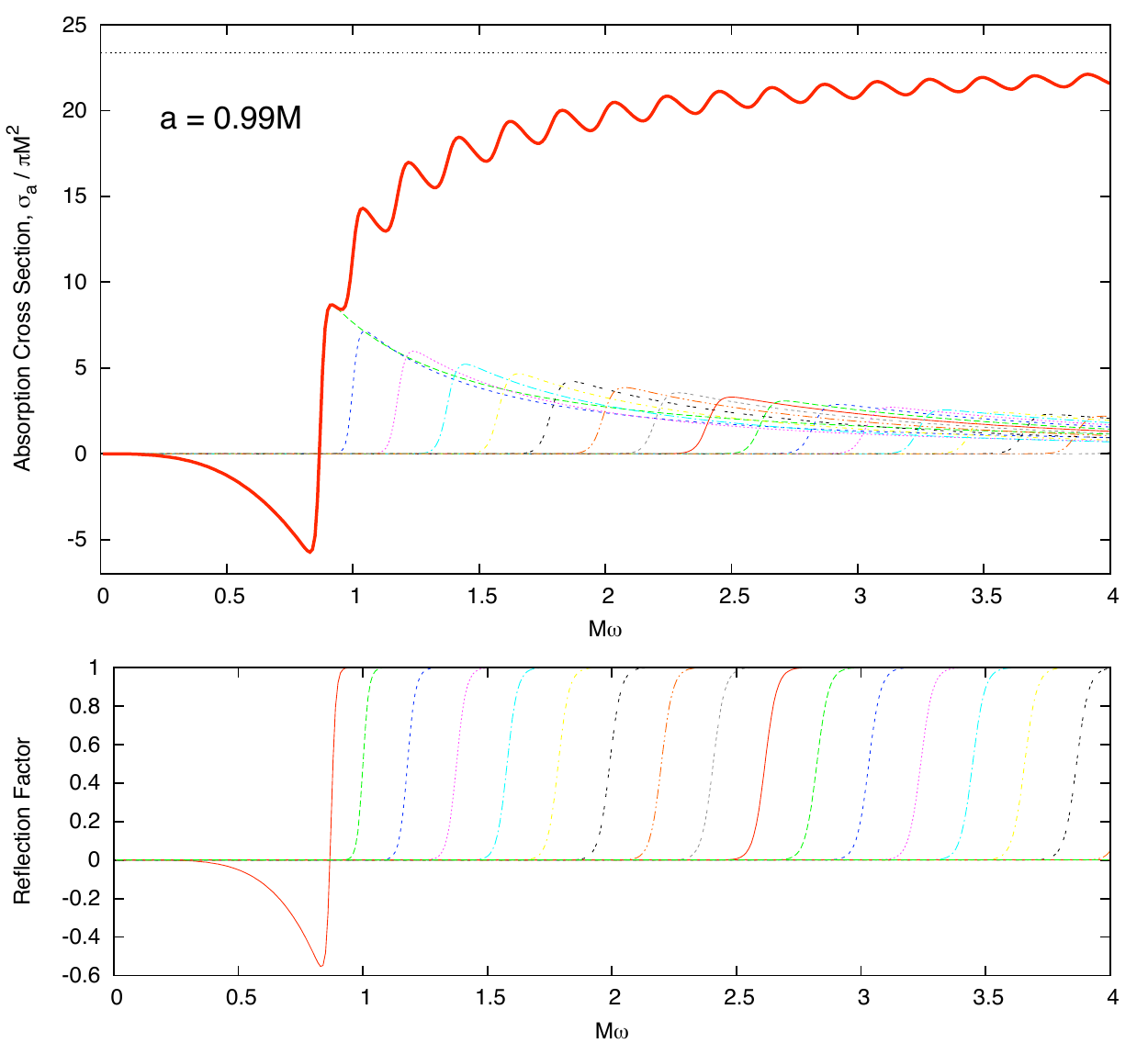}
\end{center}
\caption[]{\emph{Absorption cross section for a co-rotating wave ($\omega > 0$) incident on a fast-rotating ($a = 0.99M$) black hole along the axis of rotation.} The top plot shows the gravitational absorption cross section $\sig_a$ as a function of gravitational coupling $M\omega$. The line at $\sig_a \approx 23.4 \pi M^2$ shows the geometric-optics value. The broken curves show the contributions $\sig_l$ from each angular mode, from $l=2$ onwards. The bottom plot shows the reflection factor for each angular mode. Note the superradiance (induced emission) exhibited by the $l = 2$, $m = 2$ mode.}
\label{fig-abs-kerr1}
\end{figure}

Figure \ref{fig-abs-kerr1} shows the absorption cross section for an incident wave propagating parallel to the axis of a rotating (Kerr) black hole. In this plot, the incident wave is circularly polarized in the same sense as the rotation (i.e. $\omega > 0$). In the low-$M\omega$ regime, the cross section is \emph{negative}. This is due to the superradiance effect which enhances the amplitude of the lowest corotating modes (particularly $l=2$, $m=2$). Hence, an ingoing wave may actually stimulate net emission from a rotating hole.

In the high-coupling limit ($M\omega \gg 1$), the cross section approaches the geometric-optics value, $\sigma_a = \pi b_c^2$. For $a = 0.99M$, the geometric cross section is approximately $\sigma_a \approx 23.4 \pi M^2$. The critical impact parameter $b_c$ is found by solving (\ref{R-Theta}) numerically; values for a range of $\as$ are listed in Table \ref{tbl-rcbc}.

\begin{figure}
\begin{center}
\includegraphics[width=12cm]{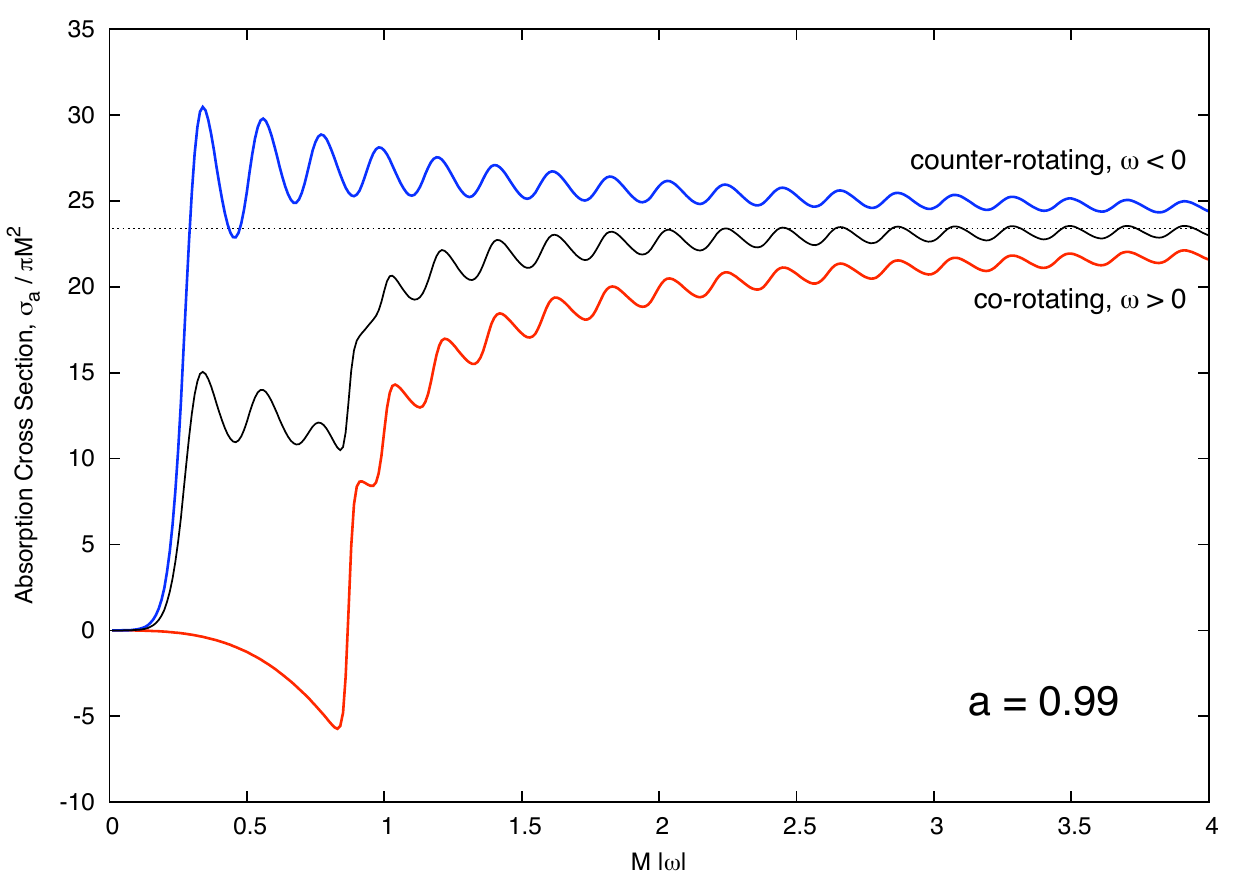}
\end{center}
\caption[]{\emph{Absorption cross section for co-rotating and counter-rotating waves incident on a fast-rotating ($\as = 0.99M$) hole along the axis of rotation.} This plot shows that more of the counter-rotating wave is absorbed. The middle curve shows the absorption cross section of a linearly-polarized incident wave. The line at $\sigma_a \approx 23.4 \pi M^2$ shows the geometric limit \cite{Chandrasekhar-1983}.}
\label{fig-abs-kerr2}
\end{figure}

Figure \ref{fig-abs-kerr2} compares the absorption cross sections of co- and counter-rotating circularly polarized incident waves at $\as = 0.99$. It shows that a greater proportion of the counter-rotating helicity ($\omega < 0$) is absorbed than the co-rotating helicity ($\omega > 0$). Superradiance is not stimulated by the counter-rotating helicity. In the limit $M\omega \gg 1$, both cross sections approach the geometric-optics limit.

\begin{figure}
\begin{center}
\includegraphics[width=8.0cm]{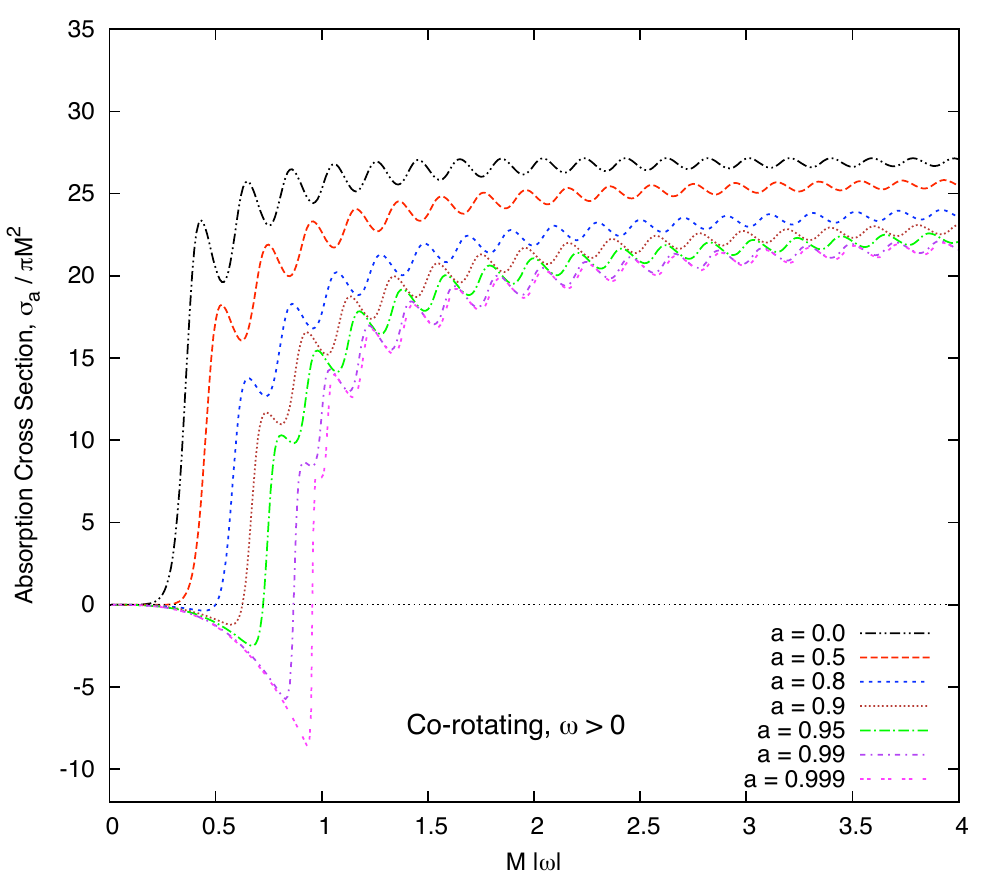}
\includegraphics[width=8.0cm]{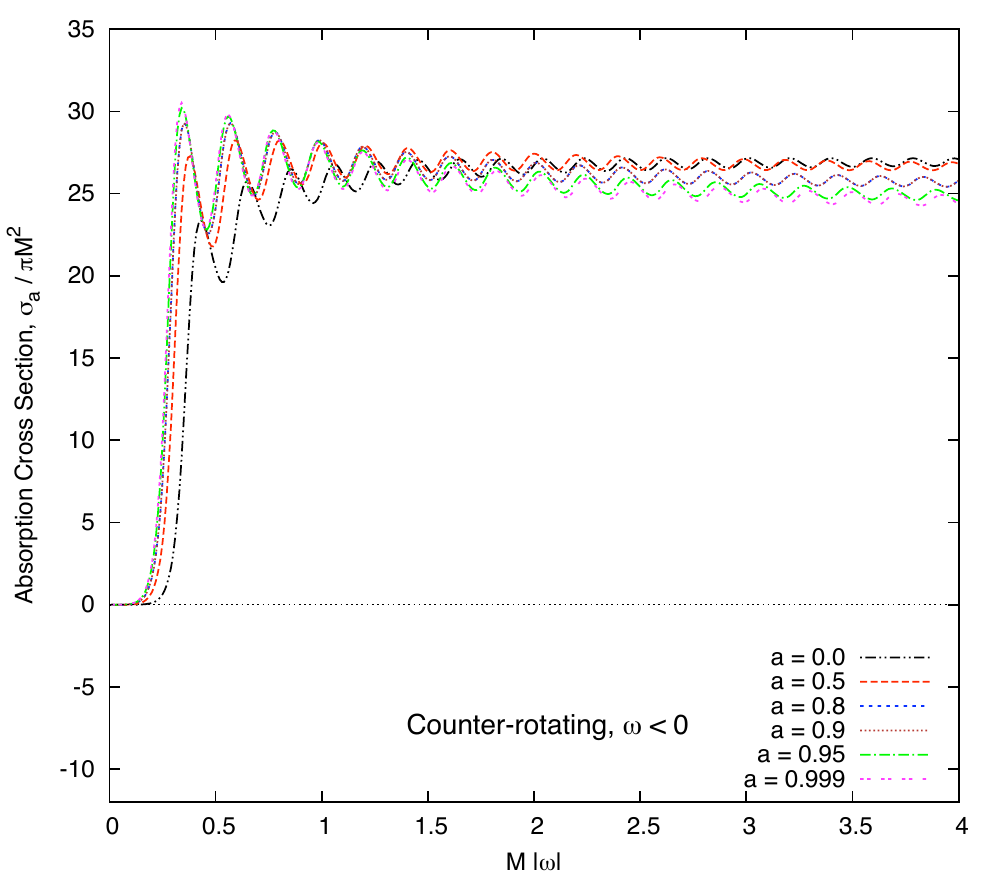}
\end{center}
\caption[]{\emph{Absorption cross section for a range of black hole rotation rates, $\as = 0$, $0.5$, $0.8$, $0.9$, $0.95$,  $0.99$ and $0.999$.} The left plot shows the cross section for the co-rotating helicity ($\omega > 0$) and the right plot shows the counter-rotating helicity ($\omega < 0$). Superradiance ($\sig_a < 0$) is present only in the co-rotating case.}
\label{fig-abs-kerr-a0to1}
\end{figure}

Figure \ref{fig-abs-kerr-a0to1} illustrates how the absorption cross section depends on rotation rate $\as$ for both co- and counter-rotating circular polarizations. For the former case, we find that the amount of stimulated emission created by superradiance increases as $\as \rightarrow 1$. Above $M \omega = 1$, superradiance is prohibited, and the absorption cross section is strictly positive. For the prograde wave ($\omega > 0$) it seems to be the case that $\sig_a(a_1, M\omega) < \sig_a(a_2, M\omega)$ if $a_1 > a_2$, for any $M\omega > 0$. This is not the case for the retrograde polarization ($\omega < 0$); the ordering of the curves is reversed as $M|\omega|$ goes from zero to infinity.

   \subsection{Scattering}
In this section we present plots of the differential scattering cross section $d\sigma / d\Omega$ across a range of couplings $M \omega$, and compare the non-rotating and rotating cases. 

\subsubsection{Long wavelengths,  $M\omega \ll 1$.}
In sections \ref{subsec-approx} and \ref{sec-long-wavelength} we reviewed the analytic results available in the long-wavelength regime. Here, we test expressions (\ref{csec-low-approx}), (\ref{csec-final}) and (\ref{phase-shift-negative}) against the results of our numerical code.

\begin{figure}
\begin{center}
\includegraphics[width=13cm]{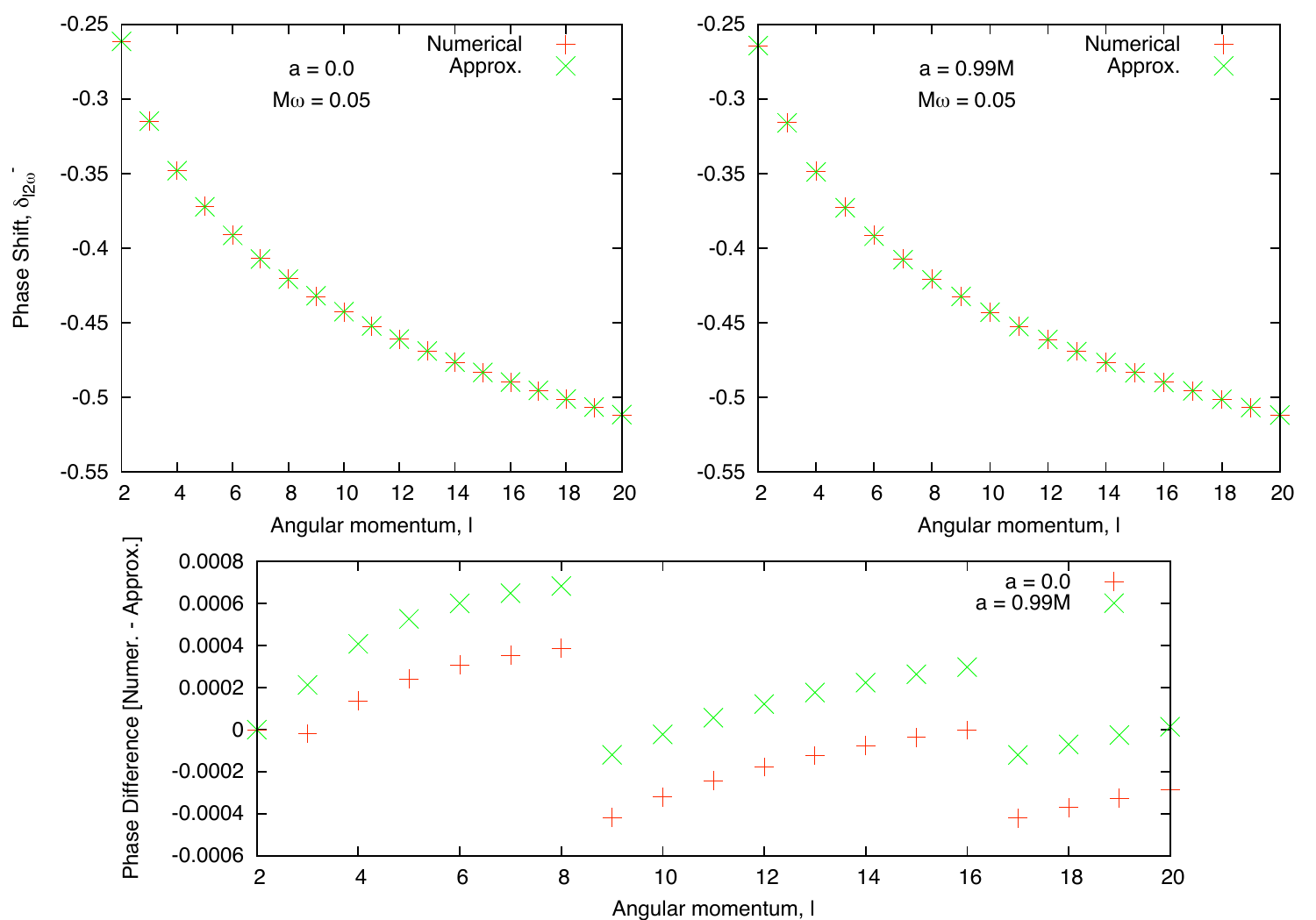}
\end{center}
\caption[]{\emph{Phase shifts ($P=-1$) at low coupling $M\omega =0.05$}. The top plots compare the numerically-determined phase shifts $\delta_{l2\omega}^{-}$ (defined in Eq. \ref{eq-phaseshift1}) with the low-$M\omega$ approximation (Eq. \ref{phase-shift-negative}). The left panel shows the $\as = 0$ phase shifts and the right panel shows $\as = 0.99$. It is no surprise to find the phase shifts are only weakly dependent on $\as$ in the low-frequency regime, since (\ref{phase-shift-negative}) predicts $\delta_{lm\omega}^{-}(a = 0) - \delta_{lm\omega}^{-}(a = \as M) \sim \mathcal{O}(\as (M \omega)^2)$. The lower panel shows the difference between the numerical and approximate values; the difference is consistent with the error $\mathcal{O}( M^3 \omega^3 )$ in approximation (\ref{phase-shift-negative}). }
\label{fig-phase-csec}
\end{figure}

In Fig.~\ref{fig-phase-csec} the numerically-determined phase shifts are compared with approximation (\ref{phase-shift-negative}) for $M\omega = 0.05$. As expected, there is good agreement. The lower panel shows the difference between the numerical and approximate values. It certainly seems plausible that the approximation (\ref{phase-shift-negative}) is valid up to $\mathcal{O}(\eps^3) \sim \mathcal{O}(0.001)$ in this case. Note that there is some ambiguity in the overall ($l$-independent) phase factor, which arises from the freedom to choose a constant of integration when defining $\rstar$ from (\ref{tortoise}). In these plots, the overall phase term has been fixed by matching approximate and numerical phases at $l=2$. 

\begin{figure}
\begin{center}
\includegraphics[width=10cm]{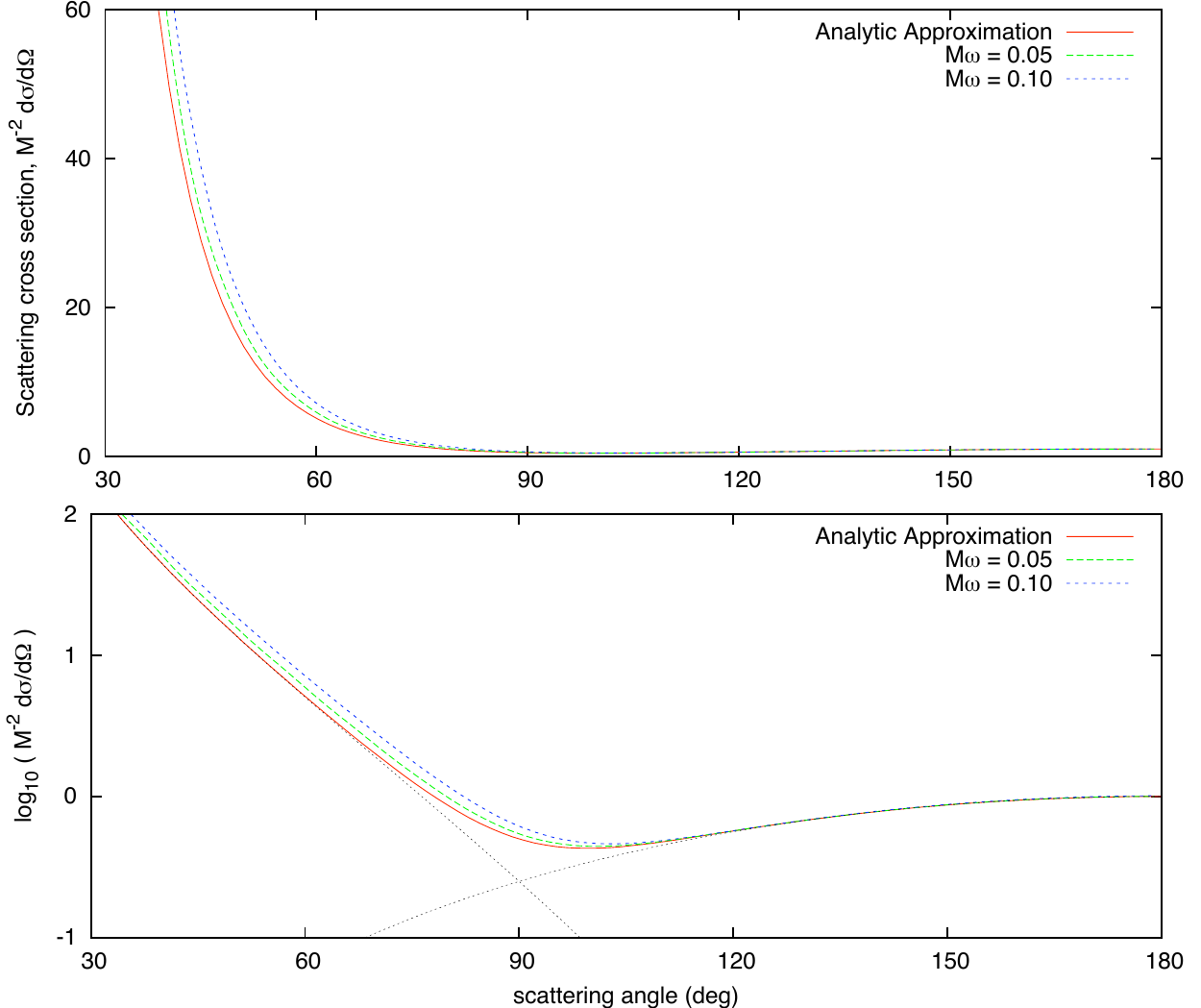}
\end{center}
\caption[]{\emph{Schwarzschild scattering cross sections at low couplings}. These plots compare the numerically-determined scattering cross sections at $M\omega = 0.05$ and $M\omega = 0.1$ with the approximation (\ref{csec-low-approx}). Note the logarithmic scale on the y-axis of the lower plot. The thin black lines show the helicity-conserving and helicity-reversing contributions.}
\label{fig-csec-schw-approx}
\end{figure}

Figure \ref{fig-csec-schw-approx} compares the numerically-determined Schwarzschild cross sections at $M\omega = 0.05$ and $M\omega = 0.1$ with approximation (\ref{csec-low-approx}). Again, as expected, we find good agreement. Note that the helicity-reversing cross section $M^{-2} |g|^2$ tends to unity in the backwards direction ($\theta = \pi$). 

\begin{figure}
\begin{center}
\includegraphics[width=10cm]{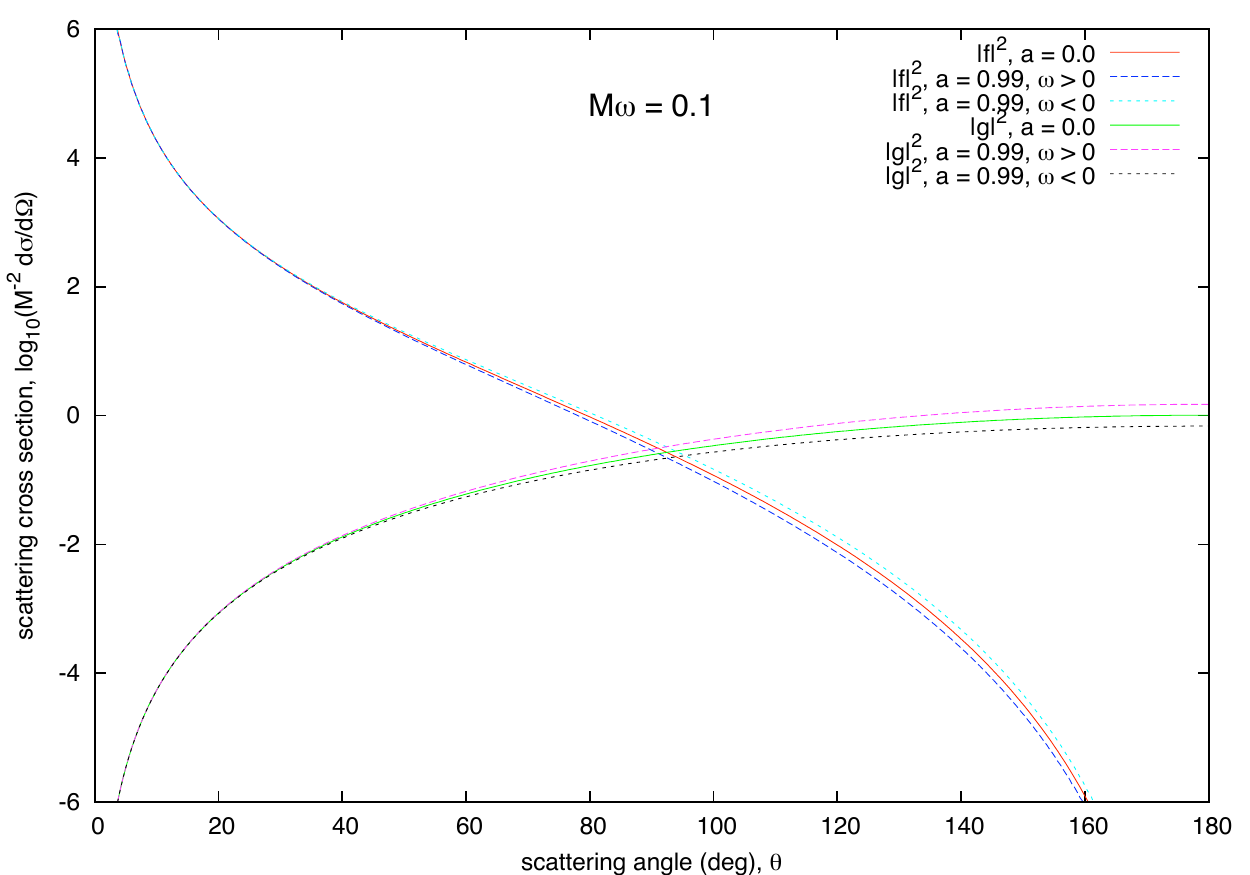}
\end{center}
\caption[]{\emph{Kerr scattering cross sections at low coupling}. This plot shows the helicity-preserving $|f|^2$ and helicity-reversing $|g|^2$ cross sections at $M\omega = 0.1$ on a logarithmic scale for three cases: (i) $a = 0$ [solid], (ii) $a = 0.99M$, co-rotating helicity $\omega > 0$ [dashed], and (iii) $a = 0.99M$, counter-rotating helicity $\omega < 0$ [dotted]. The cross sections were numerically computed using phase shifts (\ref{phase-shift-negative}) and two iterations of the series reduction method (\ref{subsec-reduction}).}
\label{fig-csec-kerr-approx}
\end{figure}

Figure \ref{fig-csec-kerr-approx} illustrates the polarizing effect of black hole rotation in the long-wavelength regime. The cross sections were calculated by summing the partial wave series up to $l_{\text{max}} = 500$ using two iterations of the series reduction technique (\ref{subsec-reduction}) and the asymptotic phase shifts (\ref{phase-shift-negative}). It is clear that there is a helicity-rotation coupling which splits the co-rotating and counter-rotating helicities. The numerically-determined cross sections are in excellent agreement with the asymptotic result (\ref{csec-final}).

\subsubsection{Schwarzschild scattering cross sections.}
Let us now examine scattering from a non-rotating hole in the intermediate regime $\lambda \sim r_S$ ($M\omega \sim 1$). In the following plots, the cross section at small scattering angles $\theta \le 30^\circ$ is suppressed. At small angles, the series reduction technique (see Sec. \ref{subsec-reduction}) is sensitive to small numerical errors in the phase shifts at large-$l$. This sensitivity is due to the divergence on-axis which arises because as $l \rightarrow \infty$ the phase shift diverges logarithmically, $\exp(2i\delta_l) \propto l^{4iM\omega}$. Since large-$l$ partial waves don't `see' the rotation of the hole, rotation does not influence the form of the divergence as $\theta \rightarrow 0$. At sufficiently small angles, approximation (\ref{csec-final}) is valid.  

\begin{figure}
\begin{center}
\includegraphics[width=14cm]{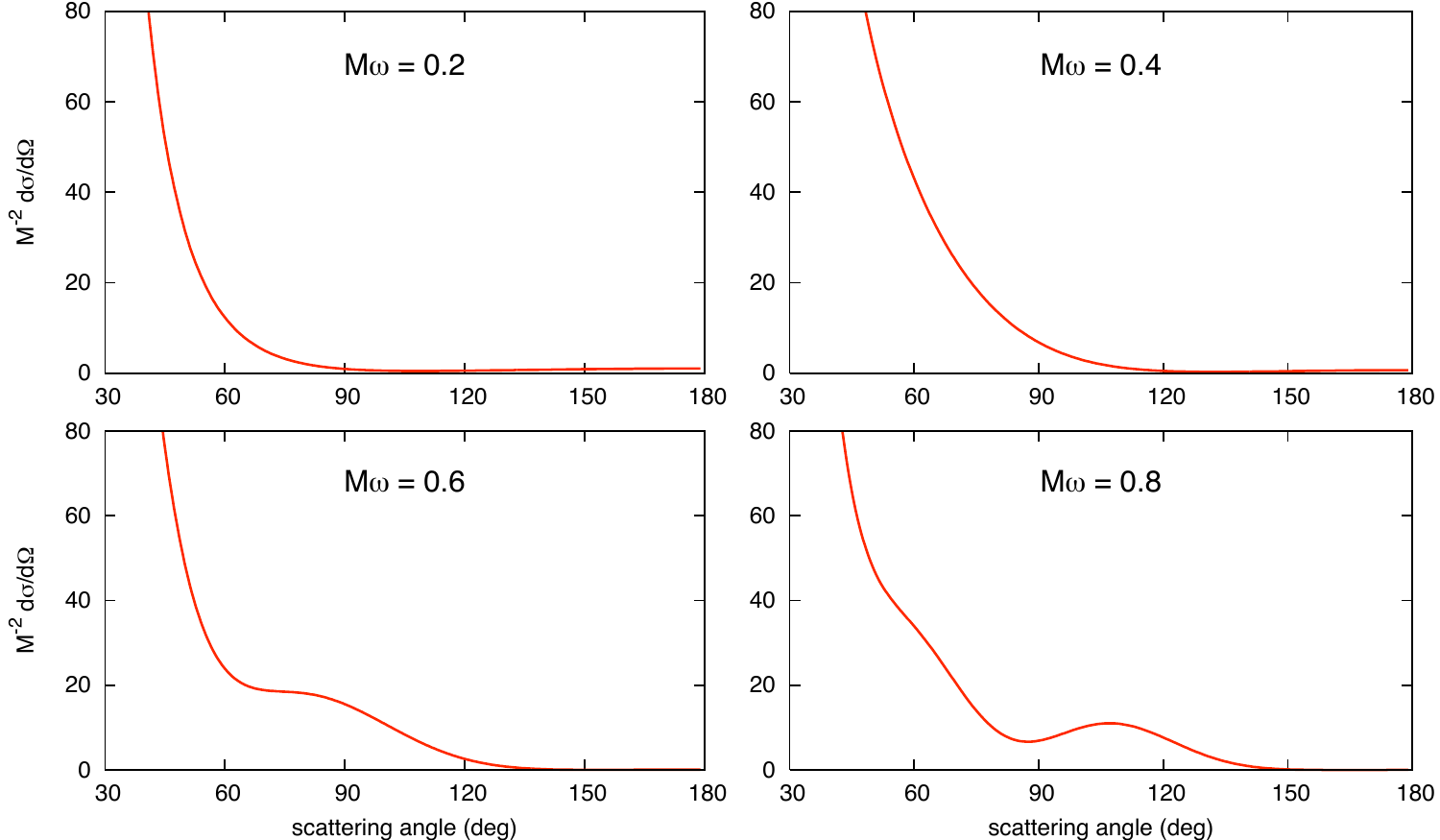}
\end{center}
\caption[]{\emph{Schwarzschild scattering cross sections ($a=0$) at intermediate couplings, $0.2 \le M\omega \le 0.8$}.}
\label{fig-csec-schw1}
\end{figure}

Figure \ref{fig-csec-schw1} shows Schwarzschild scattering cross sections up to $M \omega = 0.8$. As the coupling increases, we observe the development of additional structure at large scattering angles. A glory halo develops in the backward direction, and at large scattering angles approximation (\ref{csec-low-approx}) is no longer sufficient.

\begin{figure}
\begin{center}
\includegraphics[width=14cm]{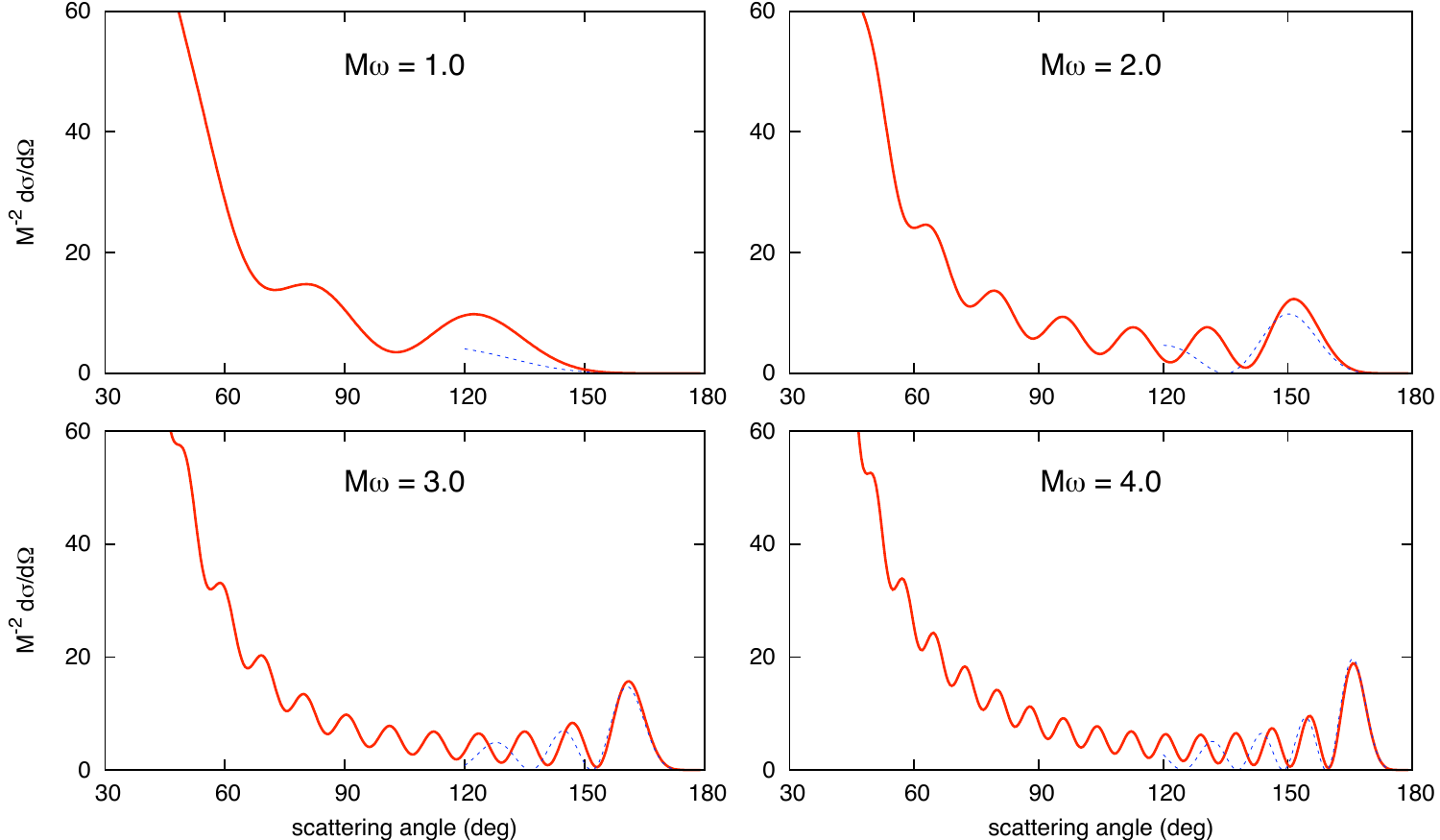}
\end{center}
\caption[]{\emph{Schwarzschild scattering cross sections ($a=0$) at higher couplings $1 \le M\omega \le 4$}. The dotted blue line shows the glory approximation of Eq. \ref{glory-approx}.}
\label{fig-csec-schw2}
\end{figure}

Figure \ref{fig-csec-schw2} shows Schwarzschild scattering cross sections up to $M \omega = 4$. The regular oscillations at large angles seen in this figure are predicted by a WKB analysis and semi-classical arguments \cite{Zhang-1984, Anninos-1992}. The angular width of the oscillations is inversely proportional to $M \omega$. As expected, the accuracy of the glory scattering approximation (\ref{glory-approx}) improves as $M \omega$ increases.

\begin{figure}
\begin{center}
\includegraphics[width=14cm]{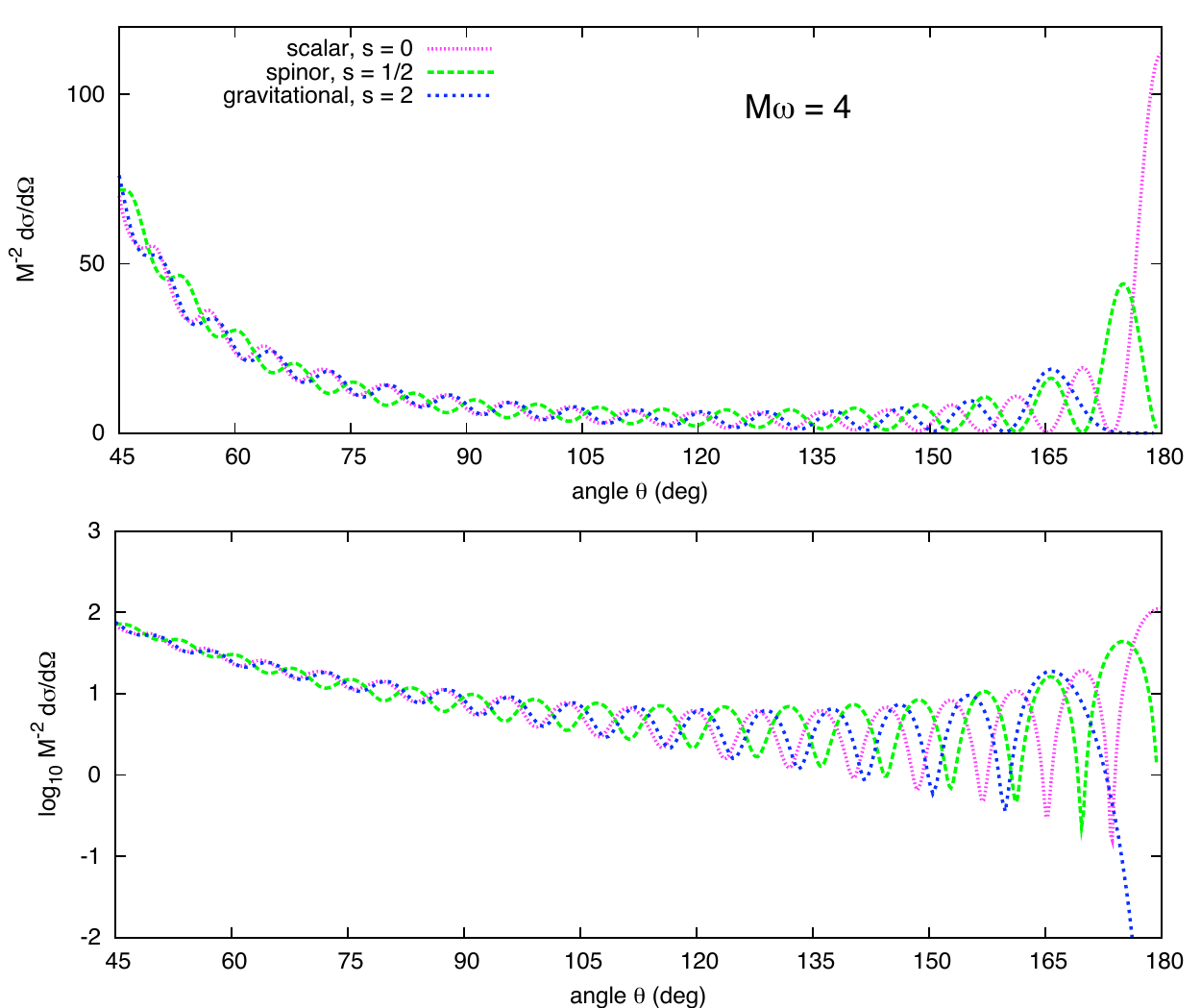}
\end{center}
\caption[]{\emph{Scattering cross sections for massless waves of spin $|s|=0$ (solid), $|s|=1/2$ (dashed) and $|s|=2$ (dotted) at $M\omega = 4$, $a = 0$}. Note the log scale on the y-axis of the lower plot.}
\label{fig-csec-schw-spins}
\end{figure}

Figure \ref{fig-csec-schw-spins} compares the Schwarzschild scattering cross section of massless waves of spin $s = 0$ (scalar), $|s| = 1/2$ (spinor/neutrino) \cite{Dolan-2006} and $|s|=2$ (gravitational). For waves with non-zero spin ($s > 0$), the cross section is zero in the backward direction (note that the amplitude $g(\theta)$ is negligible for large $M\omega$). The scalar wave has a glory maximum in the backward direction. The spiral scattering oscillations are of similar width. The first peak of the gravitational wave near $\theta \sim \pi$ coincides with a peak of the spinor wave. At intermediate angles ($45^\circ$ -- $120^\circ$), the gravitational peaks approximately coincide with the scalar peaks.

 \subsubsection{Kerr scattering: long wavelengths $M\omega \lesssim 1$.}
Let us now present results for the scattering of waves with a well-defined helicity by a rapidly-rotating ($a=0.99M$) black hole. The incident wave is travelling along the rotation axis of the hole. Hence, the helicity of the initial wave is either aligned ($\omega > 0$, `up', co-rotating, prograde) or anti-aligned ($\omega < 0$, `down', counter-rotating, retrograde) with the rotation of the hole. We will see that the cross section depends on the initial helicity, and that the scattering process induces a net polarization in an unpolarized beam.

\begin{figure}
\begin{center}
\includegraphics[width=13cm]{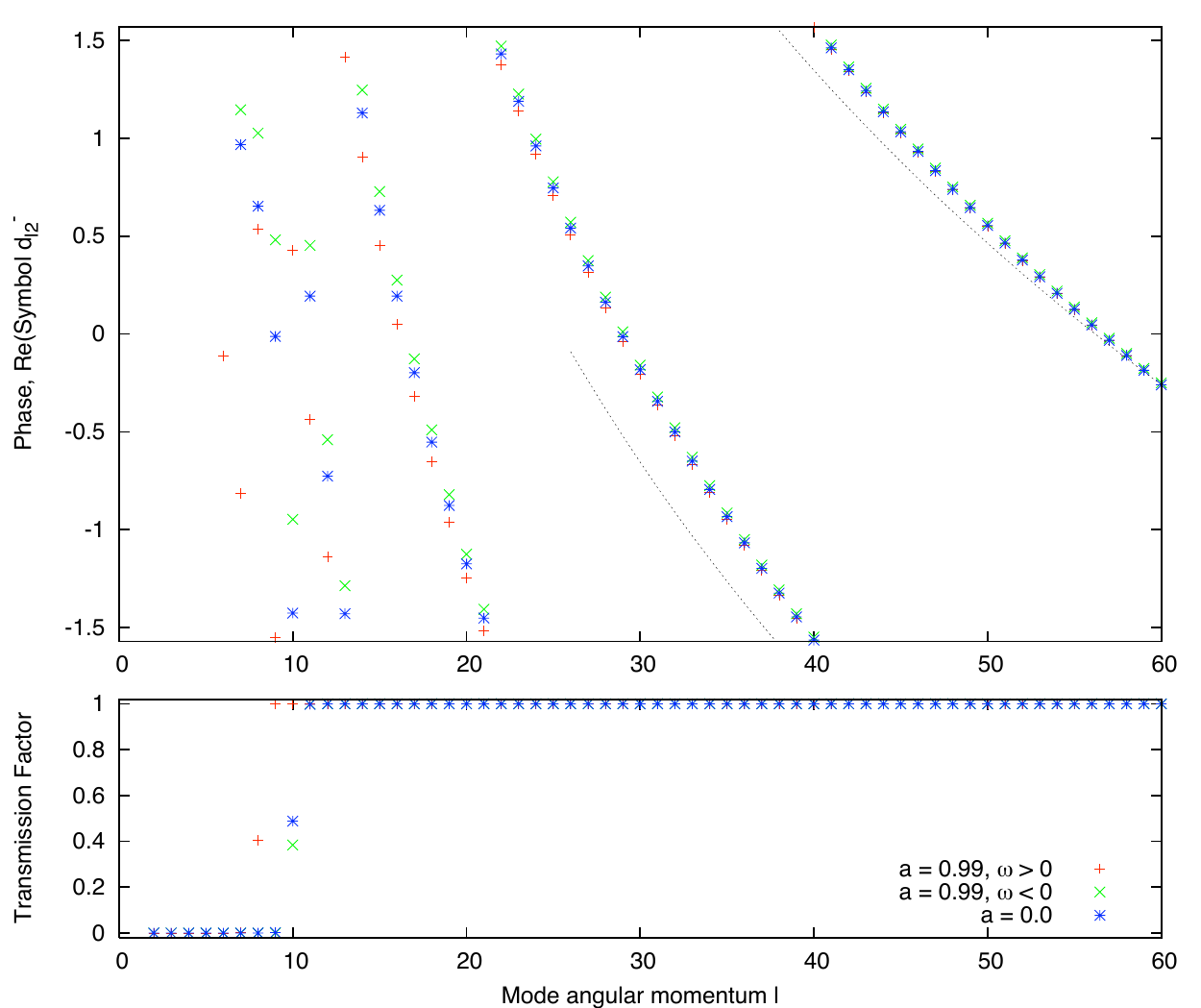}
\end{center}
\caption[]{\emph{Phase shifts and reflection factors for $2 \le l \le 60$ at $M\omega = 2$}. The upper plot compares the phase shifts for co-rotating [$\text{Re}(\delta_{l2\omega > 0}^-)$, $+$ symbol] and counter-rotating helicities [$-\text{Re}(\delta_{l2\omega < 0}^-)$, $\times$ symbol] for a fast-rotating hole ($a=0.99M$) with the Schwarzschild phase shifts [$\ast$ symbol]. The line shows the approximation (\ref{phase-shift-negative}) (with an arbitrary overall phase). The lower plot shows the proportion of each mode which is reflected.}
\label{fig-kerr-phases}
\end{figure}

Figure \ref{fig-kerr-phases} compares the (real part of) the phase shifts of `up' and `down' helicities, for a range of angular modes $2 \le l \le 60$ at $M\omega = 2$. The lower plot shows the proportion of each $l$-mode that is scattered rather than absorbed. The plot makes it clear that there is a critical value $\lc$: below, $l < \lc$, modes are absorbed; above, $l > l_c$, modes are reflected. Here, $\lc$ depends primarily on $M\omega$, but also on $\as$ and the helicity of the incident wave. $\lc$ is smaller for the co-rotating helicity than the counter-rotating helicity. The most significant differences in the `up' and `down' phase shifts occur close to $\lc$. At high $l \gg \lc$, the difference between the phase shifts is negligible. In the short-wavelength regime, the critical value $\lc$ may be estimated via semi-classical arguments \cite{Ford-1959}
\begin{equation}
b \approx \left( l + \frac{1}{2} \right) \frac{1}{\omega}  \quad \quad \Rightarrow \quad l_c \approx \omega b_c - \frac{1}{2}
\end{equation}
where $b_c$ is the critical impact parameter defined in (\ref{tbl-rcbc}). Numerical values for $b_c$ are listed in Table \ref{tbl-rcbc}. The value of $\lc$ determines the angular width of the spiral scattering and glory oscillations \cite{Glampedakis-2001}.

\begin{figure}
\begin{center}
\includegraphics[width=14cm]{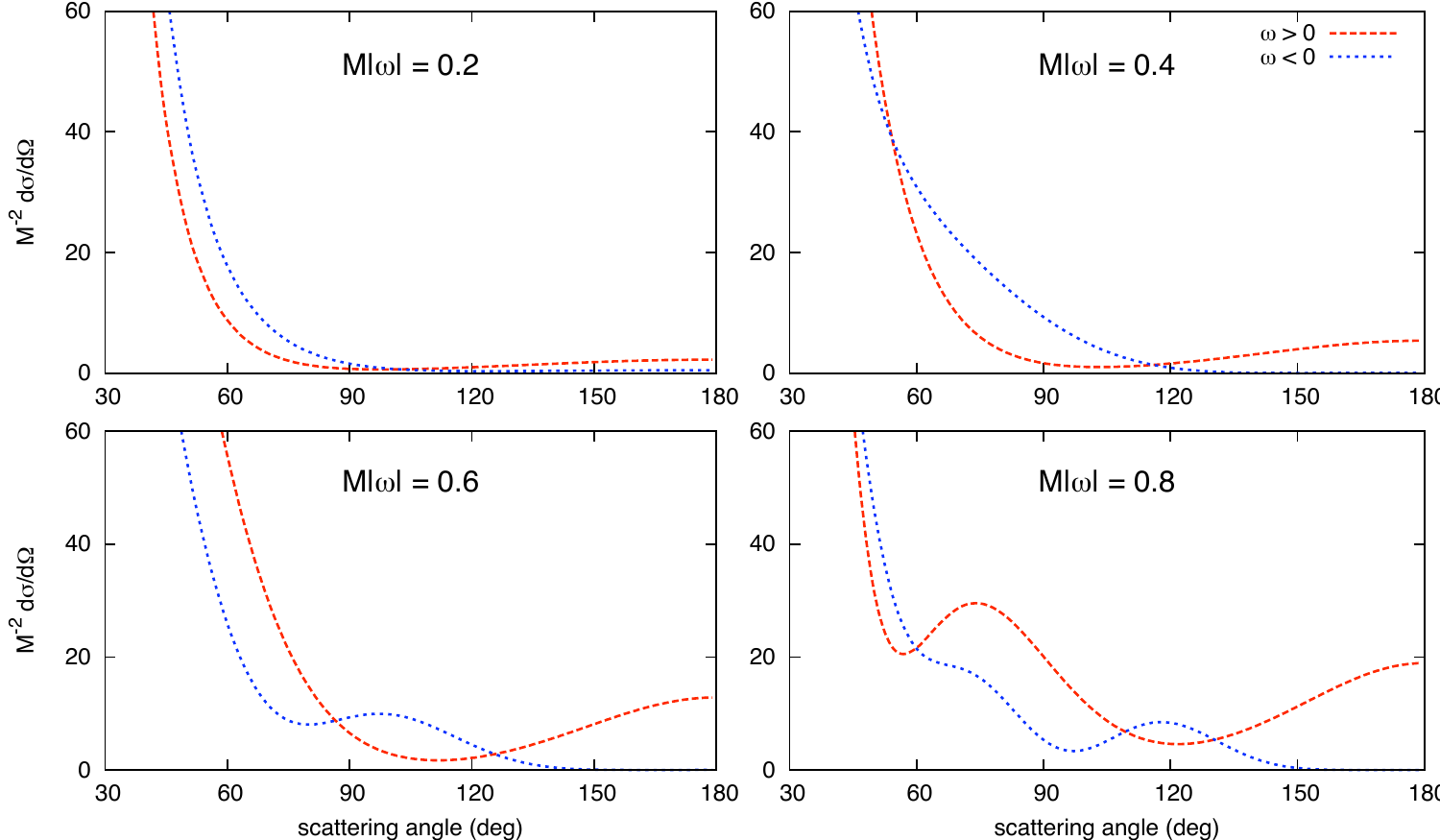}
\end{center}
\caption[]{\emph{Kerr scattering cross sections ($a = 0.99M$) at couplings, $0.2 \le M\omega \le 0.8$}. The dashed red line shows the cross section for the co-rotating helicity ($\omega > 0$), whereas the dotted blue line shows the cross section for the counter-rotating helicity ($\omega < 0$).}
\label{fig-csec-kerr-lowE}
\end{figure}

Figure \ref{fig-csec-kerr-lowE} compares the `up' (red) and `down' (blue) scattering cross sections across a range of couplings $0.2 \le |M\omega| \le 0.8$. Helicity has a substantial effect on the scattering pattern, particularly at large angles $> 30^\circ$. Superradiance from the $l = 2$, $m = 2$ mode enhances the amount of co-rotating flux that is scattered in the backward direction. 

\subsubsection{Flux in the Backward Direction $\theta = 180^\circ$.}
The backward-scattered flux arises entirely from the helicity-reversing amplitude $g(\theta = \pi)$ (Eq. \ref{g-def}), and hence the backscattered component has the opposite helicity to the incident wave. Rapid rotation $a \sim 1$ enhances the effect. A question naturally arises: what is the maximum possible cross section in the antipodal direction $\theta = 180^\circ$?

\begin{figure}
\begin{center}
\includegraphics[width=15cm]{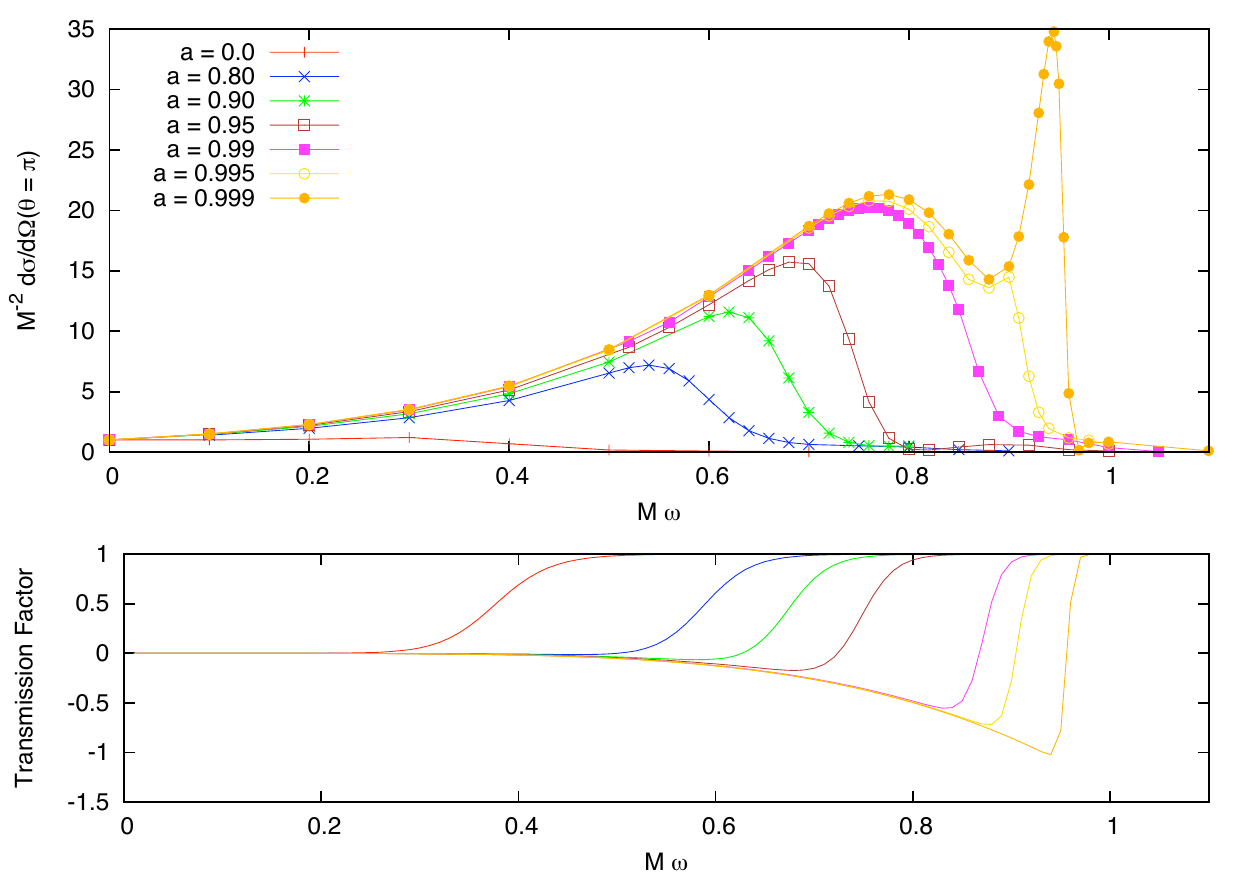}
\end{center}
\caption[]{\emph{Scattering cross section in the backwards direction for prograde wave.} The top plot shows the flux scatted in the direction opposite to incidence, $M^{-2} \frac{d \sig}{d \Omega}(\theta = \pi)$ as a function of $M\omega$, for a co-rotating wave. A range of rotation rates $a = 0, \ldots 0.999$ are plotted. The bottom plot shows the transmission factor for the $l = 2, m = 2$ mode.}
\label{fig-backflux}
\end{figure}

Figure \ref{fig-backflux} shows the cross section in the antipodal direction $M^{-2} |g(\pi)|^2$ as a function of $M \omega$, for a range of rotation rates $\as = 0.0 \ldots 0.999$. We observe an enhancement of $\sim 20.3$ times at $M \omega \approx 0.76$ for $a = 0.99M$, and an enhancement of $\sim 34.8$ times at $M \omega \approx  0.945$ for $a = 0.999M$. 

The transmission factor for the $l = 2$, $m = 2$ mode is also shown in Fig. \ref{fig-backflux}. It is clear that superradiance plays a role in increasing the back-scattered flux. Nevertheless, the maximum of $M^{-2} |g(\pi)|^2$ does not necessarily coincide with the superradiant peak. Intriguingly, for $a = 0.999M$, the backscattered component has two maxima, at $M\omega = 0.78$ and $M\omega = 0.945$. This behaviour can be understood by examining the phase relationship between the $l=2$ mode and the higher modes $l >2$. 

\begin{figure}
\begin{center}
\includegraphics[width=7.6cm]{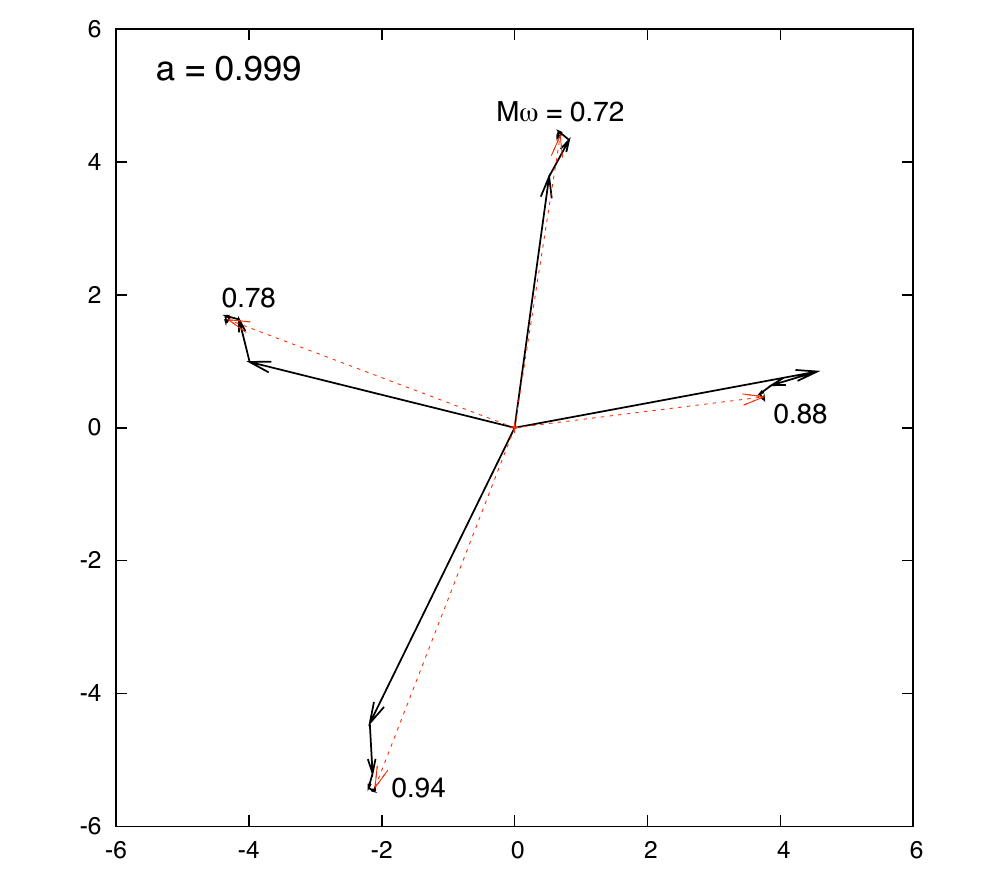}
\includegraphics[width=7.6cm]{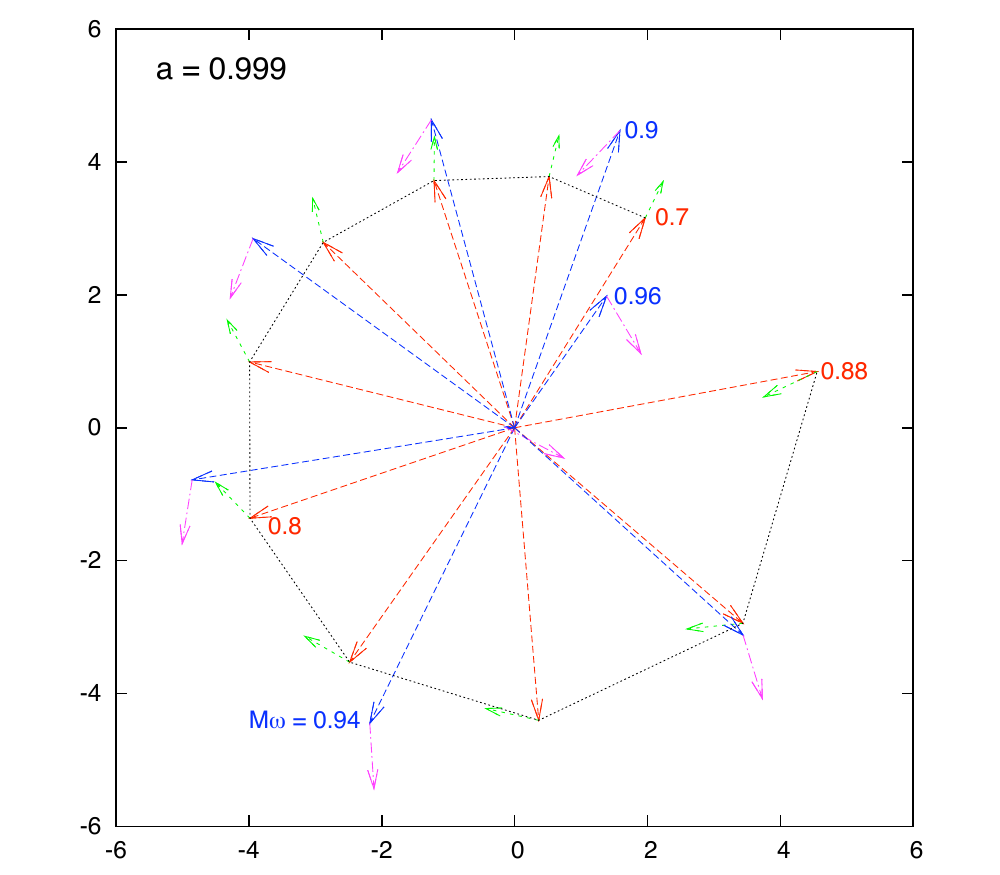}
\end{center}
\caption[]{\emph{Back-scattering amplitudes for $a = 0.999M$}. The complex amplitude $g(\theta = \pi)$ arises from a sum over modes, $g = \sum_{l=2} g_l$. The left plot shows $g$ (the dotted red arrow), and $g_2$ (the long black arrow) and $g_{l > 2}$ (shorter black arrows) in the Argand diagram for $M\omega = 0.72$, $0.78$, $0.88$ and $0.94$. The right plot shows $g_2$ (red and blue arrows) and $\sum_{l=3} g_l$ (green and purple arrows) for a range of couplings $M\omega = 0.7 \ldots 0.98$. The phase of $g_2$ changes more rapidly than the phase of higher modes $g_{l > 2}$, leading to interference effects and the double-peak structure in Fig. \ref{fig-backflux}.   }
\label{fig-arrows}
\end{figure}

Figure \ref{fig-arrows} illustrates how the helicity-reversed backscattering amplitude $g(\pi)$  is constructed from a sum over $l$, as $g(\theta = \pi) = \sum_{l=2} g_l$. For $a = 0.999M$, the contribution from $g_2$ is largest; nevertheless, the contribution from the higher modes $g_{l>2}$ is non-negligible, and the phase relationship between $l=2$ and the higher modes is responsible for the dual-peak structure seen in Fig. \ref{fig-backflux}. At $M\omega = 0.78$ and $M\omega = 0.945$ the contributions are roughly in-phase (i.e.  the arrows in Fig. \ref{fig-arrows} are roughly aligned), whereas at $M\omega = 0.88$, the contributions are roughly out-of-phase (i.e. arrows roughly anti-aligned). The transition from constructive ($M\omega = 0.78$) to destructive ($M\omega = 0.88$) interference and back again ($M\omega = 0.94$) occurs because the phase of $g_2$ changes more rapidly with $M \omega$ than the phase of $g_{l > 2}$. This is shown clearly in the right plot of Fig. \ref{fig-arrows}.

\subsubsection{Kerr scattering: short wavelengths $M\omega \gtrsim 1$.}
\begin{figure}
\begin{center}
\includegraphics[width=14cm]{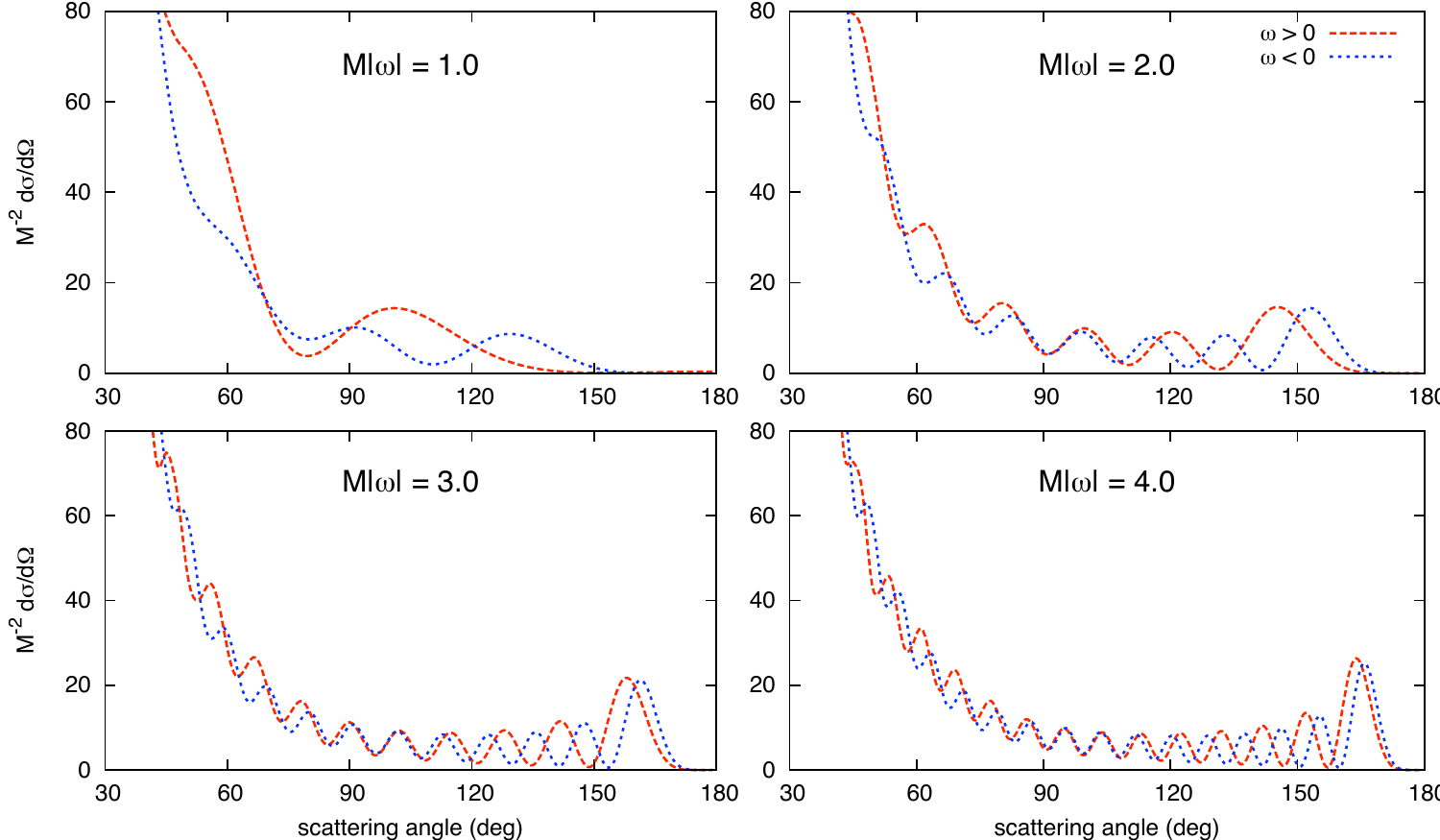}
\end{center}
\caption[]{\emph{Kerr scattering cross sections ($a = 0.99M$) at higher couplings ($1 \le M\omega \le 4$)}. The dashed red (dotted blue) line shows the cross section for the co-rotating (counter-rotating) helicity with $\omega > 0$ ($\omega < 0$).
}
\label{fig-csec-kerr-highE}
\end{figure}
Cross sections at higher couplings $1 \le M\omega \le 4$ are shown in Figure \ref{fig-csec-kerr-highE}. Above the critical superradiant frequency, $M\omega_c = a / r_+$, the flux scattered in the exact backward direction is negligible. The angular width of the spiral scattering oscillations is proportional to $1 / l_c$; hence the `up' helicity leads to wider-angle oscillations than the `down' helicity. The interaction between BH rotation and wave helicity decreases in significance as $M\omega \rightarrow \infty$.

\begin{figure}
\begin{center}
\includegraphics[width=14cm]{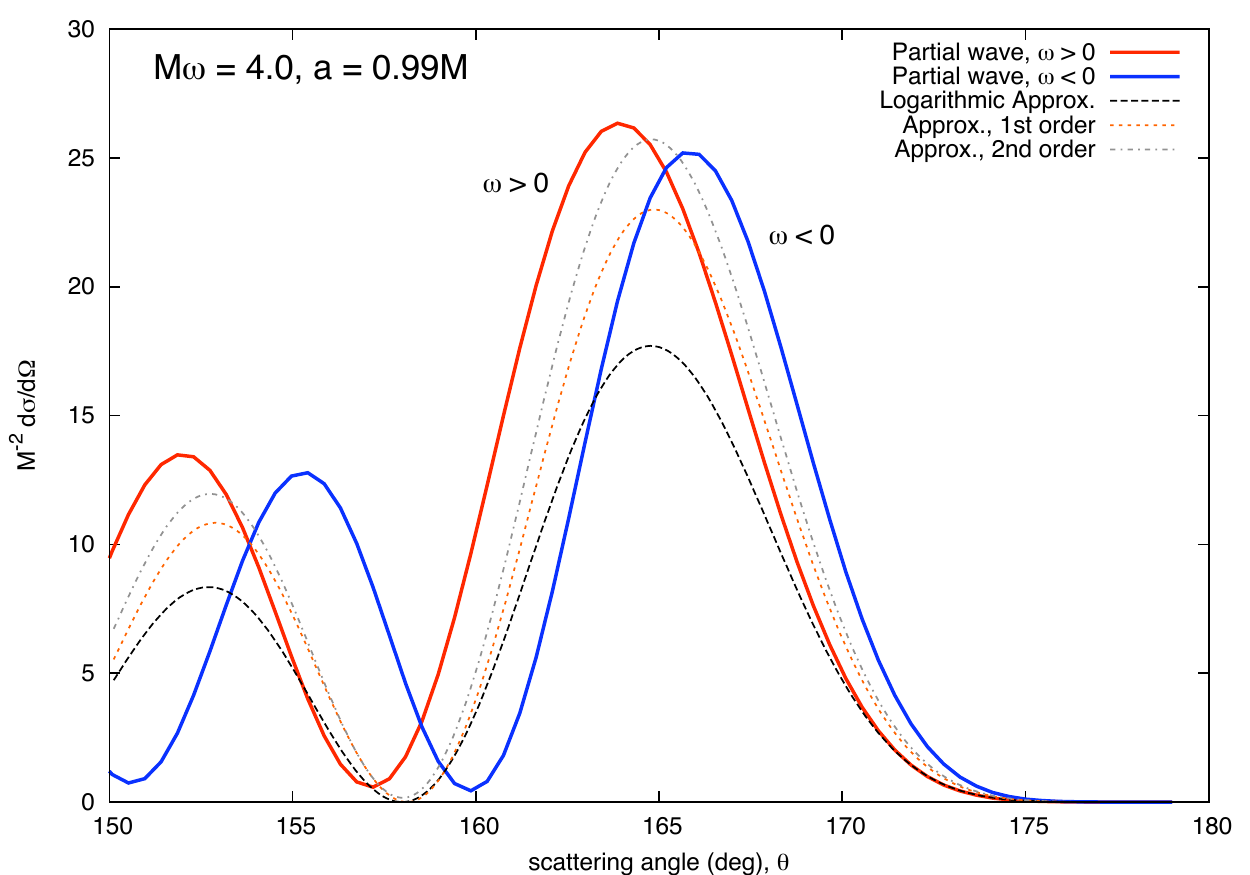}
\end{center}
\caption[]{\emph{Glory peaks for $M|\omega| = 4$ and $a = 0.99M$}. The solid lines show the partial-wave cross sections for the co-rotating ($\omega > 0$) and counter-rotating ($\omega < 0$) helicities. The broken lines show the semi-classical approximations derived in Sec. \ref{subsec-glory}. The most accurate approximation (labelled ``2nd order'') is found by including the additional contribution from rays which are scattered by $\theta \sim 3\pi$.}
\label{fig-glory-approx}
\end{figure}

In Figure \ref{fig-glory-approx} we examine the glory more closely for the special case $M \omega = 4.0$, $a = 0.99M$. The numerically-determined co-rotating and counter-rotating cross sections are shown as solid lines. The broken lines show the semi-classical glory approximations (\ref{glory-approx}). The dotted line was found using the logarithmic approximation to the scattering angle (\ref{log-approx}); the dashed line using exact solutions to the geodesic equations (\ref{R-Theta}). The dot-dash line shows that the glory approximation is improved if we include the second-order contribution from rays which orbit the hole one-and-a-half times. A significant deficiency of the approximation is that it does not take into account the helicity-rotation coupling that splits the co- and counter-rotating cases. In future, it would be interesting to model how geodesics are deflected by the effect of a spin-rotation coupling.  

\subsubsection{The extremal rotation limit, $\as \rightarrow 1$.} Numerical accuracy becomes harder to obtain as $\as \rightarrow 1$. The main source of error is in the numerical calculation of phase shifts. A small parameter $\nu = (r_+ - r_-)/2 \approx M \sqrt{2 (1-\as)}$ appears in the denominator in the near-horizon series expansions. To test the accuracy of the phase shifts, we tried changing the arbitrary parameters in the numerical code: the matching radii $r_m$, $r_{\text{mid}}$ and $r_\infty$. We found the change in phase to be invariant to at least one part in $10^{-5}$ for non-extremal $\as$. The error was confirmed to be less than the thickness of the lines in all the plots we have presented here. However, we found results beyond $\as > 0.9995$ to be less reliable. A separate analysis of the extremal case $\as = 1$ would be of interest.

\section{Discussion and Conclusions\label{sec-discussion}}
In this study we have shown that it is possible to compute accurate gravitational-wave cross sections using a partial-wave series approach. Although the key formulae (\ref{f-def}--\ref{T-trans}) were derived three decades ago by Matzner \emph{et al.} \cite{Matzner-1978, Handler-1980, Futterman-1988}, we believe this is the first time that accurate results for the rotating case have been presented.

Early studies \cite{Matzner-1978, Handler-1980, Futterman-1988} encountered two obstacles to progress. Firstly, the Teukolsky equation has a long-ranged potential for perturbations with spin $|s| > 1/2$. This leads to so-called `peeling' behaviour in the asymptotic solutions which renders the determination of phase shifts numerically difficult. Secondly, the partial wave series (\ref{f-def}) has an infinite number of terms which increase in magnitude with $l$, and it is divergent at $\theta = 0$. 

The first obstacle is overcome by transforming the radial equation. Matzner and Ryan \cite{Matzner-1978} used Press and Teukolsky's transformation \cite{Press-1973}; we used a Sasaki-Nakamura transformation \cite{Sasaki-1982}. Finding the phase shifts is then relatively straightforward -- although of course we benefit from the rapid advance in computing power over the last thirty years.

It is overcoming the second obstacle -- approximating an infinite, divergent series -- that is the key to numerical progress. Regardless of computing power, the series must be truncated at some $l = l_{\text{max}}$. Yet, the magnitude of the coefficients in the series (\ref{f-def}) increases with $l$. Matzner \emph{et al.} tried decomposing the phase shifts into a long-range `Newtonian' contribution (that varies logarithmically with $l$) and a short-range remainder (that decays with $l$). This approach works well for scalar waves \cite{Glampedakis-2001}, but it does not seem to work for higher spins \cite{Dolan-2006}. In truncating the series, Matzner \emph{et al.} introduce numerical error with an angular frequency proportional to $1 / l_\text{max}$. See for example Fig. 14 in \cite{Handler-1980}. In this work, we circumvented the obstacle by employing a series reduction method to improve the convergence properties of the series \cite{Yennie-1954}.

The results presented here shed new light on three intriguing scattering phenomena: polarization, helicity violation, and glory/spiral diffraction. Below we briefly recap the important points.

The coupling between the rotation of the black hole and the helicity of the incident wave means that a rotating black hole has a polarizing effect. In other words, since waves of co- and counter-rotating helicities are distinguished by a rotating black hole, a partial polarization is induced in an initially unpolarized beam. The polarization is a function of scattering angle. In the long-wavelength limit, the partial polarization is given by (\ref{pol-final}). At shorter wavelengths the polarization it is a more complicated function of scattering angle (see e.g. Fig. \ref{fig-csec-kerr-highE}), due to diffractive effects. 

The parity-dependence of the phase shifts leads to a scattered component $M^{-2} |g(\theta)|^2$ (Eq. \ref{g-def}) which has the opposite helicity to the incident flux. The helicity-reversed flux is largest in the antipodal direction ($\theta = 180^\circ \equiv \pi$). Conversely, the helicity-conserved flux $|f(\theta)|^2$ is zero in the backwards direction.  In the long-wavelength limit we showed that $M^{-2} |g(\pi)|^2 = 1 + 4 a \omega + \mathcal{O}(\omega^2)$.  In the non-rotating case, we find $|g(\pi)|^2$ has a maximum of just $M^{-2} |g(\pi)|^2 \sim 1.2$ at $M \omega = 0.3$ (Fig. \ref{fig-csec-schw-approx}), and it is suppressed at shorter wavelengths (Fig. \ref{fig-csec-schw2}). For a co-rotating incident helicity, the back-scattered flux is enhanced by superradiance in the $l = 2$, $m = 2$ mode, which occurs for frequencies $\omega < \omega_c = a / M r_+$. We find the backscattered flux may be enhanced by an order of magnitude or more as the extremal rotation limit is approached ($a \rightarrow 1$). In particular, for $\as = 0.999$ we observe a maximum enhancement of $M^{-2} |g(\pi)|^2 \sim 35$ at $M\omega \approx 0.945$. This contrasts with the scalar-wave ($s=0$) case, for which superradiance has a negligible effect \cite{Glampedakis-2001}.

At short wavelengths, $\lambda \lesssim r_S$ ($M\omega \gtrsim 1$), we observe regular peaks and troughs in the scattering cross sections (Figs. \ref{fig-csec-schw2}, \ref{fig-csec-schw-spins} and \ref{fig-csec-kerr-highE}). This is a diffraction effect, due to the interference of paths scattered through angles $\theta, 2\pi - \theta, 2\pi + \theta$ \cite{Anninos-1992}. The angular width of the oscillations is (approximately) proportional to the ratio of the inner unstable orbit radius to the incident wavelength. Hence, the angular width is roughly proportional to $M\omega$, but also depends on rotation $\as$ and incident helicity. Again, since co- and counter-rotating helicities are distinguished by a rotating black hole, the scattered wave acquires a partial polarization which depends on scattering angle.

Unlike light, gravitational radiation is able to `see through' the complicated and messy astrophysical region surrounding a black hole (e.g. accretion disk, jets, synchrotron radiation, etc) to probe the near-horizon geometry. Is there any prospect of detecting the gravitational radiation that is scattered by black holes? Since gravitational radiation has yet to be detected \emph{at all}, the short answer is: not yet! The best hope in the short term is that experimentalists detect the strongest signals generated by highly dynamic, strong-field interactions, such as the final merger of black holes. Dynamic (i.e. time-dependent) perturbations will excite black hole `quasinormal modes' (QNMs) \cite{Vishveshwara-1970b}. The frequencies and decay times of QNMs would provide a clear signature of the underlying black hole parameters $M$ and $\as$. 

Diffraction patterns like those shown in Figs. \ref{fig-csec-schw2}, \ref{fig-csec-schw-spins} and \ref{fig-csec-kerr-highE} may provide an alternative signature for future detectors. Let us imagine a supermassive black hole (such as Sag.~A*) illuminated by a strongly-collimated and long-lasting beam of gravitational radiation. Let us assume the incident radiation has a short wavelength $\lambda \ll r_s$ and a wide-band spectrum $\Delta \omega_0 \gg 1 / M$. Our angle of observation $\theta$ is effectively fixed, but we are free to vary the observation frequency. Changing the observation frequency by $\delta \omega \sim 1 / M$ would bring us from an interference maximum to a minimum \cite{Futterman-1988}, and, if the BH is rotating, the polarization of the signal would be modified. Hence the characteristic signature of BH scattering would be a signal whose intensity and polarization oscillates rapidly over a frequency $\delta \omega$. 

Two extensions of this work suggest themselves. Firstly, one could treat waves approaching from an arbitary angle of incidence relative to the BH rotation axis. The relevant partial wave formulae, given in \cite{Futterman-1988}, involve an additional sum over the azimuthal numbers $m$. We would expect to find maximal polarization for waves impinging along the axis of rotation, and zero polarization for waves approaching in the equatorial plane. Secondly, it would be relatively straightforward to apply the numerical method detailed here to the scattering of electromagnetic waves from a rotating hole, since the partial wave formulae are supplied in \cite{Futterman-1988}, and an extension of the Sasaki-Nakamura to $|s| = 1$ is available in \cite{Hughes-2000b}. In addition, it should also be straightforward to extend the long-wavelength approximation of section \ref{sec-long-wavelength} to the electromagnetic case. We hope this would shed light on the physical origin of the factor of two difference between (\ref{pol-final}) and the result (\ref{pol-guadagnini}) of Barbieri and Guadagnini \cite{Barbieri-2005} for scattering from classical rotating matter. We hope to undertake this calculation in the near future. 

\ack 
Thanks to Bahram Mashhoon for email correspondence, Lu\'is Crispino and Ednilton Oliveira for interesting discussions, and Marc Casals for proof-reading the manuscript. Thanks also to Calvin Smith, Paul Watts and Adrian Ottewill for helpful comments. Financial support from the Irish Research Council for Science, Engineering and Technology (IRCSET) is gratefully acknowledged.

\appendix

\section{Scattering Angle Approximation\label{appendix-darwin}}

The left-hand side of (\ref{R-Theta}) may be expressed in terms of elliptic integrals. For deflection angles close to $\theta = \pi$, we may expand in powers of $a/b$,
\begin{eqnarray}
b \Theta &= 2 K( a/b ) - F(\cos(\theta/2), a/b) \\ &\approx \frac{\theta + \pi}{2}  +  \left( \frac{\pi}{4}\frac{a^2}{b^2} + \frac{9 \pi}{64}\frac{a^4}{b^4} +  \ldots \right) + \frac{(\theta - \pi)^3}{48}\frac{a^2}{b^2} + \ldots  \label{Theta-expansion}
\end{eqnarray}
where $K(k)$ and $F(z, k)$ are complete and incomplete elliptic integrals of the first kind.

The right-hand side of (\ref{R-Theta}) may be expressed as 
\begin{equation}
- \int_{\infty}^{r_3} \frac{dr}{\sqrt{R}} = \frac{2 F(z, k) }{\sqrt{r_3 r_2 - r_3 r_0 - r_1 r_2 + r_1 r_0}}
\end{equation}
where
\begin{equation}
z^2 = \frac{r_2-r_0}{r_3-r_0}, \quad \text{and} \quad k^2 = \left( \frac{r_2-r_1}{r_3-r_1} \right) \left( \frac{r_3-r_0}{r_2-r_0} \right).
\end{equation}

We are interested in the strong-deflection limit $b \sim b_c$. In this case, the largest two roots $r_2$ and $r_3$ lie close to $r_c$. To derive a logarithmic approximation similar to Darwin's, let
\begin{eqnarray}
& r_2 = r_c(1 - \delta), \quad r_3 = r_c(1+\delta), \\ 
\Rightarrow \; & k \approx 1 - \frac{r_c}{r_c-r_0} \delta , \quad z \approx 1 - \frac{r_c(r_1-r_0)}{(r_c-r_1)(r_c-r_0)} \delta
\end{eqnarray}
where
\begin{equation}
\delta^2 \approx  \frac{(2r_c^2 - 4Mr_c + 2a^2)}{(6r_c^2 + 2a^2 - b_c^2)} \frac{b_c^2}{r_c^2} \left( \frac{b - b_c}{b_c} \right) + \mathcal{O}((b-b_c)^2) .
\end{equation}
Let us now make use of the asymptotic expansion
\begin{eqnarray}
F(z, k) \sim \ln \left( \frac{4z / z^\prime}{ 1 + (1 + k^{\prime 2} z^2 / z^{\prime 2})^{1/2}} \right)
\end{eqnarray}
where $z^\prime = \sqrt{1 - z^2}$ and $k^\prime = \sqrt{1 - k^2}$. Then the right-hand side of (\ref{R-Theta}) is approximately
\begin{eqnarray}
- \int_{\infty}^{R_3} \frac{dr}{\sqrt{R}} \approx \frac{\ln \left[ \frac{(1+\sqrt{1+\alpha})^4 r_c^2 \delta^2}{64(r_c-r_0)^2}  \right]}{2 \sqrt{(r_c-r_0)(r_c-r_1)}}  \label{R-expansion}
\end{eqnarray}
where $\alpha = (k^\prime/z^\prime)^2 = (r_1 - r_0) / (r_c - r_1)$.
Combining (\ref{Theta-expansion}) and (\ref{R-expansion}) we reach the conclusion that the scattering angle can be expressed in the logarithmic form (\ref{log-approx}), where
\begin{equation}
C(a) \approx \frac{b_c}{(r_c - R_0)(r_c - R_1)}
\label{Ca-eq}
\end{equation}
and
\begin{equation}
\fl \frac{1}{D(a)} \approx  \frac{(1+\sqrt{1+\alpha})^4 r_c^2 }{64(r_c-R_0)^2} \frac{(2r_c^2 - 4Mr_c + 2a^2)}{(6r_c^2 + 2a^2 - b_c^2)} \frac{Mb_c}{r_c^2} \exp\left( \frac{\pi}{C(a)} \left[ 1 + a^2/2b_c^2 + 9a^4/32b_c^4 \right] \right)
\label{Da-eq}
\end{equation}

\section{Calculation of Spin-Weighted Spherical Harmonics\label{appendix-spherical-harmonics}}
This section describes a stable method for accurate computation of the spherical harmonics of spin weight $s=-2$ using recurrence relations \cite{Hughes-2000}. 
The zero-spin-weight spherical harmonics are written
\begin{equation}
{}_0 Y_{lm}(\theta) = A(l,m) P_{lm}(\cos \theta), 
\end{equation}
where 
\begin{equation}
A(l,m) = \sqrt{\frac{2 l + 1}{4 \pi}} \sqrt{\frac{(l-m)!}{(l+m)!}} \, .
\end{equation}
The associated Legendre polynomials $P_{lm}(x)$ can be computed via the recurrence relation
\begin{equation}
(l - m) \, P_{lm}(x) = x\, (2l-1) P_{l-1,m} - (l+m-1)P_{l-2,m}  \label{LegP-1}
\end{equation}
with initial values
\begin{eqnarray}
P_{mm}(x)    &= (-1)^m (2m - 1)!! \, (1-x^2)^{m/2} \nn \\
P_{m+1,m}(x) &= x (2m + 1) P_{mm} . \label{LegP-2}
\end{eqnarray}
The operator $\check{\delta}$ lowers the spin weight of a function \cite{Newman-1966}, as follows
\begin{eqnarray}
\check{\delta}_s \, Y_{lm}(\theta) &\equiv -(\sin\theta)^{-s} \left[\frac{\partial}{\partial \theta} + \frac{m}{\sin \theta} \right] (\sin\theta)^s \, {}_sY_{lm}(\theta) \nn \\
 &= - \left[ (l+s)(l-s+1) \right]^{1/2} \, {}_{(s-1)}Y_{lm}(\theta) .
\end{eqnarray}
Hence the $s=-2$ functions are found by acting twice with a spin-lowering operator $\check{\delta}$ on the zero-spin-weight spherical harmonics ${}_0Y_{lm}$. This yields
\begin{equation}
\fl{}_{-2} Y_{lm}(\theta) = \frac{A(l,m)}{\sqrt{(l-1)l(l+1)(l+2)}} \left[(1-x^2) \frac{d^2}{dx^2} - 2m \frac{d}{dx} + \frac{m^2 - 2mx}{1-x^2} \right] P_{lm}(x) .
\end{equation}
To compute these harmonics, we need recurrence relations for $dP_{lm}/dx$ and $(1-x^2) d^2P_{lm} / dx$. These are straightforward to derive from (\ref{LegP-1}) and (\ref{LegP-2}):
\begin{eqnarray}
\fl\quad\quad\frac{d P_{lm}}{dx} &= \frac{1}{l-m} \left[ (2l-1)\left( P_{l-1,m} + x \frac{dP_{l-1,m}}{dx} \right) - (l+m-1) \frac{dP_{l-2,m}}{dx} \right] , \label{LegP-d} \\
\fl\quad\quad\frac{d P_{mm}}{dx} &= -mx(-1)^m (2m-1)!! (1-x^2)^{(m-2)/2} , \\
\fl\quad\quad\frac{d P_{m+1,m}}{dx} &= (2m+1) \left( P_{mm} + x \frac{dP_{mm}}{dx} \right), 
\end{eqnarray}
and
\begin{eqnarray}
\fl\quad\quad\frac{d^2 P_{lm}}{dx^2} = \frac{1}{l - m} \left[ (2l - 1) \left( 2 \frac{dP_{l-1,m}}{dx} + x \frac{d^2 P_{l-1,m}}{dx^2} \right) - (l+m-1) \frac{d^2 P_{l-2,m}}{dx^2} \right] , \label{LegP-dd} \\
\fl\quad\quad(1-x^2) \frac{d^2 P_{mm}}{dx^2} = m(-1)^m(2m-1)!!(1-x^2)^{(m-2)/2} \left( x^2(m-2) - (1-x^2) \right) , \\
\fl\quad\quad(1-x^2) \frac{d^2 P_{m+1,m}}{dx^2} = (2m+1) \left( 2(1-x^2) \frac{dP_{mm}}{dx} + x(1-x^2) \frac{d^2 P_{mm}}{dx^2} \right) .
\end{eqnarray}

\section{Coefficients in the Mano-Suzuki-Takasugi formalism\label{appendix-mst}}
The coefficients $a_n^{\nu}$ appearing in the MST formalism (\ref{Am-eq}--\ref{Ap-eq}) are 
\begin{eqnarray}
\fl a_{-2}^\nu &= - \frac{ (l-1+s)^2 (l+s)^2 \left[ (l - 1)\kappa - i m \astar \right] \left[ l\kappa - i m \astar \right]}{4(l-1)l^2(2l-1)^2(2l+1)} \, \eps^2 + \mathcal{O}(\eps^3), \label{a-nu-m2} \\
\fl a_{-1}^\nu &= i \frac{(l+s)^2 \left[l\kappa - i m \astar\right]}{2l^2(2l+1)} \, \eps - \frac{(l+s)^2}{2l^2(2l+1)} \left[ 1 + i \frac{ l \kappa - i m \astar }{(l-1) l^2 (l+1)} m \astar s^2 \right] \eps^2 + \mathcal{O}(\eps^3) , \\
\fl a_0^\nu &= 1, \\
\fl a_1^\nu &= i \frac{(l+1-s)^2 \left[(l+1) \kappa + i m \astar\right]}{2(l+1)^2(2l+1)} \, \eps + \frac{(l+1-s)^2}{2(l+1)^2(2l+1)} \left[1 - i \frac{(l+1)\kappa + i m \astar}{l(l+1)^2(l+2)} m \astar s^2 \right] \eps^2 , \\
\fl a_2^\nu &= - \frac{(l+1-s)^2(l+2-s)^2 \left[(l+1)\kappa + i m \astar\right] \left[(l+2)\kappa + i m \astar \right]}{4 (l+1)^2 (l+2) (2l+1) (2l+3)^2} \, \eps^2 + \mathcal{O}(\eps^3)  \label{a-nu-2} .
\end{eqnarray}

With the Cordon-Shortley phase convention, the Clebsch-Gordan coefficients appearing in (\ref{deqs1}--\ref{deqs2}) are
\begin{eqnarray}
\left< l, 1, 2, 0 | l + 1, 2 \right> &= \left[ (l-1)(l+3)/(2l+1)(l+1) \right]^{1/2} , \label{cleb1} \\
\left< l, 1, 2, 0 | l , 2 \right> &=  \left[4 / l(l+1) \right]^{1/2},  \\
\left< l, 1, 2, 0 | l - 1, 2 \right> &=  -\left[(l-2)(l+2) / (2l+1)l \right]^{1/2}  ,
\end{eqnarray}
and
\begin{eqnarray}
\left< l, 2, 2, 0 | l+2, 2 \right> &= \left[ 6 \frac{(2l)!}{(2l+4)!} \frac{(l+4)!}{(l+2)!} \frac{(l!}{(l-2)!} \right]^{1/2} ,\\
\left< l, 2, 2, 0 | l+1, 2 \right> &= \left[ \frac{12}{(2l+1)} \frac{(l-1)(l+3)}{l (l+1)(l+2)}  \right]^{1/2} ,\\
\left< l, 2, 2, 0 | l, 2 \right> &=  \frac{-2(l-3)(l+4)}{\left[ (2l-1)(2l)(2l+2)(2l+3) \right]^{1/2} } ,  \\
\left< l, 2, 2, 0 | l-1, 2 \right> &= - \left[ \frac{12}{(2l+1)} \frac{(l-2)(l+2)}{(l-1)l(l+1)} \right]^{1/2}, \\ 
\left< l, 2, 2, 0 | l-2, 2 \right> &= \left[ 6 \frac{(2l-3)!}{(2l+1)!} \frac{(l+2)!}{l!} \frac{(l-2)!}{(l-4)!} \right]^{1/2}  .\label{cleb2}
\end{eqnarray}

The expansion coefficients $\{ d_{-2}^{(0)},  d_{-1}^{(0)}, d_{-1}^{(1)}, d_{0}^{(2)}, d_{+1}^{(0)}, d_{+1}^{(1)}, d_{+2}^{(0)}  \}$ appearing in (\ref{bdef}) are
\begin{eqnarray}
d_{+1}^{(0)}(l) &= 2 \left[ (2l+1) (2l+3) \right]^{-1/2}  \frac{(l-1)(l+3)}{(l+1)^2} , \label{d-explicit-1}  \\
d_{-1}^{(0)}(l)  &= -2 \left[ (2l+1) (2l-1) \right]^{-1/2} \frac{(l-2)(l+2)}{l^2} , \\
d_{+1}^{(1)}(l) &= 4 \left[ (2l+1) (2l+3) \right]^{-1/2} \frac{(l-1)(l+3)((l+1)^2 - 8)}{l(l+1)^4(l+2)} , \\
d_{-1}^{(1)}(l) &= -4 \left[ (2l+1) (2l-1) \right]^{-1/2} \frac{(l-2)(l+2)(l^2 - 8)}{(l-1)l^4(l+1)} , \\
d_{+2}^{(0)}(l) &= \frac{1}{2} \left[ (2l+1) (2l+5) \right]^{-1/2} \frac{ (l-1)l(l+3)(l+4)((l+9)}{(2l+3)^2 (l+1)^2 (l+2)} , \\
d_{-2}^{(0)}(l) &= -\frac{1}{2} \left[ (2l+1) (2l-3) \right]^{-1/2} \frac{(l-3)(l-2)(l+1)(l+2)(l-8)}{(2l-1)^2 (l-1) l^2} .  \label{d-explicit-2}
\end{eqnarray}

\section{Functions in the Sasaki-Nakamura equation\label{appendix-sas}}
The function $F(r)$ appearing in (\ref{sas-eqn}) is given by
\begin{equation}
F(r) = \frac{d\eta / dr}{\eta} \frac{\Delta}{r^2 + a^2}
\end{equation}
where
\begin{equation}
\eta(r) = c_0 + c_1 / r + c_2 / r^2 + c_3 / r^3 + c_4 / r^4
\end{equation}
and
\begin{eqnarray}
c_0 &= -12i\omega M + \lambda(\lambda + 2) - 12 a \omega (a \omega - m), \nn \\
c_1 &= 8 i a \left[ 3 a \omega - \lambda(a \omega - m) \right],  \nn \\
c_2 &= -24 i a M (a \omega - m) + 12 a^2 \left[ 1-2(a \omega - m)^2 \right], \nn \\
c_3 &= 24 i a^3(a \omega - m) - 24M a^2, \nn \\
c_4 &= 12 a^4 .   \label{c-co-defn}
\end{eqnarray}
The function $U(r)$ appearing in (\ref{sas-eqn}) is
\begin{equation}
U(r) = \frac{\Delta U_1(r)}{(r^2+a^2)^2} + G(r)^2 + \frac{\Delta \, \frac{dG}{dr}}{r^2 + a^2} - F(r)G(r)
\end{equation}
where
\begin{eqnarray}
G(r)   &= - \frac{2 (r-M)}{r^2 + a^2} + \frac{ r \Delta }{(r^2 + a^2)^2} ,\\
U_1(r) &= V(r) + \frac{\Delta^2}{\beta} \left[ \frac{d}{dr} \left( 2 \alpha + \frac{d \beta / dr}{\Delta} \right) - \frac{ d\eta / dr}{\eta} \left( \alpha + \frac{d \beta / dr}{\Delta} \right) \right] .
\end{eqnarray}
The function $K(r)$ was defined in Eq. (\ref{K-defn}), and $V(r)$ is the Teukolsky potential,
\begin{equation}
V(r) = - \frac{K^2 + 4i(r-M)K}{\Delta} + 8 i \omega r + \lambda_{lm} .
\end{equation}


\begin{thebibliography}{10}

\bibitem{Lorimer-1998}
D.~R. Lorimer.
\newblock Binary and Millisecond Pulsars.
\newblock {\em Living Rev. Relativity} {\bf 1}, 10 (1998).

\bibitem{Frey-2007}
R.~E. Frey {\em et al.}
\newblock LIGO: Status and recent results.
\newblock {\it AIP Conf. Proc.} {\bf 928}, 11 (2007).

\bibitem{LIGO-2007}
{LIGO Scientific Collaboration} and K.~{Hurley}.
\newblock Implications for the origin of {GRB 070201} from {LIGO} observations.
\newblock [arXiv:0711.1163] (2007).

\bibitem{Danzmann-2003}
K. Danzmann and A. R\"udiger.
\newblock LISA technology -- concept, status, prospects.
\newblock {\it Class. Quantum Grav.} {\bf 20}, S1 (2003). 

\bibitem{Pretorius-2005}
F. Pretorius.
\newblock Evolution of binary black hole spacetimes.
\newblock {\em Phys. Rev. Lett.}, {\bf{95}}:121101 (2005). [gr-qc/0507014].

\bibitem{Hildreth-1964}
W.~W. Hildreth.
\newblock {\em The Interaction of Scalar Gravitational Waves with the
  {Schwarzschild} Metric}.
\newblock PhD thesis, Princeton University (1964).

\bibitem{Matzner-1968}
R.~A. Matzner.
\newblock Scattering of massless scalar waves by a {Schwarzschild}
  ``singularity''.
\newblock {\em J. Math. Phys.}, {\bf{9}}:163--170 (1968).

\bibitem{Chrzanowski-1976}
P.~L. Chrzanowski, R.~A. Matzner, M.~P. Ryan, and V.~D. Sandberg.
\newblock Zero-mass plane waves in nonzero gravitational backgrounds.
\newblock {\em Phys. Rev. D}, {\bf{14}}:317--326 (1976).

\bibitem{Matzner-1977}
R.~A. Matzner and M.~P. Ryan.
\newblock Low-frequency limit of gravitational scattering.
\newblock {\em Phys. Rev. D}, {\bf{16}}:1636--1642 (1977).

\bibitem{Matzner-1978}
R.~A. Matzner and M.~P. Ryan.
\newblock Scattering of gravitational radiation from vacuum black holes.
\newblock {\em Astrophys. J. Suppl.}, {\bf{36}}:451--481 (1978).

\bibitem{Handler-1980}
F.~A. Handler and R.~A. Matzner.
\newblock Gravitational wave scattering.
\newblock {\em Phys. Rev. D}, {\bf{22}}:2331--48 (1980).

\bibitem{Sanchez-1976}
N.~G. S\'anchez.
\newblock Scattering of scalar waves from a {S}chwarzschild black hole.
\newblock {\em J. Math. Phys.}, {\bf{17}}:688 (1976).

\bibitem{Sanchez-1977}
N.~G. S\'anchez.
\newblock Wave scattering and absorption problem for a black hole.
\newblock {\em Phys. Rev. D}, {\bf{16}}:937--945 (1977).

\bibitem{Sanchez-1978a}
N.~G. S\'anchez.
\newblock Absorption and emission spectra of a {Schwarzschild} black hole.
\newblock {\em Phys. Rev. D}, {\bf{18}}:1030---1036 (1978).

\bibitem{Sanchez-1978b}
N.~G. S\'anchez.
\newblock Elastic scattering of waves by a black hole.
\newblock {\em Phys. Rev. D}, {\bf{18}}:1798--1804 (1978).

\bibitem{Mashhoon-1973}
B.~Mashhoon.
\newblock Scattering of Electromagnetic Radiation from a Black Hole.
{\it Phys.\ Rev.}\ D {\bf 7}, 2807--2814 (1973). 

\bibitem{Mashhoon-1974}
B.~Mashhoon.
\newblock Electromagnetic scattering from a black hole and the glory effect.
{\it Phys.\ Rev.}\ D {\bf 10}, 1059--1063 (1974). 

\bibitem{Mashhoon-1975}
B.~Mashhoon.
\newblock Influence of gravitation on the propagation of electromagnetic radiation.
{\it Phys.\ Rev.}\ D {\bf 11}, 2679--2684 (1975). 

\bibitem{Dolan-2006}
S.~R. Dolan, C.~J.~L. Doran, and A.~N. Lasenby.
\newblock Fermion scattering by a {S}chwarzschild black hole.
\newblock {\em Phys. Rev. D}, {\bf{74}}:064005 (2006). [gr-qc/0605031].

\bibitem{Fabbri-1975}
R. Fabbri.
\newblock Scattering and absorption of electromagnetic waves by a
  {Schwarzschild} black hole.
\newblock {\em Phys. Rev. D}, {\bf{12}}:933--942 (1975).

\bibitem{Futterman-1981}
J.~A.~H. Futterman.
\newblock {\em The scattering of massless plane waves by rotating black holes}.
\newblock PhD thesis, {University of Texas, Austin} (1981).

\bibitem{DeLogi-1977}
W.~K. de~Logi and S.~J. Kov\'acs.
\newblock Gravitational scattering of zero-rest-mass plane waves.
\newblock {\em Phys. Rev. D}, {\bf{16}}:237 (1977).

\bibitem{Guadagnini-2002}
E. Guadagnini.
\newblock Gravitational deflection of light and helicity asymmetry.
{\it Phys.\ Lett.}\ B {\bf 548}, 19--23 (2002). [gr-qc/0207036].

\bibitem{Barbieri-2004}
A. Barbieri and E. Guadagnini.
\newblock Gravitational optical activity.
{\it Nucl.\ Phys.}\ B {\bf 703}, 391--399 (2004). 

\bibitem{Barbieri-2005}
A. Barbieri and E. Guadagnini. 
\newblock Gravitational helicity interaction.
{\it Nucl. Phys.}\ B {\bf 719}, 53--66 (2005). [gr-qc/0504078]. 

\bibitem{Guadagnini-2008}
E. Guadagnini. 
\newblock Gravitons scattering from classical matter.
\newblock {\it Class. Quantum Grav.} {\bf 25}, 095012 (2008). [arXiv:0803.2855].

\bibitem{DeWitt-Morette-1984}
C. DeWitt-Morette and B.~L. Nelson.
\newblock Glories -- and other degenerate points of the action.
\newblock {\em Phys. Rev. D}, {\bf{29}}:1663--1668 (1984).

\bibitem{Zhang-1984}
T-R. Zhang and C. DeWitt-Morette.
\newblock {WKB} cross section for polarized glories of massless waves in curved
  space-times.
\newblock {\em Phys. Rev. Lett.}, {\bf{52}}:2313--2316 (1984).

\bibitem{Matzner-1985}
R.~A. Matzner, C. DeWitt-Morette, B. Nelson, and T-R. Zhang.
\newblock Glory scattering by black holes.
\newblock {\em Phys. Rev. D}, {\bf{31}}:1869--1878 (1985).

\bibitem{Anninos-1992}
P. Anninos, C. DeWitt-Morette, R.~A. Matzner, P. Yioutas, and T-R. Zhang.
\newblock Orbiting cross sections: Application to black hole scattering.
\newblock {\em Phys. Rev. D}, {\bf{46}}:4477--4494 (1992).

\bibitem{Futterman-1988}
J.~A.~H. Futterman, F.~A. Handler, and R.~A. Matzner.
\newblock {\em Scattering from black holes}.
\newblock Cambridge University Press (1988).

\bibitem{Andersson-1995}
N. Andersson.
\newblock Scattering of massless scalar waves by a {Schwarzschild} black hole:
  A phase-integral study.
\newblock {\em Phys. Rev. D}, {\bf{52}}:1808--1820 (1995). 

\bibitem{Glampedakis-2001}
K. Glampedakis and N. Andersson.
\newblock Scattering of scalar waves by rotating black holes.
\newblock {\em Class. Quantum Grav.}, {\bf{18}}:1939--1966 (2001). [gr-qc/0102100].

\bibitem{Westervelt-1971}
P.~J. Westervelt.
\newblock Scattering of electromagnetic and gravitational waves by a static
  gravitational field: Comparison between the classical (general-relativistic)
  and quantum field-theoretic results.
\newblock {\em Phys. Rev. D}, {\bf{3}}:2319 (1976).

\bibitem{Peters-1976}
P.~C. Peters.
\newblock Differential cross sections for weak-field gravitational scattering.
\newblock {\em Phys. Rev. D}, {\bf{13}}:775--777 (1976).

\bibitem{Doran-2002}
C.~J.~L. Doran and A.~N. Lasenby.
\newblock Perturbation theory calculation of the black hole elastic scattering
  cross section.
\newblock {\em Phys. Rev. D}, {\bf{66}}:024006 (2002). [gr-qc/0106039].

\bibitem{Dolan-2008}
S.~R. Dolan.
\newblock Scattering of long-wavelength gravitational waves.
\newblock {\em Phys. Rev. D}, {\bf{77}}:044004 (2008). [arXiv:0710.4252]

\bibitem{Laven-2005}
P. Laven.
\newblock How are glories formed?
\newblock {\em Appl. Opt.}, {\bf{44}}:5675-5683 (2005).

\bibitem{Darwin-1959}
C. Darwin.
\newblock The gravity field of a particle. 
\newblock {\em Proc. R. Soc. London A} {\bf{249}}, 180 (1959).

\bibitem{Chandrasekhar-1983}
S. Chandrasekhar.
\newblock {\em The Mathematical Theory of Black Holes}.
\newblock Oxford University Press (1983).

\bibitem{Berti-2005}
E. Berti, V. Cardoso, and M. Casals.
\newblock Eigenvalues and eigenfunctions of spin-weighted spheroidal harmonics
  in four and higher dimensions.
\newblock {\em Phys. Rev. D}, {\bf{73}}:024013 (2006). [gr-qc/0511111].

\bibitem{Teukolsky-1972}
S.~A. Teukolsky.
\newblock Rotating black holes: separable wave equations for gravitational and
  electromagnetic perturbations.
\newblock {\em Phys. Rev. Lett.}, {\bf{29}}:1114--1118 (1972).

\bibitem{Press-1973}
W.~H. Press and S.~A. Teukolsky.
\newblock Perturbations of a rotating black hole. {II}. {D}ynamical stability
  of the {Kerr} metric.
\newblock {\em Astrophys. J.}, {\bf{185}}:649--673 (1973).

\bibitem{Teukolsky-1973}
S.~A. Teukolsky.
\newblock {Perturbations of a rotating black hole. I. {F}undamental equations
  for gravitational, electromagnetic and neutrino-field perturbations}.
\newblock {\em Astrophys. J.}, {\bf{185}}:635--647 (1973).

\bibitem{Teukolsky-1974}
S.~A. Teukolsky and W.~H. Press.
\newblock Perturbations of a rotating black hole. {III}. {I}nteraction of the
  hole with gravitational and electromagnetic radiation.
\newblock {\em Astrophys. J.}, {\bf{193}}:443--461 (1974).

\bibitem{Newman-1962}
E. Newman and R. Penrose.
\newblock An approach to gravitational radiation by a method of spin
  coefficients.
\newblock {\em J. Math. Phys.}, {\bf{3}}:566--578 (1962).

\bibitem{Seidel-1989}
E. Seidel.
\newblock A comment on the eigenvalues of spin-weighted spheroidal functions.
\newblock {\em Class. Quantum Grav.}, {\bf{6}}:1057--1062 (1989).

\bibitem{Casals-2004}
M. Casals and A.~C. Ottewill.
\newblock High frequency asymptotics for the spin-weighted spheroidal equation.
\newblock {\em Phys. Rev. D}, {\bf{71}}:064025 (2005). [gr-qc/0409012].

\bibitem{Leaver-1985}
E.~W. Leaver.
\newblock An analytic representation for the quasi-normal modes of {K}err black
  holes.
\newblock {\em Proc. R. Soc. London A}, {\bf{402}}:285 (1985).

\bibitem{Sasaki-1982}
M.~Sasaki and T.~Nakamura.
\newblock Gravitational radiation from a {Kerr} black hole. {I}. {F}ormulation
  and a method for numerical analysis.
\newblock {\em Prog. Theor. Phys.}, {\bf{67}}:1788 (1982).

\bibitem{Hughes-2000}
S.~A. Hughes.
\newblock Evolution of circular, non-equatorial orbits of {Kerr} black holes
  due to gravitational wave emission.
\newblock {\em Phys. Rev. D}, {\bf{61}}:084004 (2000). [gr-qc/9910091]. Errata: {\em ibid.} {\bf 63}, 049902 (2001); {\bf 65}, 069902 (2002); {\bf{67}}, 089901 (2003).

\bibitem{Hughes-2000-erratum}
S.~A. Hughes.
\newblock Erratum: Evolution of circular, nonequatorial orbits of Kerr black holes due to gravitational-wave emission [Phys. Rev. D 61, 084004 (2000)]. 
\newblock {\em Phys. Rev. D}, in press.

\bibitem{Abramowitz-1965}
M.~Abramowitz and I.~A.~Stegun.
\newblock {\em Handbook of Mathematical Functions with Formulas, Graphs and Mathematical Tables.}
\newblock Dover, New York (1965).

\bibitem{MST}
S. Mano, H. Suzuki and E. Takasugi.
\newblock Analytic Solutions of the Teukolsky equation and their low frequency expansions.
{\it Prog. Theor. Phys.} {\bf 95}, 1079-1096 (1996). [gr-qc/9603020].

\bibitem{MT}
S. Mano and E. Takasugi.
\newblock Analytic Solutions of the Teukolsky Equation and Their Properties.
{\it Prog. Theor. Phys.} {\bf 97}, 213--232 (1997). [gr-qc/9611014].

\bibitem{Sasaki-2003}
M.~Sasaki and H.~Tagoshi.
\newblock Analytic black hole perturbation approach to gravitational radiation. 
{\it Living Rev. Relativity} {\bf 6}, 6 (2003). [http://www.livingreviews.org/lrr-2003-6].

\bibitem{Poisson-1995}
E.~Poisson and M.~Sasaki.
\newblock Gravitational radiation from a particle in a circular orbit around a
  black hole. {V}. {B}lack-hole absorption and tail corrections.
\newblock {\em Phys. Rev. D}, {\bf{51}}:5753--5767 (1995). [gr-qc/9412027].

\bibitem{Goldberg-1967}
J.~Goldberg, A.~Macfarlane, E.~Newman, F.~Rohrlich and E.~Sudarshan,
\newblock Spin-s Spherical Harmonics and Edth. 
\newblock {\em J. Math. Phys.}, {\bf{8}}:2155, 1967.

\bibitem{Regge-1957}
T. Regge and J.~A. Wheeler.
\newblock Stability of a {Schwarzschild} singularity.
\newblock {\em Phys. Rev.}, {\bf{108}}:1063--1069 (1957).

\bibitem{Yennie-1954}
D.~R. Yennie, D.~G. Ravenhall, and R.~N. Wilson.
\newblock Phase-shift calculation of high-energy electron scattering.
\newblock {\em Phys. Rev.}, {\bf{95}}:500 (1954).

\bibitem{Unruh-1976-absorption}
W.~G. Unruh.
\newblock Absorption cross section of small black holes.
\newblock {\em Phys. Rev. D}, {\bf{14}}:3251---3259 (1976).

\bibitem{Crispino-2007}
L.~C.~B. Crispino, E.~S. Oliveira, A. Higuchi and G.~E.~A. Matsas.
\newblock Absorption cross section of electromagnetic waves for Schwarzschild black holes.
\newblock {\em Phys. Rev. D}, {\bf{75}}:104012 (2007).

\bibitem{Ford-1959}
K.~W. Ford and J.~A. Wheeler.
\newblock Applications of semiclassical scattering analysis.
\newblock {\it Annals of Physics} {\bf 7} (1959).

\bibitem{Vishveshwara-1970b}
C.~V. Vishveshwara.
\newblock Scattering of gravitational radiation by a {Schwarzschild} black
  hole.
\newblock {\em Nature}, {\bf{227}}, 936 (1970).

\bibitem{Hughes-2000b}
S.~A. Hughes.
\newblock Computing radiation from {Kerr} black holes: Generalization of the
  {Sasaki-Nakamura} equation.
\newblock {\em Phys. Rev. D}, {\bf{62}}:044029 (2000). [gr-qc/0002043]. Erratum: {\em Phys. Rev. D} {\bf{67}}:089902 (2003).

\bibitem{Newman-1966}
E.~T. Newman and R. Penrose.
\newblock Note on the {Bondi-Metzner-Sachs} group.
\newblock {\em J. Math. Phys.}, {\bf{3}}:566 (1966).




\end{thebibliography}

\section*{References}
\bibliographystyle{unsrt}

\end{document}